\documentclass{aa}  
\usepackage{graphicx}
\usepackage{txfonts}
\usepackage{enumitem}
\usepackage{tabularx}

\usepackage{fixltx2e}
\usepackage{color}
\usepackage{lscape}
\usepackage{longtable}
\usepackage{multirow}

\begin{document}

\title{J-PLUS: 2-D analysis of the stellar population in NGC 5473 and NGC 5485}

\author{I.~San Roman\inst{\ref{a1}}\fnmsep\thanks{e-mail: isanroman@cefca.es}, P. S\'{a}nchez-Bl\'{a}zquez\inst{2},  A.~J.~Cenarro\inst{1}, L.~A.~D\'{i}az-Garc\'{i}a\inst{1}, C.~L\'{o}pez-Sanjuan\inst{1}, J. Varela\inst{1}, G. Vilella-Rojo\inst{1}, S.~Akras\inst{\ref{a6}}, S. Bonoli\inst{1}, A.~L.~Chies Santos\inst{\ref{a7}}, P.~Coelho\inst{\ref{a3}}, A.~Cortesi\inst{\ref{a3}},  A.~Ederoclite\inst{1}, Y. Jim\'{e}nez-Teja\inst{\ref{a6}}, R. Logro\~{n}o-Garc\'{i}a\inst{1}, R. Lopes de Oliveira\inst{\ref{a4},\ref{a6},\ref{a5},\ref{a8}},  J.~P.~Nogueira-Cavalcante\inst{\ref{a6}}, A. Orsi\inst{1}, H. V\'{a}zquez Rami\'{o}\inst{1}, K. Viironen\inst{1}, D. Crist\'{o}bal-Hornillos\inst{1}, R. Dupke\inst{\ref{a6}}, A. Mar\'{i}n-Franch\inst{1}, C. Mendes de Oliveira\inst{\ref{a3}}, M. Moles\inst{1} and  L. Sodr\'{e}\inst{\ref{a3}}}

    \institute{Centro de Estudios de F\'{i}sica del Cosmos de Arag\'{o}n (CEFCA), Unidad Asociada al CSIC, Plaza San Juan 1, E-44001 Teruel, Spain\label{a1}
 \and
 Departamento de F\'{i}sica Te\'{o}rica, Universidad Autonoma de Madrid (UAM-CSIC), 28049 Cantoblanco, Madrid, Spain\label{a2}
 \and
Observat\'{o}rio Nacional/MCTI, Rua Gal. Jos\'{e} Cristino, 77, S\~ao Crist\'ov\~ao 20921-400 Rio de Janeiro, RJ, Brazil\label{a6}
\and
Departamento de Astronomia, Instituto de F\'{i}sica, Universidade Federal do Rio Grande do Sul (UFRGS), Porto Alegre, RS, Brazil\label{a7}
\and
Instituto de Astronomia, Geof\'{i}sica e Ci\^{e}ncias Atmosf\'{e}ricas (IAG), Universidade de S\~{a}o Paulo (USP), R. do Mat\~{a}o 1226, 05508-090, S\^{a}o Paulo, Brazil\label{a3}
\and
Departamento de F\'{i}sica, Universidade Federal de Sergipe, Av. Marechal Rondon, S/N, 49000-000 S\~ao Crist\'ov\~ao, SE, Brazil\label{a4}
\and
X-ray Astrophysics Laboratory, NASA Goddard Space Flight Center, Greenbelt, MD 20771, USA\label{a5}
\and
Department of Physics, University of Maryland, Baltimore County, 1000 Hilltop Circle, Baltimore, MD 21250, USA\label{a8}
}

\titlerunning {2D study of J-PLUS galaxies} 
\authorrunning {San Roman et al.}               

\abstract{ The spatial variations of stellar population properties within a galaxy are intimately related to their formation process. 
 Therefore, spatially resolved studies of galaxies are essential to uncover their formation  and assembly. Although the arrival of integral field unit (IFU) surveys has brought a significant breakthrough in the field,  recent techniques that combine photometric multi-filter surveys  with spectral fitting diagnostics have opened a new  way  to  disentangle   the   stellar   population   of   spatially-resolved galaxies with a relatively low-cost compared to IFU surveys. 
}     
         {The Javalambre Photometric Local Universe Survey (J-PLUS) is a dedicated multi-filter designed to observed $\sim$ 8500 deg$^{2}$ of the Northern sky using twelve narrow-, intermediate- and broad-band filters in the optical range. In this study, we test the potential of the multi-filter observation carried out with J-PLUS to investigate the properties of spatially-resolved nearby galaxies.}
         {We present detailed 2D maps of stellar population properties (age, metallicity, extinction, and stellar mass surface density) for two early-type galaxies observed in both, J-PLUS and CALIFA surveys: NGC 5473 and NGC 5485. Radial structures are also compared and luminosity- and mass-weighted profiles are derived. We use MUFFIT to process the J-PLUS photometric multi-filter observations, and two different techniques (STARLIGHT and STECKMAP) to analyze IFU CALIFA data.} 
         {
        We demonstrate that this novel technique delivers radial stellar population gradients in good agreement with the IFU technique CALIFA/STECKMAP although comparison of the absolute values reveals the existence of intrinsic systematic differences. Radial stellar population gradients differ when CALIFA/STARLIGHT methodology is used. Age and metallicity radial profiles derived from J-PLUS/MUFFIT are very similar when luminosity- or mass-weighted properties are used, suggesting that the contribution of a younger component is small and the star formation history of these early-type galaxies are well represented by mainly an old SSP component.}
         {We present the potential of J-PLUS to explore the unresolved stellar populations of spatially-extended local galaxies. Comparison between the three methodologies reveals some discrepancies suggesting that the specific characteristics of each method causes important differences. We conclude that the ages, metallicities and extinction derived for individual galaxies not only depend on the chosen models but also depend on the method used. Future work is required to evaluate in detail the origin of these differences and to quantify the impact that different fitting routines have on the derived stellar population properties.}

\keywords{galaxies: evolution - galaxies: formation - galaxies: photometry - galaxies: elliptical}

\maketitle

\section{Introduction}

The study of the stellar content of galaxies is crucial to unveil their formation and assembly. In the last fifteen years, the field has witnessed the outbreak of integral field spectroscopy (IFS) surveys (\citealp[SAURON,][]{deZeeuwetal2002}; \citealp[VENGA,][]{Blancetal2010};  \citealp[PINGS,][]{RosalesOrtegaetal2010}; \citealp[DiskMass,][]{Bershadyetal2010}; \citealp[ATLAS$^{3D}$,][]{Cappellarietal2011}; \citealp[CALIFA,][]{Sanchezetal2012}). While large surveys of galaxies such as the Sloan Digital Sky Survey \citep[SDSS;][]{Yorketal2000}, the Galaxy and Mass Assembly project \citep[GAMA;][]{Driveretal2011}, or the 2dF Galaxy Redshift Survey \citep[2dFGRS;][]{Collessetal2001} obtain one spectrum per galaxy, IFS surveys spectrally map galaxies pixel by pixel. These IFS  surveys allow detailed spatial analyses through multiple spectra of each galaxy by creating a 2D map of the object. While these surveys are very powerful, they are still limited in number of galaxies and in redshift range (e.g., CALIFA  observes $\sim$650 galaxies with redshifts limited to $z$ < 0.03). Currently, a new generation of multiplexed IFS surveys, which can observe many galaxies simultaneously, has become a reality (\citealp[SAMI,][]{Croometal2012}; \citealp[MaNGA,][]{Bundyetal2015}).  Although this technique has allowed to increase significantly the number of galaxies, there are still limitations in terms of the redshift range probed and the galactocentric distance analyzed (i.e., few effective radii, R$_\mathrm{eff}$) . For example, MaNGA aims to obtain spatially resolved spectroscopy of 10,000 galaxies but it will be limited to resolve galaxies spatially out to R=1.5 R$_\mathrm{eff}$  (with a subsample reaching R=2.5 R$_\mathrm{eff}$) and with a median redshift of $z$ $\sim$ 0.03 \citep{Bundyetal2015}. Redshifts are limited to $z$ < 0.095 in the SAMI survey and the data typically reach 1.7 R$_\mathrm{eff}$ (2 R$_\mathrm{eff}$ for 40$\%$ of the sample).

Recent hydrodynamical simulations  find that the information content of the accretion history is retained in the stellar population profiles only at very large radii (R > 2 R$_\mathrm{eff}$) from the galactic center \citep{Cooketal2016}. The limitations of current IFU surveys at these low signal-to-noise (S/N) regimes suggest that deep photometric studies in galactic stellar halos are essential to unveil the formation and assembly of local galaxies.

On the other hand, the number of alternative techniques such as multi-filter surveys is significantly increasing (e.g., \citealp[COMBO-17,][]{Wolfetal2003}; \citealp[ALHAMBRA,][]{Molesetal2008}; \citealp[PAU,][]{Castanderetal2012}; \citealp[SHARDS,][]{PerezGonzalezetal2013}: \citealp[J-PAS,][]{Benitezetal2014}; \citealp[J-PLUS,][hereafter Paper I]{Cenarroetal2017}). These photometric surveys aim at a diversity of scientific goals but with a common characteristic: a well sampled spectral energy distribution (SED) of galaxies using broad-, intermediate- and/or narrow-band filters in the optical range. Half-way between classical photometry and standard spectroscopy, these retrieved SEDs are, effectively, spectra with a low-spectral resolution depending on the filter system (e.g., R $\sim$ 20 for ALHAMBRA; R $\sim$ 50 for J-PAS). Although multi-filter observing techniques suffer from the lack of high spectral resolution, their advantages over standard spectroscopy are multiple: 1) IFU-like character, allowing a pixel-by-pixel investigation of extended galaxies; 2) a uniform and non-biased spatial sampling that allows environmental studies; 3) larger galaxy samples than multi-object spectroscopic surveys; 4) no sample selection criteria other than the photometric depth in the detection band; and 5) analysis of lower brightness surface areas than in spectroscopy, allowing the studies of the outermost regions of the galaxies and of galaxies at higher redshifts (i. e., multi-filter surveys are generally deeper than traditional spectroscopic studies since direct imaging is more efficient than spectroscopy). 
It also allows studies of very nearby galaxies ($z$ < 0.01) that are too spatially extended to be suitable for the small field of view of current IFU surveys. In this context multi-filter surveys open a way to improve our knowledge of galaxy formation and evolution that complements standard multi-object spectroscopic surveys. 
 
\citet{Sanromanetal2018} developed a novel technique to analyze unresolved stellar populations of spatially resolved galaxies based on photometric multi-filter surveys. In that work, we applied the technique to a sample of 29 massive (M$_{\star}$ > 10$^{10.5}$ M$_{\sun}$) early-type galaxies at z < 0.3 from the ALHAMBRA survey \citep{Molesetal2008} to derive stellar population and extinction gradients out to 2 -- 3.5 R$_\mathrm{eff}$. We found, on average, flat luminosity-weighted age gradients ($\nabla$log Age$_\mathrm{L}$ = 0.02 $\pm$ 0.06 dex/R$_\mathrm{eff}$) and negative luminosity-weighted gradients in metallicity  ($\nabla$[Fe/H]$_\mathrm{L}$ = -- 0.09 $\pm$ 0.06 dex/R$_\mathrm{eff}$). Although these results are in agreement with previous long-slit analyses  \citep[e.g.][]{Mehlertetal2003, SanchezBlazquezetal2006, SanchezBlazquezetal2007, Redaetal2007, Spolaoretal2010} and also with the most recent IFU studies \citep[e.g.][]{Rawleetal2008, Rawleetal2010, Kuntschneretal2010, Wilkinsonetal2015, Goddardetal2016}, they are discrepant when compared with some recent results. Most studies in the literature have found either flat or slightly positive age gradients in early-type galaxies, however recent IFU works present disparate results \citep[see Table 3 in][for a comprehensive review]{Sanromanetal2018}.  
In particular, the results of \citet{GonzalezDelgadoetal2015} found, using a sample of 41 early-type galaxies from the CALIFA survey, very negative inner (< R$_\mathrm{eff}$) luminosity-weighted age gradients ($\sim$ -- 0.25 dex/R$_\mathrm{eff}$) that become flatter ($\sim$ -- 0.05 dex/R$_\mathrm{eff}$) at larger galactocentric distances (up to 2 R$_\mathrm{eff}$). Most recently, \citet{Boardmanetal2017} observed twelve H I-detected early-type galaxies and found median age gradients of -- 0.047 dex/dex (in log-space),  reaching approximately 3 half-light radii. IFU MaNGA studies reveal contradictory results; while \citet{Goddardetal2016} found flat age gradient inside R < R$_\mathrm{eff}$, \citet{Zhengetal2016} found a slightly negative gradient (--0.05 $\pm$ 0.01 dex/R$_\mathrm{eff}$). Both MaNGA studies analyze similar galaxy sample but using different spectral fitting techniques and stellar population models.  To shed light into this problem, in this paper we propose the analysis of common objects observed with the photometric multi-filter J-PLUS and IFU CALIFA but analyzed with different techniques. 

The Javalambre Photometric Local Universe Survey (J-PLUS, Paper I) is a photometric multi-filter survey defined to observe $\sim$8500 deg$^{2}$ of the Northern sky. Combination of J-PLUS observations with spectral fitting diagnostics will disentangle the stellar population of spatially extended galaxies.  On the other side,  the Calar Alto Legacy Integral Field Area (CALIFA) survey is a pioneer in the integral field spectroscopy legacy projects. Recent studies using data from CALIFA provided the most comprehensive results so far regarding the radial variations of the stellar population parameters and star formation histories of nearby galaxies \citep[e.g.,][]{Perezetal2013, Sanchezetal2014, SanchezBlazquezetal2014}.

The specific goals of this paper are: 1)  to illustrate the potential of J-PLUS to analyze unresolved stellar populations of spatially extended local galaxies, and 2) to compare our methodology \citep[MUFFIT,][]{DiazGarciaetal2015} applied to J-PLUS data with two different ones applied to CALIFA data: STARLIGHT \citep{CidFernandesetal2013} and STECKMAP \citep{Ocvirketal2006a, Ocvirketal2006b}. 
 
This paper is organized as follows. Section \ref{sec:data} provides a brief overview of the J-PLUS Science Verification Data (SVD) and the CALIFA survey as well as the photometric properties of our sample. In Sect. \ref{sec:method}, we describe the  technical aspects of the methodologies used to analyze the different data sets. Section \ref{sec:map}  presents the J-PLUS and CALIFA 2D maps of age, [Fe/H], A$_\mathrm{v}$, and stellar mass density and Sect. \ref{sec:rad}  presents the radial profiles and gradients. The stellar mass-to-light ratio analysis and the integrated properties of the sample are presented in Sect. \ref{sec:ML}  and Sect. \ref{sec:int}, respectively. We discuss the results in Sect. \ref{sec:discussion}. Throughout this paper we assume a $\Lambda$CDM cosmology with $H_0 = 70 $~km s$^{-1}$, $\Omega_\mathrm{M}=0.30$, and $\Omega_\mathrm{\Lambda}=0.70$.

\section{Observations and Data Reduction}\label{sec:data}
\subsection{J-PLUS SVD}
\begin{figure}
\begin{center}
\includegraphics[bb=30 40 1050 980, width=1.0\columnwidth]{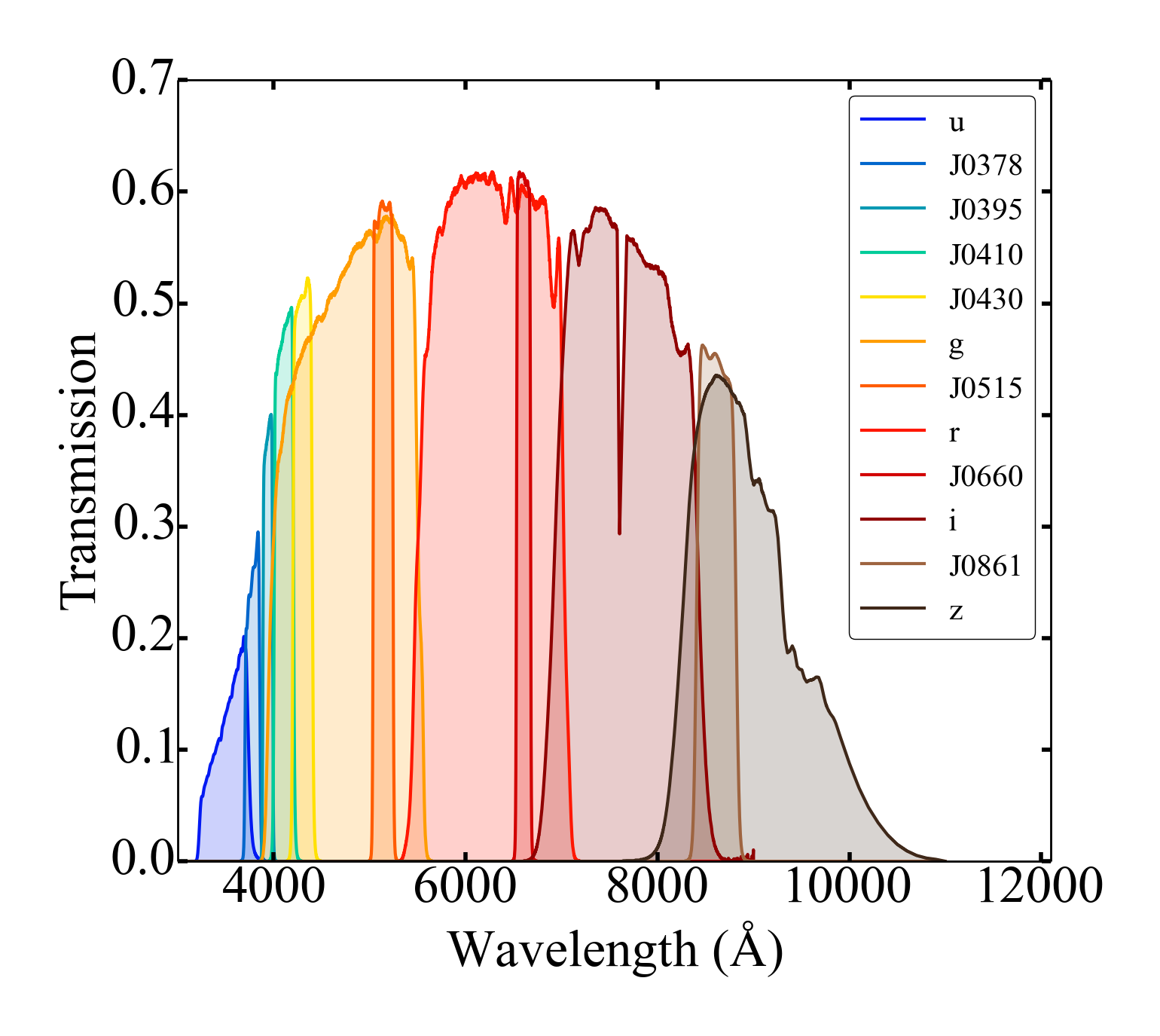}
\caption{Transmission curves of the J-PLUS filter system. The curves are computed after accounting for the effects of both the efficiency of the CCD and the atmospheric extinction.} 
\label{fig:transmission}
\end{center}
\end{figure}

J-PLUS is a multi-filter survey carried out with the Javalambre Auxiliary Survey Telescope (JAST/T80), a 0.83m telescope installed at the Observatorio Astrof\'{i}sico de Javalambre (OAJ) in Teruel, Spain. The survey uses the panoramic camera T80Cam that provides a large field of view of 2 deg$^{2}$ with a pixel scale of 0.55" pixel$^{-1}$. J-PLUS was primarily conceived to perform the calibration tasks for the main J-PAS\footnote{The Javalambre Physics of the Accelerating Universe Astrophysical Survey (J-PAS) is a very wide-field cosmological survey to be conducted from the OAJ with the 2.5m Javalambre Survey Telescope, JST/T250, and the panoramic camera JPCam (4.7 deg$^{2}$ field of view). It will cover 8500 deg$^{2}$ with an unprecedented filter set of 54 contiguous, narrow band optical filters (145 $\AA$ width each, placed $\sim$ 100 $\AA$ apart) plus two broad filters at the blue and red sides of the optical range, and 3 SDSS-like filters.} survey that will observe a contiguous area of 8500 deg$^{2}$. The specially designed filter system will cover the optical range with twelve broad-, intermediate-, and narrow-band filters. The photometric filter set is composed of 4 broad ($g$, $r$, $i$, and $z$), 2 intermediate ($u$ and $J0861$), and 6 narrow-band ($J0378$, $J0395$, $J0410$, $J0430$, $J0515$, and $J0660$) filters optimized to provide an adequate sampling of the SED. Figure \ref{fig:transmission} shows the transmission curves of the complete set of filters. The final survey parameters and scientific goals, as well as the technical requirements of the filter set, are described in Paper I. In this paper, we make use of observations collected during the science verification phase of J-PLUS (1500041, P.I.: G. Vilella). These observations are available through the J-PLUS web page\footnote{http://j-plus.es/datareleases} and are part of J-PLUS early data release (EDR). In addition to the present paper, the J-PLUS EDR and SVD have so far been used to refine the membership in nearby galaxy clusters \citep{Molinoetal2018}, analyze the globular cluster M15 \citep{Bonattoetal2018}, study the H$\alpha$ emission of several local galaxies \citep{LogronoGarciaetal2018}, and compute the stellar and galaxy number counts up to $r = 21$ \citep{LopezSanjuanetal2018}.

Data processing and calibration is carried out using an automatized pipeline developed and implemented at the Centro de Estudios del Cosmos de Arag\'{o}n (CEFCA)\footnote{http://www.cefca.es/}. The data processing includes standard steps such as overscan subtraction, flat-field correction, and rejections of  bad pixels and cosmic rays. If needed, fringe corrections are applied to the images. The pipeline makes use of the packages \texttt{Scamp} \citep{Bertin2006} and \texttt{Swarp} \citep{Bertinetal2002} to perform the astrometric calibration and image coadding. The photometric calibration is performed through a series of calibration procedures (e.g., based on SDSS observations and spectrophotometric standard stars) rather than relying on a single calibration technique. More technical details involved in the data processing and calibration procedure can be found in Paper I. Table \ref{tab:1} summarizes the journal of observations and provides basic information on the filters used. We note that due to the science verification nature of the observations, the exposure times are  different from the general observation conditions of J-PLUS. In spite of these peculiarities, SVD is representative of the whole J-PLUS survey and analysis presented in this paper are directly applicable to future 2D J-PLUS studies (see details in Paper I). 

\begin{figure*}[h]
\begin{center}
\includegraphics[width=0.45\textwidth]{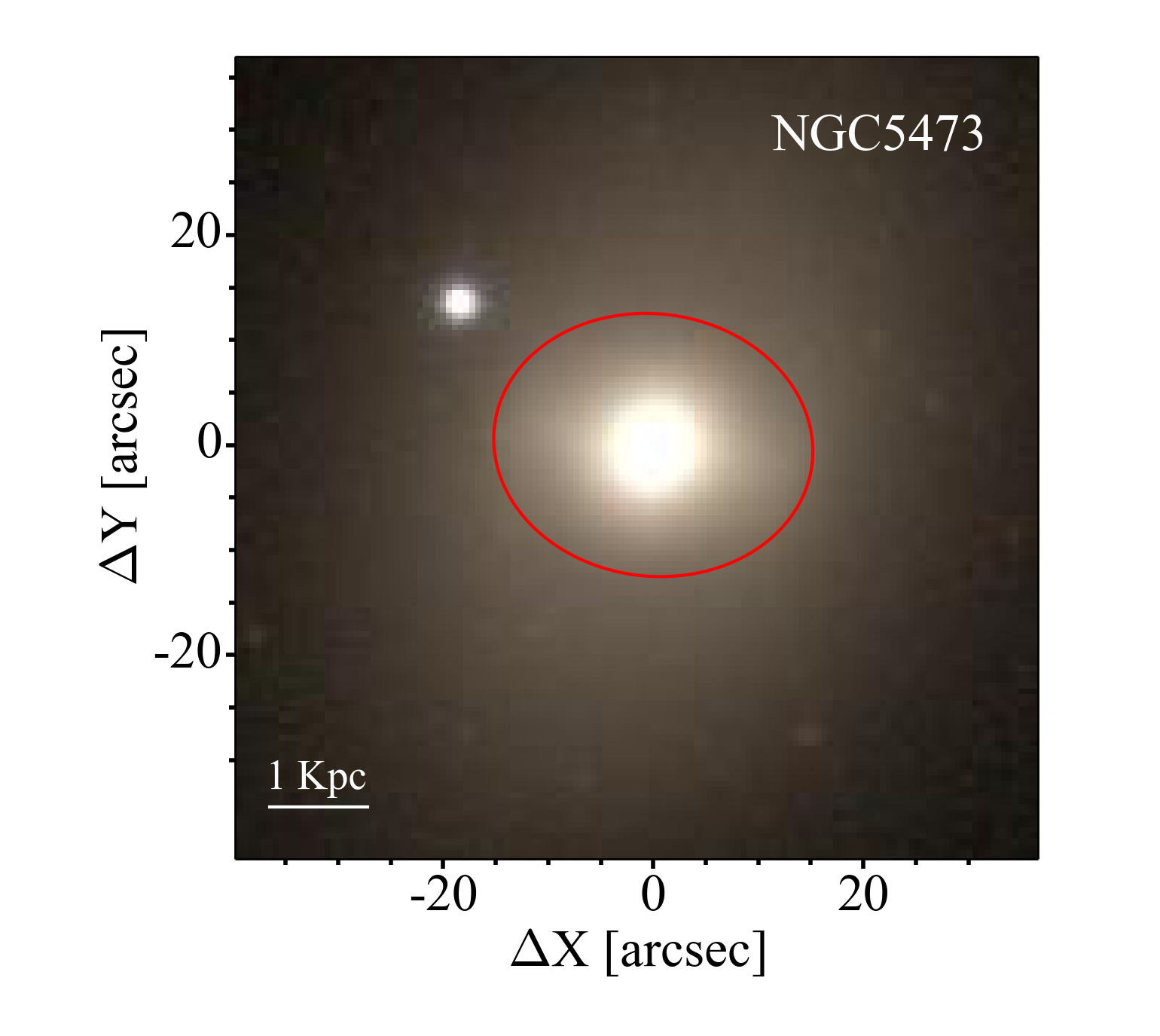}
\includegraphics[width=0.45\textwidth]{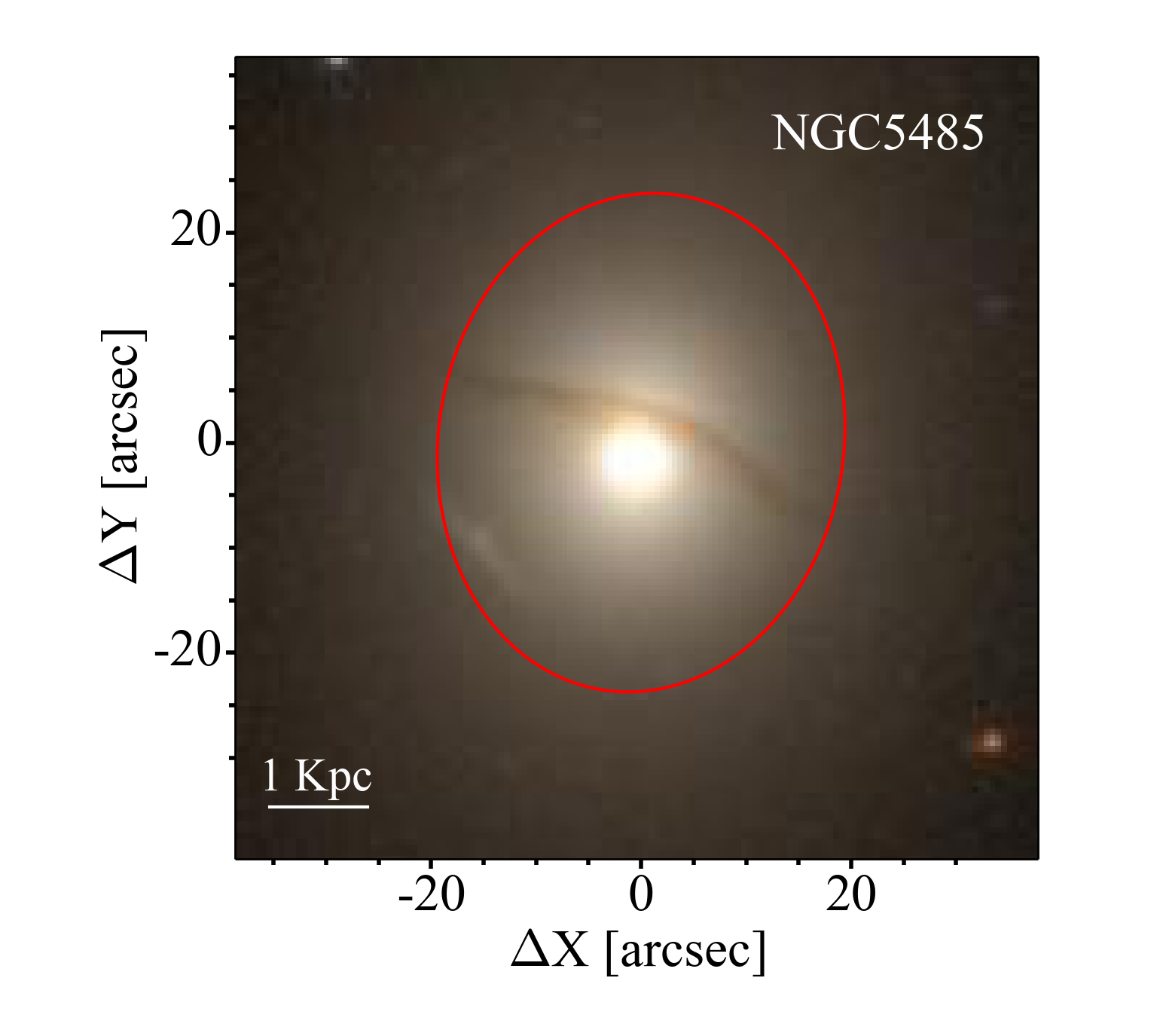}
\caption{J-PLUS colored composite images of the objects analyzed in this paper. The red solid line in each panel delimits a ellipse with the semi-major axis R=3 R$_\mathrm{eff}$.}
\label{fig:1}
\end{center}
\end{figure*}

The target sample of this paper was selected exclusively to study common objects observed with both J-PLUS and CALIFA surveys. For the purposes of this early paper, we focus on the analysis of spheroidal, large and bright galaxies. This selection criteria restrict this work to the study of the only two early-type galaxies, NGC 5473 and NGC 5485.  We note that there are no objects in common between J-PLUS and any other IFU survey at the moment (e.g., MaNGA DR14, SAMI DR1, Sauron) except for CALIFA and ATLAS$^{3D}$. The ATLAS$^{3D}$ survey also presents IFU data for the two objects analyzed in this paper.  Although no direct comparison is made with these observations due to the limited covered area (< 1 R$_\mathrm{eff}$) and the small wavelength range (480 -- 538 nm) of ATLAS$^{3D}$, a qualitative comparison is made in Sect. \ref{sec:rad}. Table \ref{tab:2} summarizes the basic properties of the two objects analyzed in this study. Figure \ref{fig:1} shows the J-PLUS color images of these objects. Visual inspection of Fig. \ref{fig:1} shows that while NGC 5473 looks like a classical early-type galaxy, NGC 5485 is a more complex galaxy with a prominent minor-axis dust lane. 

\subsection{CALIFA}

CALIFA \citep{Sanchezetal2012} is a pioneer wide-field IFS survey of  667 galaxies in the local universe.  The observations were carried out with the Potsdam Multi-Aperture Spectrometer \citep[PMAS,][]{Rothetal2005} in the PPaK mode \citep{Verheijenetal2004} at the 3.5m telescope of Calar Alto observatory. PPaK contains 382 fibers of 2.7" diameter each and a 74" x 64" field of view. Three different spectral setups are available: i) a low-resolution V500 setup covering the wavelength range 3745 -- 7500 $\AA$ with a spectral resolution of 6.0 $\AA$ (FWHM); ii) a medium-resolution V1200 setup covering the wavelength range 3650 -- 4840 $\AA$ with a spectral resolution of 2.3 $\AA$ (FWHM); and iii) the combination of the cubes from both, i) and ii), setups (called COMBO) with a spectral resolution of 6.0 $\AA$ and a wavelength range between 3700 -- 7500 $\AA$. The target sample has been selected from the photometric catalog of Sloan Digital Sky Survey \citep[SDSS,][]{Yorketal2000} as a sample limited in the apparent isophotal diameter, 45"< isoA$_{r}$ < 80", to fill the field of view of PPaK and cover the redshift range  0.005 < $z$ < 0.03. We refer to \citet{Sanchezetal2012} and \citet{Sanchezetal2016} for details on the observational strategy and data processing. The COMBO data cubes of the two objects analyzed in this paper are available through the CALIFA website\footnote{http://califa.caha.es/} and belong to the CALIFA DR3.

\begin{table}
\caption{J-PLUS SVD Observation Summary}
\label{tab:1}
\centering
\small
\begin{tabular}{rllll}

\hline 
\hline \\[-1ex]
Filter  &  $\lambda$$_\mathrm{eff}$ & $\Delta\lambda$$_\mathrm{eff}$ & Exp. Time   & FWHM$_{mean}$\\[0.5ex]

            &  		(nm)			        &        (nm)                                             &  (s) &  (arcsec)\\[0.5ex]
\hline \\[-1ex]

 $u$  	& 348.5	& 50.8	& 3 x 207	& 1.49 \\
 $J0378$   	& 378.5  	& 16.8	& 3 x 200	& 1.28\\
$J0395$    	& 395.0   	& 10.0	& 3 x 98 	& 1.23\\
$J0410$   	& 410.0 	& 20.0	& 3 x 39	 & 1.25 \\
$J0430$   	& 430.0 	& 20.0	& 3 x 37 	& 1.69 \\
$g$   	& 480.3 	& 140.9	& 3 x 52 	& 1.15 \\
$J0515$   	& 515.0 	& 20.0	& 3 x 41 	& 1.13\\
$r$ 	& 625.4 	& 138.8	& 3 x 80 	& 1.16 \\
$J0660$  	& 660.0 	& 13.8	& 3 x 270	 & 1.15\\
$i$ 	&  766.8 	& 153.5	& 3 x 26 	& 1.23\\
$J0861$ 	& 861.0 	& 40.0	& 3 x 270 & 1.13 \\
 $z$  	& 911.4 	& 140.9 	& 3 x 36 	& 1.05\\

\hline
\hline
\multicolumn{5}{p{0.75\columnwidth}}{Notes. Col. 1: Filter name; Col. 2: Central wavelength; Col. 3: Effective pass band; Col. 4: Exposure time; Col. 5: Mean Full width at half maximum.}\\
\end{tabular}
\end{table}

\begin{table*}
\caption{Objects General Properties}
\label{tab:2}
\centering
\small
\begin{tabular}{ccccccccc}

\hline 
\hline \\[-1ex]
 Object & CALIFA ID & R.A. (J2000.0)  &  Dec. (J2000.0) &  Hubble Type   &  Environment & M$_{B}$$\textsuperscript{a}$ & Redshift$\textsuperscript{b}$ & log M$_{JAM}$$\textsuperscript{c}$ (M$_{\sun}$)  \\[0.5ex]
 
\hline \\[-1ex]

NGC 5473 & 703 & 14:04:43.22        & +54:53:33   & E-S0  & Isolated  & --20.21  & 0.006 & 11.09\\
NGC 5485 & 708 & 14:07:11.37        & +55:00:06   &     S0  & Isolated  & --19.89    & 0.006  & 11.05\\
\hline
\hline
\multicolumn{9}{l}{Notes. Basic parameters from HyperLeda.}\\
\multicolumn{9}{l}{$\textsuperscript{a}$ B-band absolute magnitude.}\\
\multicolumn{9}{l}{$\textsuperscript{b}$ Spectroscopic redshift from SDSS.}\\
\multicolumn{9}{l}{$\textsuperscript{c}$ Dynamical mass inferred from ATLAS$^{3D}$ survey \citep{Cappellarietal2013}.}\\
\end{tabular}
\end{table*}

\section{Method}\label{sec:method}
Three different methodologies are used throughout this work: one single method to process the photometric multi-filter observations (J-PLUS/MUFFIT), and two different techniques to analyze IFU CALIFA data (CALIFA/STARLIGHT and CALIFA/STECKMAP).

Single stellar population (SSP) models are a key ingredient to disentangle physical properties of galaxies stellar populations. They are the basis to transform observed quantities to physical properties and involve choices among different IMFs, stellar libraries, and isochrones. Although several studies show a general good agreement when using different model sets, some evidence indicates systematic differences associated with the different SSP models used \citep[e.g.,][]{Coelhoetal2009, Diasetal2010, CidFernandesetal2014, DiazGarciaetal2015, Sanromanetal2018}. To minimize the differences due to the methodologies, we will perform the comparison between the three methods described above using the same stellar population models \citep[][BC03 hereafter]{BruzualCharlot2003} except for CALIFA/STARLIGHT that uses BC03 updated version Charlot \& Bruzual (2007, private communication)\footnote{The Charlot \& Bruzual (2007) models are available at http://www.bruzual.org/~gbruzual/cb07)}. These SSPs are an update of BC03 models, where STELIB \citep{LeBorgneetal2003} is replaced by the MILES \citep{SanchezBlazquezetal2006} and GRANADA \citep{Martinsetal2005, GonzalezDelgadoetal2005} stellar libraries. In addition, the updated version incorporates an improved TP-AGB treatment \citep{Marigoetal2007}.  \citet{Marastonetal2006} have shown that the treatment of the TP-AGB phase of the stellar evolution is a source of discrepancy in the determination of the spectroscopic age and mass of high-z (1.4 < $z$ < 2.7) galaxies. Although the results inferred using different prescriptions show significant differences in the infrared, major discrepancies are not expected in the optical regime \citep[e.g.,][]{BruzualCharlot2007, Rocketal2016}. We selected the Padova 1994 tracks \citep{Bressan1993, Fagotto1994a,Fagotto1994b,Girardi1996} that cover a range of ages from 0.001 to 14 Gyr and metallicities [Fe/H]= --2.3, --1.7, --0.7, --0.4, 0.0, +0.4. A \cite{Chabrier2003} initial mass function (IMF) has been used in all the cases.

\subsection{J-PLUS/MUFFIT} \label{sec:muffit}

The method used in this analysis has been extensively described and tested in \citet{Sanromanetal2018}. It can be summarized in three main steps: the homogenization of the point-spread functions (PSF), the spatial binning of the images through a centroidal Voronoi tessellation \citep[CVT,][]{Cappellarietal2003}, and the SED fitting of each bin. For the SED fitting, we used the code \textrm{MUFFIT} \citep[MUlti-Filter FITting for stellar population diagnostics; ][]{DiazGarciaetal2015}. \textrm{MUFFIT} is a generic code optimized to retrieve the main stellar population parameters of galaxies in photometric multi-filter surveys. 

To perform good-quality multicolor photometry of such wide field of view, PSF homogenization processes are required. PSF variations cause that the light inside a given aperture is redistributed differently across the field of view and from filter-to-filter. These effects may produce artificial structures that could bias our results \citep{Bertin2011}. To avoid this problem, we performed a PSF homogenization in the twelve bands. \texttt{SExtractor} \citep{BertinArnouts1996} and \texttt{PSFEx} \citep{Bertin2013} were used in every image to generate an homogenization kernel, where the worst (widest) PSF value of the image set was chosen as a target PSF.  A 2D Moffat model is used as a homogenization kernel. The images were convolved with their corresponding kernels using a fast Fourier transform, bringing the images of all the bands to the same circular PSF. Finally, we need to take into account that the homogenization process has consequences in the image noise, producing pixel-by-pixel correlations. To correct for this, we recalculate the noise model of the images using the procedure described in \citet{Labbeetal2003} and \citet{Molinoetal2014}. This recalculated noise model is then used for computing the photometric errors.

To ensure a reliable determination of the stellar population parameters we perform a CVT imposing a minimum S/N of 10 in the $J0378$ filter. We chose this filter because, for the selected targets, it is the one with the lowest S/N.  We note that this choice is a conservative limit and that slightly lower S/N could extended the analysis to larger galactocentric radii (e.g., sacrificing the S/N of 1 out of 12 filters). As mentioned in the introduction, multi-filter techniques allow the analysis of galaxy profiles at larger galactocentric radii and at higher redshift than spectroscopic surveys. Although overall this idea is true and IFU-like photometric techniques can map fainter surface brightness levels, for this specific study we have been conservative and only the pixels inside the Kron radius of the blue filter J0378, R$_\mathrm{Kron}$, were included in the analysis. R$_\mathrm{Kron}$ is defined by SExtractor as a flexible elliptical aperture that confines most of the flux from an object and has been empirically tested to enclose > 90$\%$ of the object light. A comprehensive study of MUFFIT performance on J-PLUS and the dependency with the S/N of each filter is out of the scope of this paper and will be presented in a future work. The tessellation was then applied to the images in all the filters, and finally, the photometry of every region in all the filters was determined. For details about the tessellation method see \citet{Sanromanetal2018}. J-PLUS images are already background subtracted. This subtraction is done globally over the entire image. To assure a good background subtraction, we further performed a local sky subtraction considering an area of 100 x 100 pixels (55" x 55") around each target galaxy.

After performing the CVT and the photometry of every region in every filter is determined, we run \textrm{MUFFIT} to obtain 2D maps of different stellar population properties. The code compares the multi-filter fluxes of galaxies with the synthetic photometry of mixtures of two SSPs for a range of redshifts and extinctions through an error-weighted $\chi^{2}$ approach.  Several studies have shown that the mixture of two SSPs is a suitable and reliable approach to describe the stellar populations of early-type galaxies \citep{Rogersetal2010, Ferrerasetal2000, Kavirajetal2007, Lonoceetal2014}. More recently, \citet{LopezCorredoiraetal2017} fit a set of 20 red galaxies with models of a single-burst SSP, a combination of two SSPs, and an extended star formation history.  They concluded that exponentially decaying extended star formation models ($\tau$-models) improve slightly the fits (they have lower average reduced $\chi$$^{2}$) with respect to single-burst models, but they are considerably worse than the two SSPs based fits ($\chi$$^{2}$ > 20$\%$ larger). They conclude that the models with 2 SSPs represent better red galaxies. Based on these studies, we consider a 2-SSP model fitting approach the best method for our specific work. 

 Throughout this work the Fitzpatrick reddening law has been used \citep{Fitzpatrick1999} with extinctions values A$\textsubscript{V}$ in the range of 0 to 3.1. This extinction law is suitable for dereddening any photospectroscopic data, such as J-PLUS \citep[further details in][]{Fitzpatrick1999}. To minimize the number of free fitting parameters we provide \textrm{MUFFIT} with a fixed redshift value. We have used the spectroscopic redshifts determined by SDSS (Table \ref{tab:2}). We note that, in the future, J-PLUS will provide accurate photo-redshifts values for local galaxies (see details in Paper I).

\subsection{CALIFA/STARLIGHT} \label{sec:califa}

The first method for extracting stellar population information from the CALIFA data cubes is based on a full spectral synthesis approach using the \textrm{STARLIGHT} code \citep{CidFernandesetal2005}. Previous to the spectral fitting, all spaxels containing light from spurious sources (e.g., foreground stars and background galaxies) are masked. The spaxels with S/N < 3 are also masked. The cubes are then segmented into Voronoi zones using the routine described in \citet{Cappellarietal2003}. The target S/N is set to 20, which leaves most of the spaxels inside one half light radius unbinned.

\textrm{STARLIGHT} fits an observed spectrum in terms of a model built by a linear combination of SSPs from a base spanning different ages and metallicities. Dust effects are modeled as a foreground screen with a \citet{Cardellietal1989} reddening law with R$_\mathrm{v}$=3.1. The spectral fits were performed in the rest-frame 3700 -- 6850 $\AA$ interval where the main optical emission lines were masked ([OII], H$\gamma$, H$\beta$, [OIII], HeI, [OI], H$\alpha$, [NII], [SII]).  Because of its interstellar absorption component, the NaI doublet was also masked. A more detailed explanation about CALIFA/STARLIGHT process can be found in \citet{CidFernandesetal2013, CidFernandesetal2014}. The spectra were then processed through PYCASSO\footnote{http://picasso.ufsc.br/}  \citep[the Python CALIFA Starlight Synthesis Organizer;][]{deAmorimetal2017} producing the results discussed in the next sections. 

The star formation histories are derived using two different stellar population models, the Granada \citep{Martinsetal2005, GonzalezDelgadoetal2005} and the Charlot \& Bruzual (2007, private communication). We chose to compare the results with the latter as these models are very similar to those used in J-PLUS/MUFFIT and CALIFA/STECKMAP.

\begin{sidewaysfigure*}[htbp]
\begin{center}
\includegraphics[width=0.15\columnwidth, angle=90]{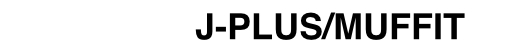}
\includegraphics[width=0.20\columnwidth]{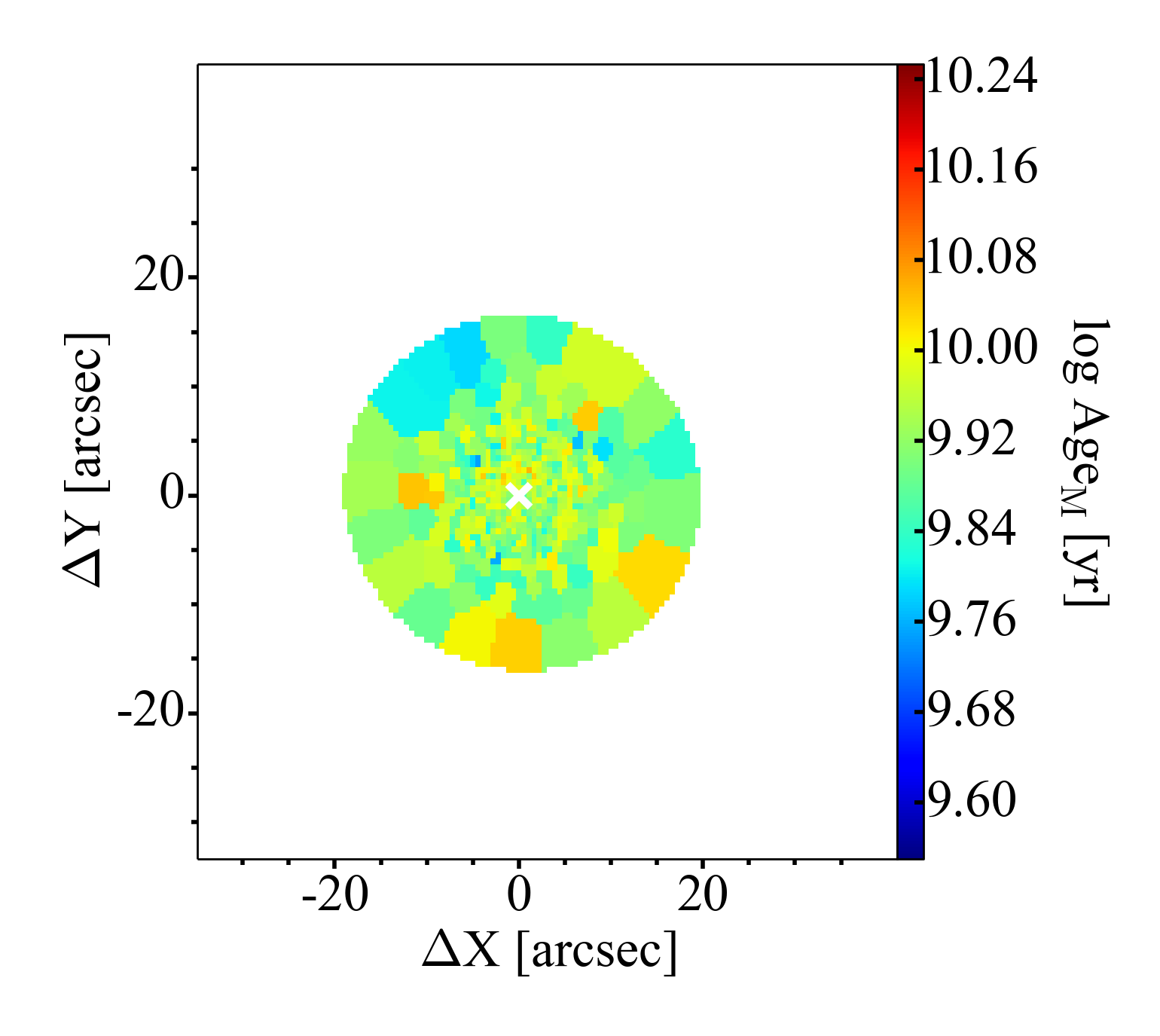}
\includegraphics[width=0.20\columnwidth]{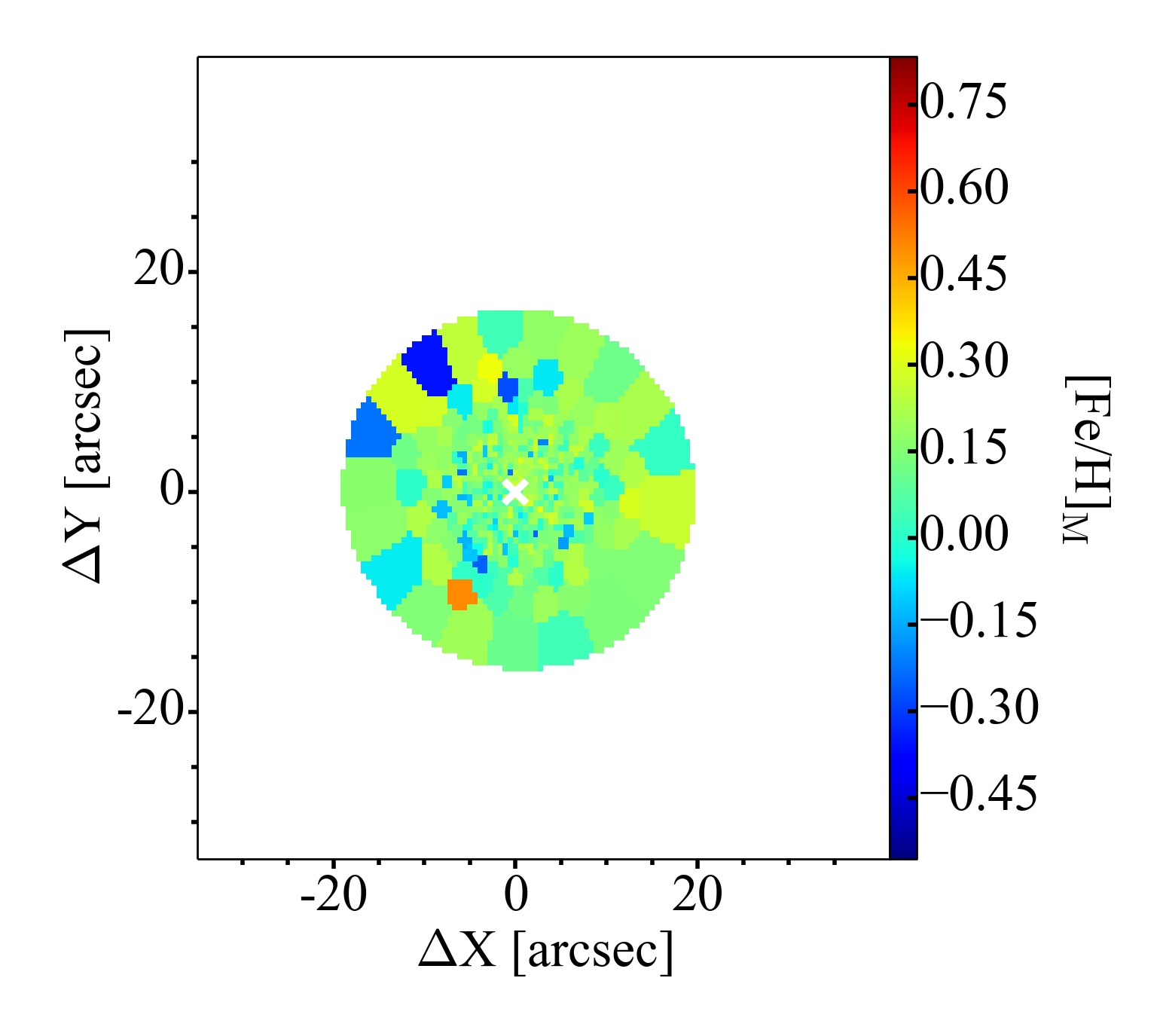}
\includegraphics[width=0.20\columnwidth]{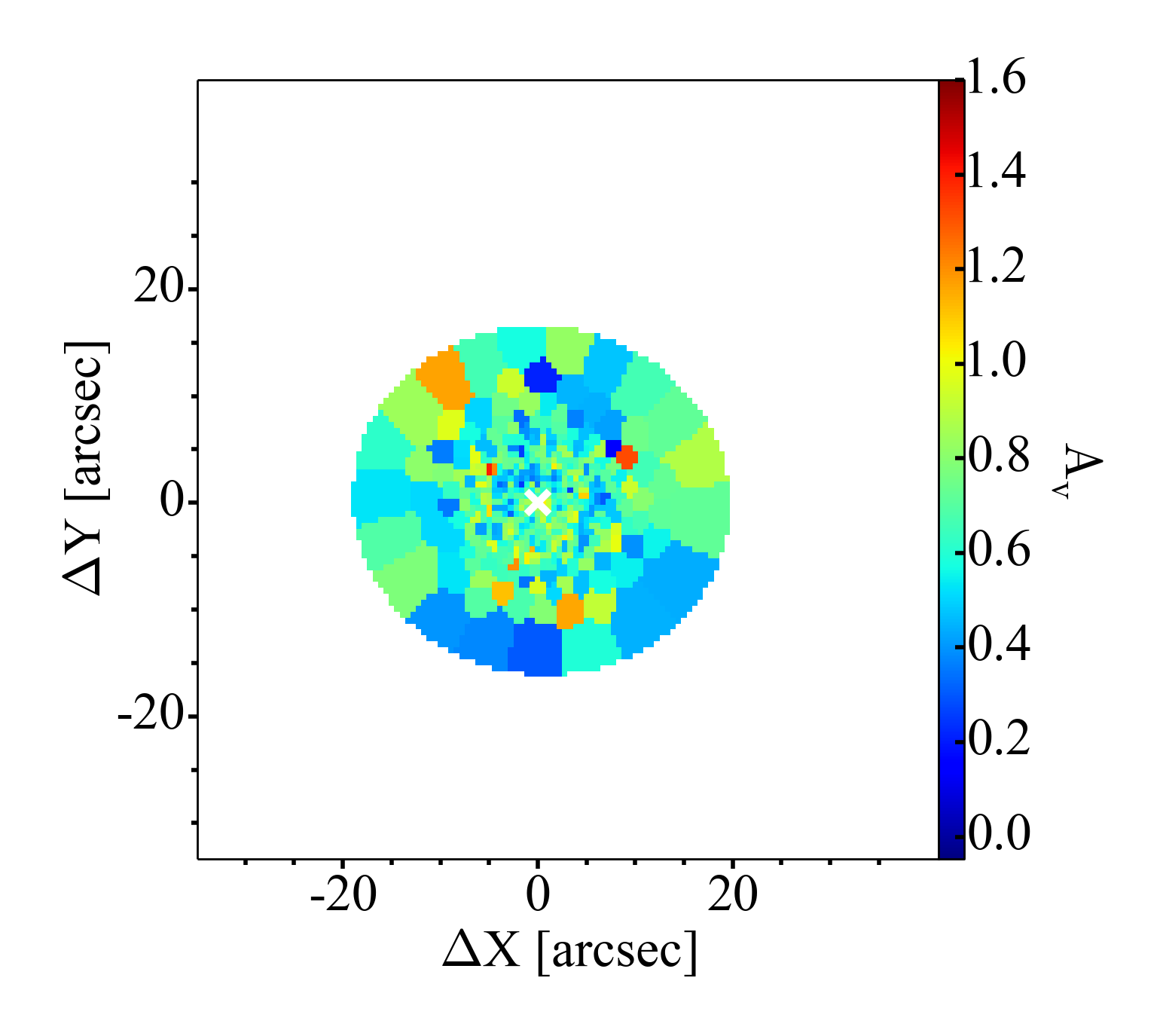}
\includegraphics[width=0.20\columnwidth]{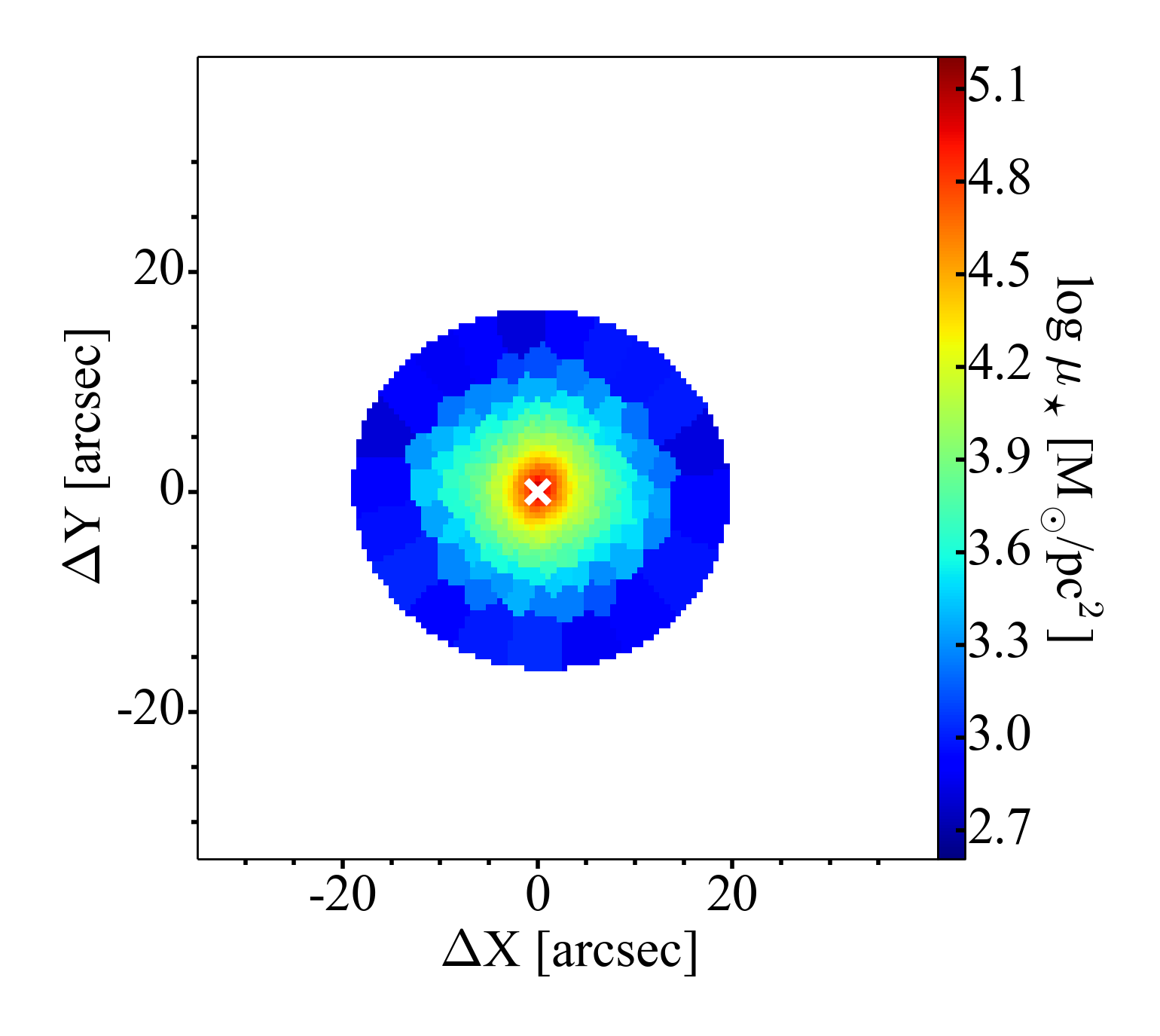}

\includegraphics[width=0.15\columnwidth, angle=90]{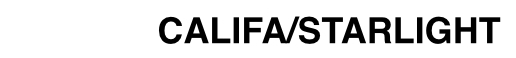}
\includegraphics[width=0.20\columnwidth]{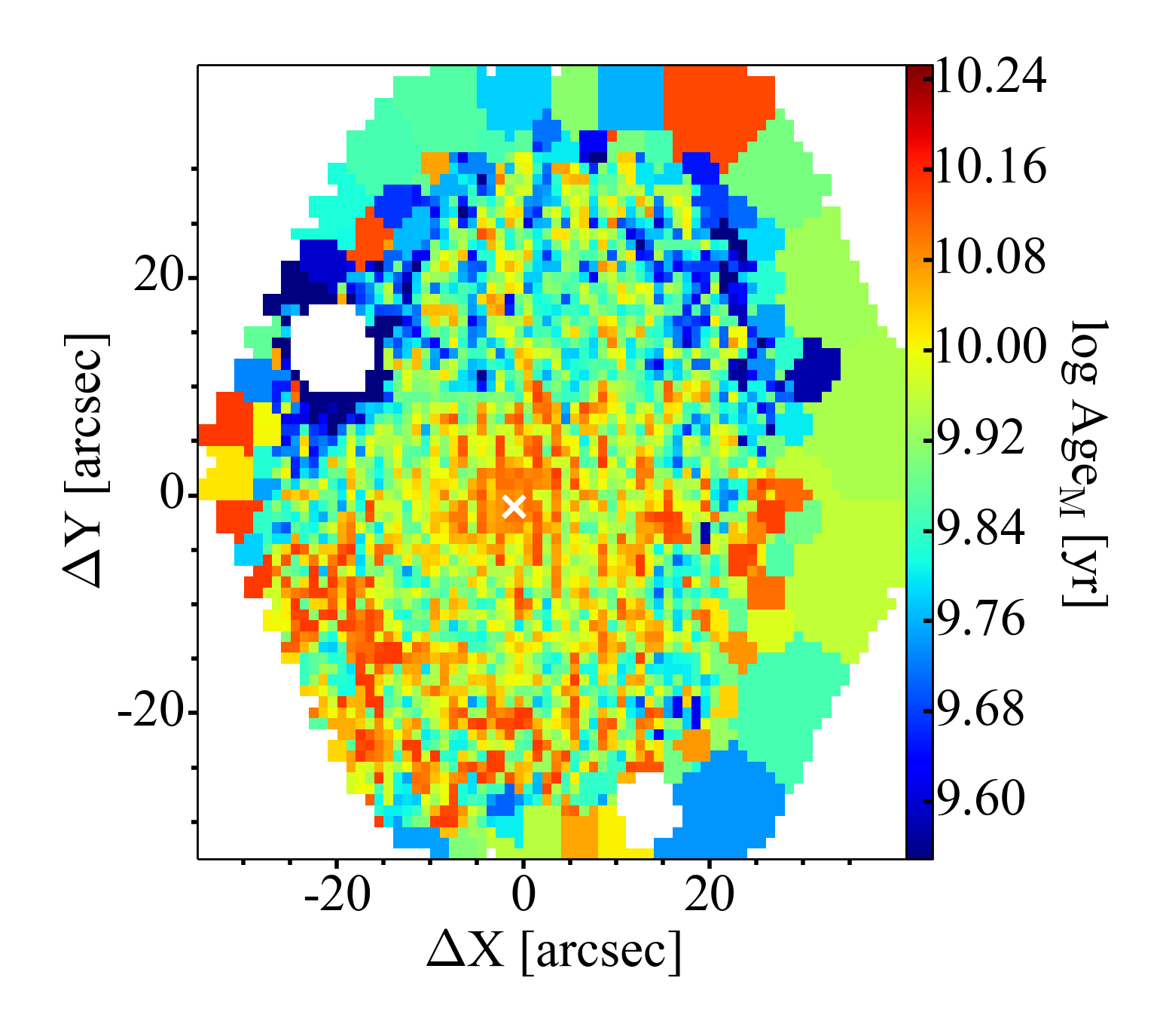}
\includegraphics[width=0.20\columnwidth]{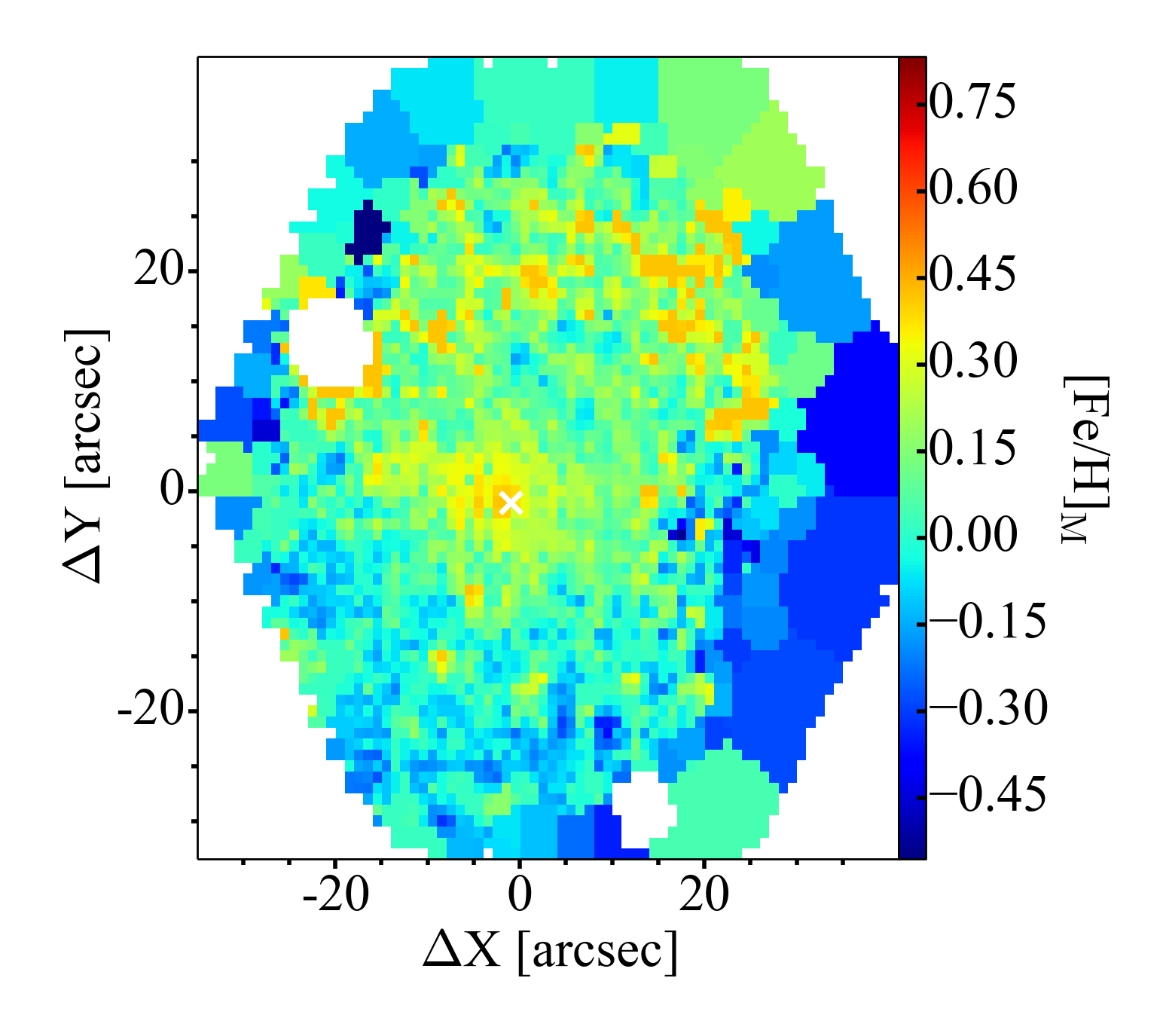}
\includegraphics[width=0.20\columnwidth]{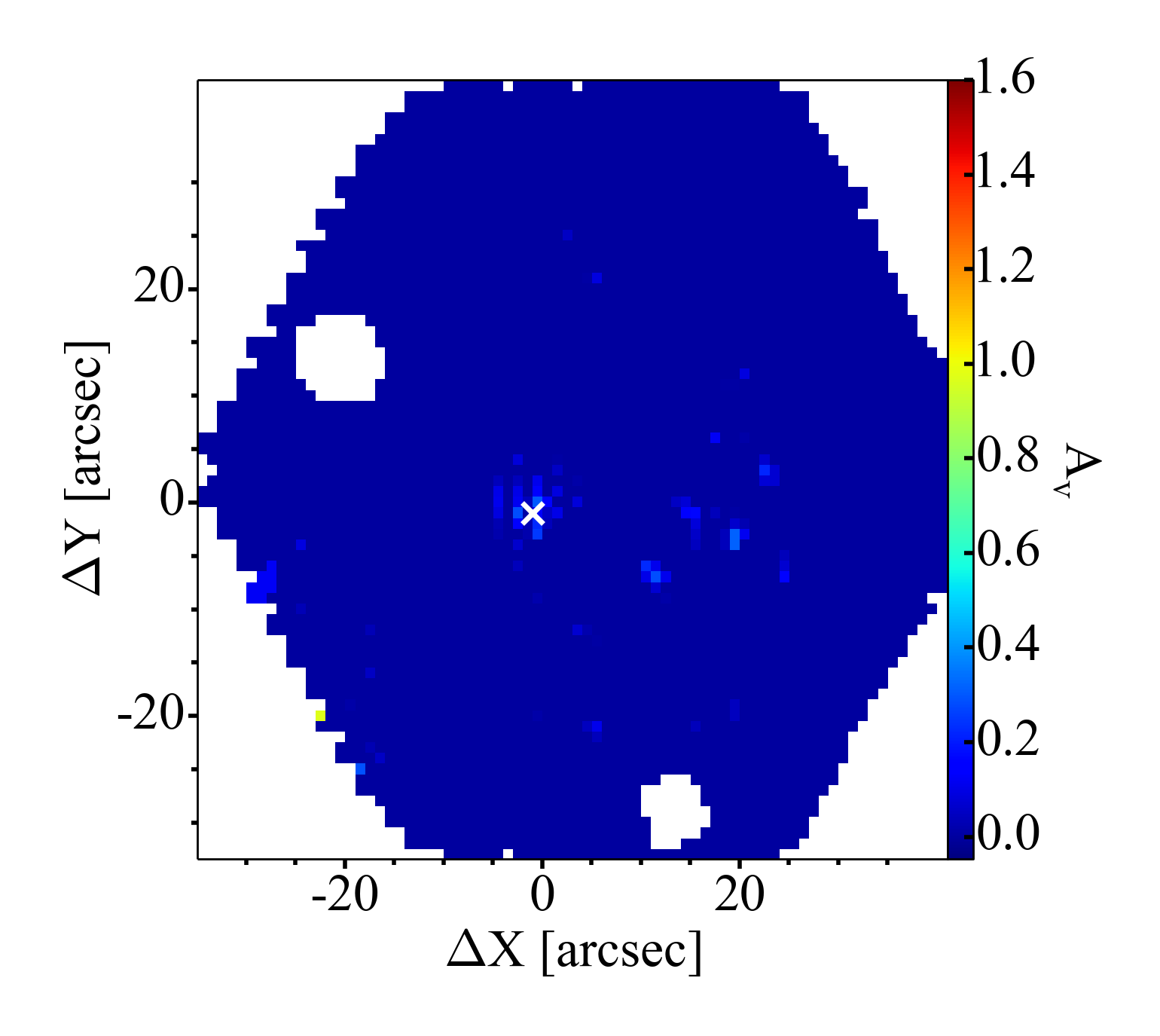}
\includegraphics[width=0.20\columnwidth]{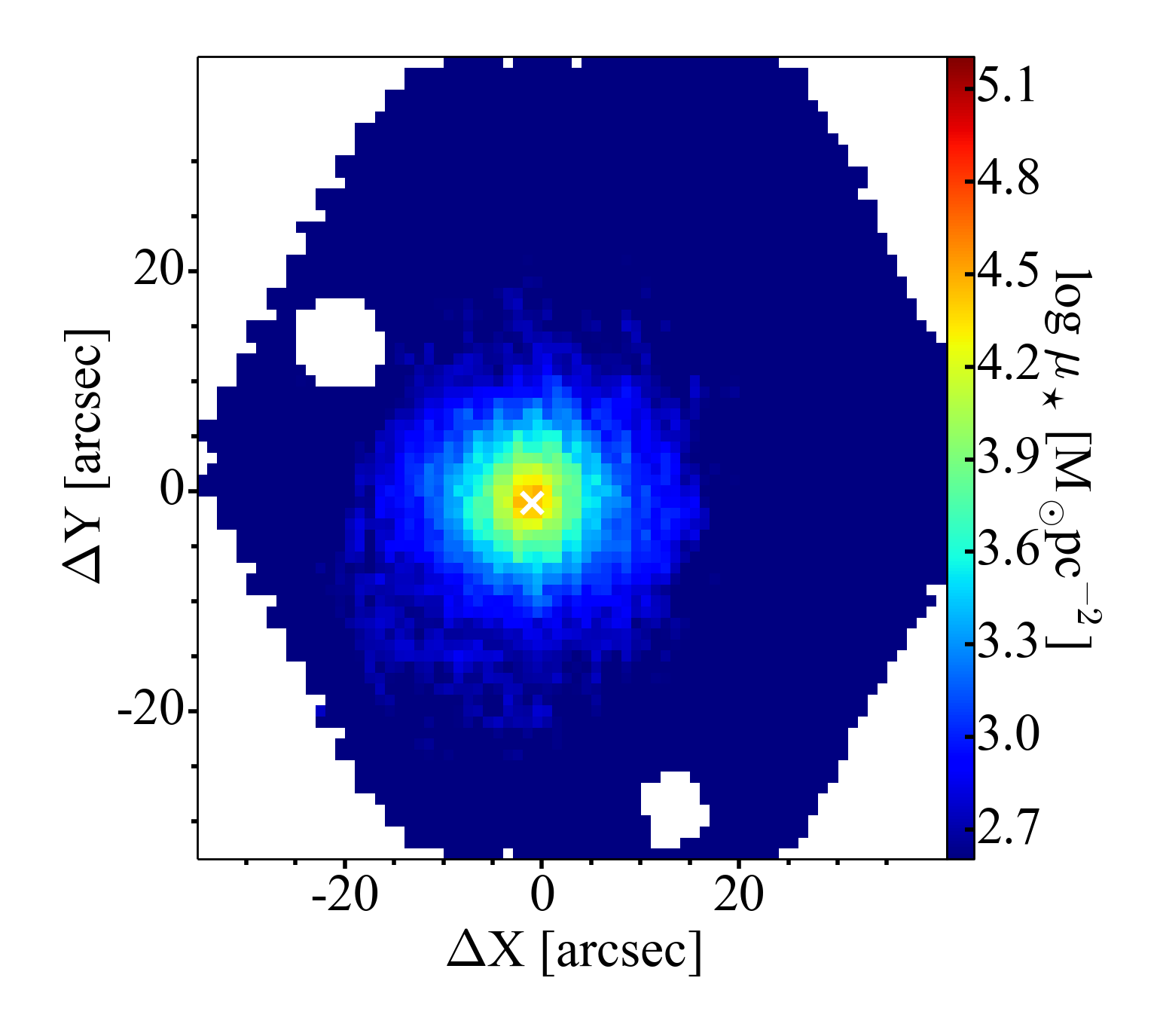}

\includegraphics[width=0.15\columnwidth, angle=90]{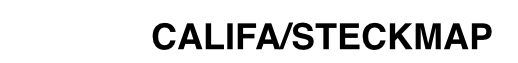}
\includegraphics[width=0.20\columnwidth]{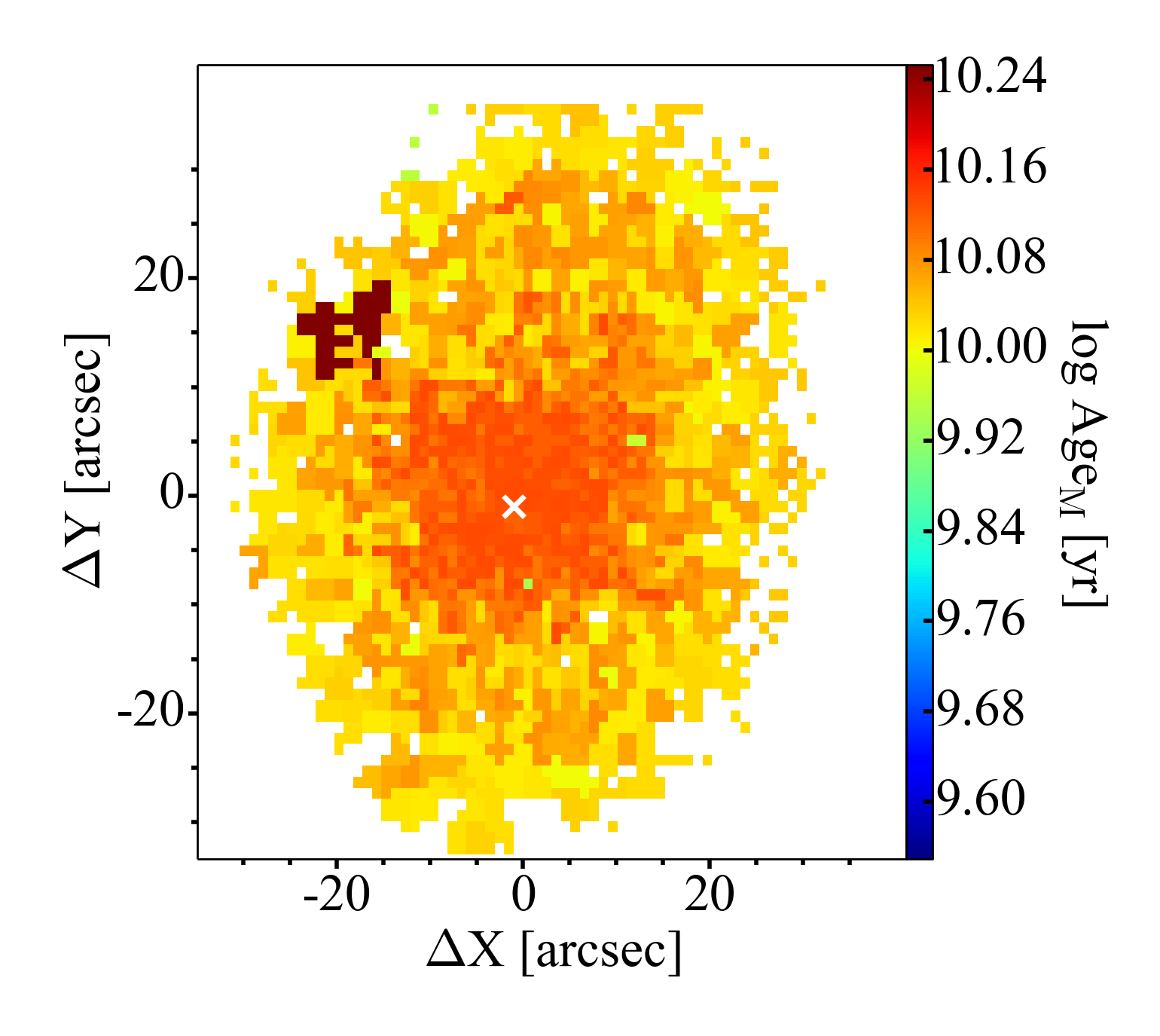}
\includegraphics[width=0.20\columnwidth]{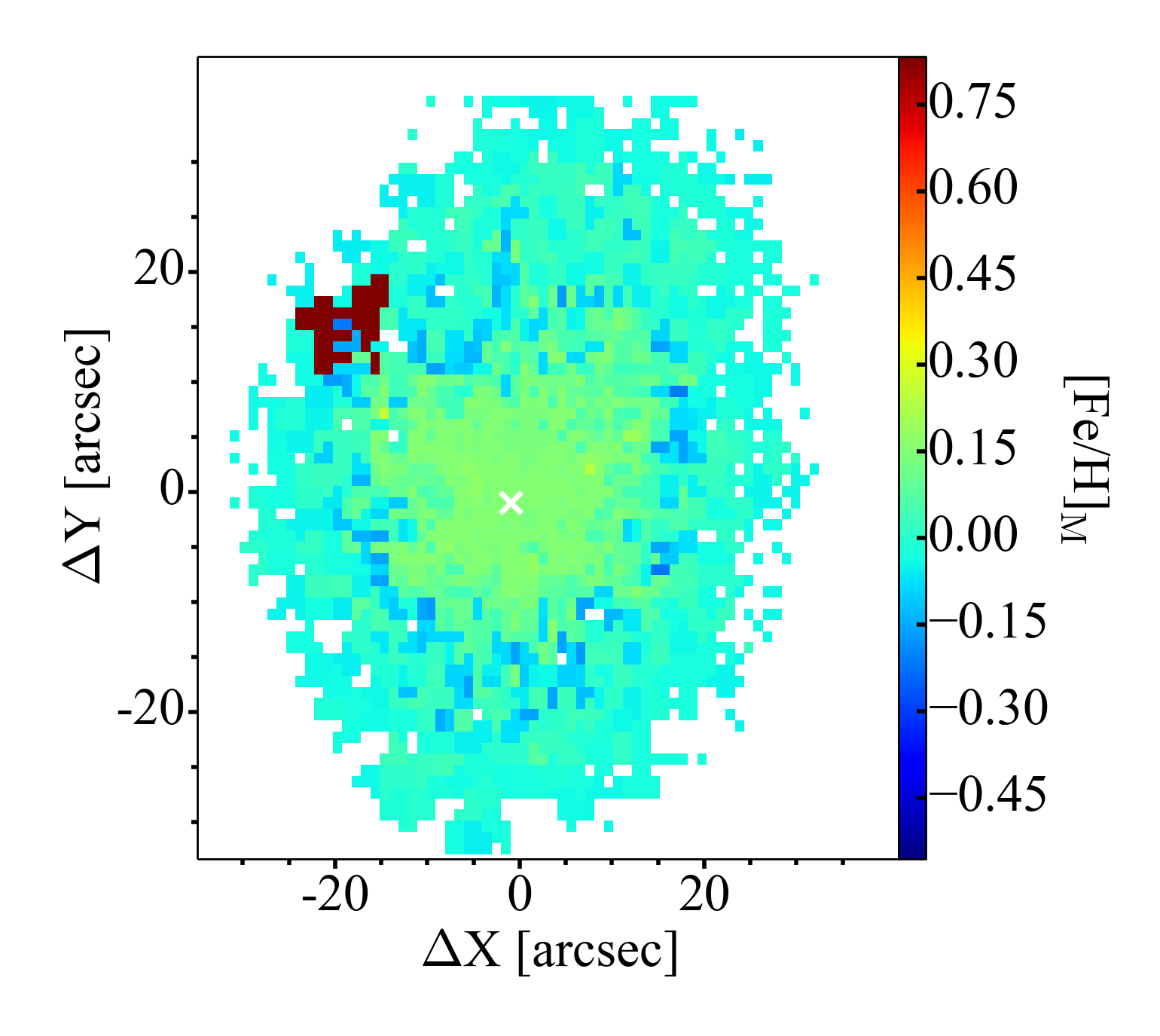}
\includegraphics[width=0.20\columnwidth]{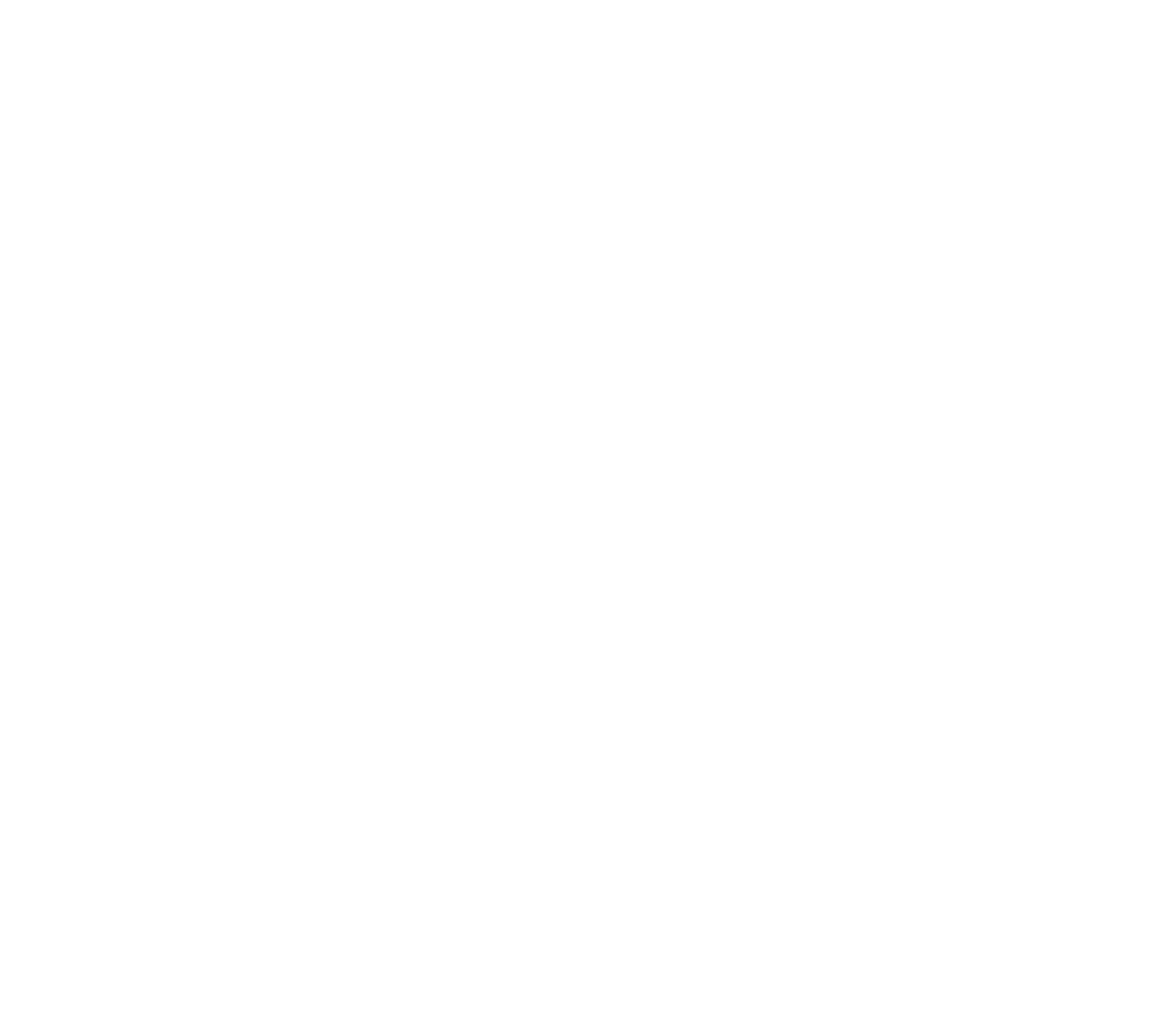}
\includegraphics[width=0.20\columnwidth]{blank.png}

\caption{Mass-weighted stellar population properties maps for NGC 5473 determined by J-PLUS/MUFFIT (first row), CALIFA/STARLIGHT (second row) and CALIFA/STECKMAP (third row). Each column corresponds to 2D maps for age (log Age), metallicity ([Fe/H]), extinction parameter (A$_\mathrm{v}$) and stellar mass surface density (log $\mu$$_{\star}$). The color range is the same for the different methods. The center of the galaxy is marked with a white cross in each panel.} 
\label{fig:2Dmapsa}
\end{center}
\end{sidewaysfigure*}

\begin{sidewaysfigure*}[htbp]
\begin{center}
\includegraphics[width=0.15\columnwidth, angle=90]{JPLUS_MUFFIT.jpg}
\includegraphics[width=0.20\columnwidth]{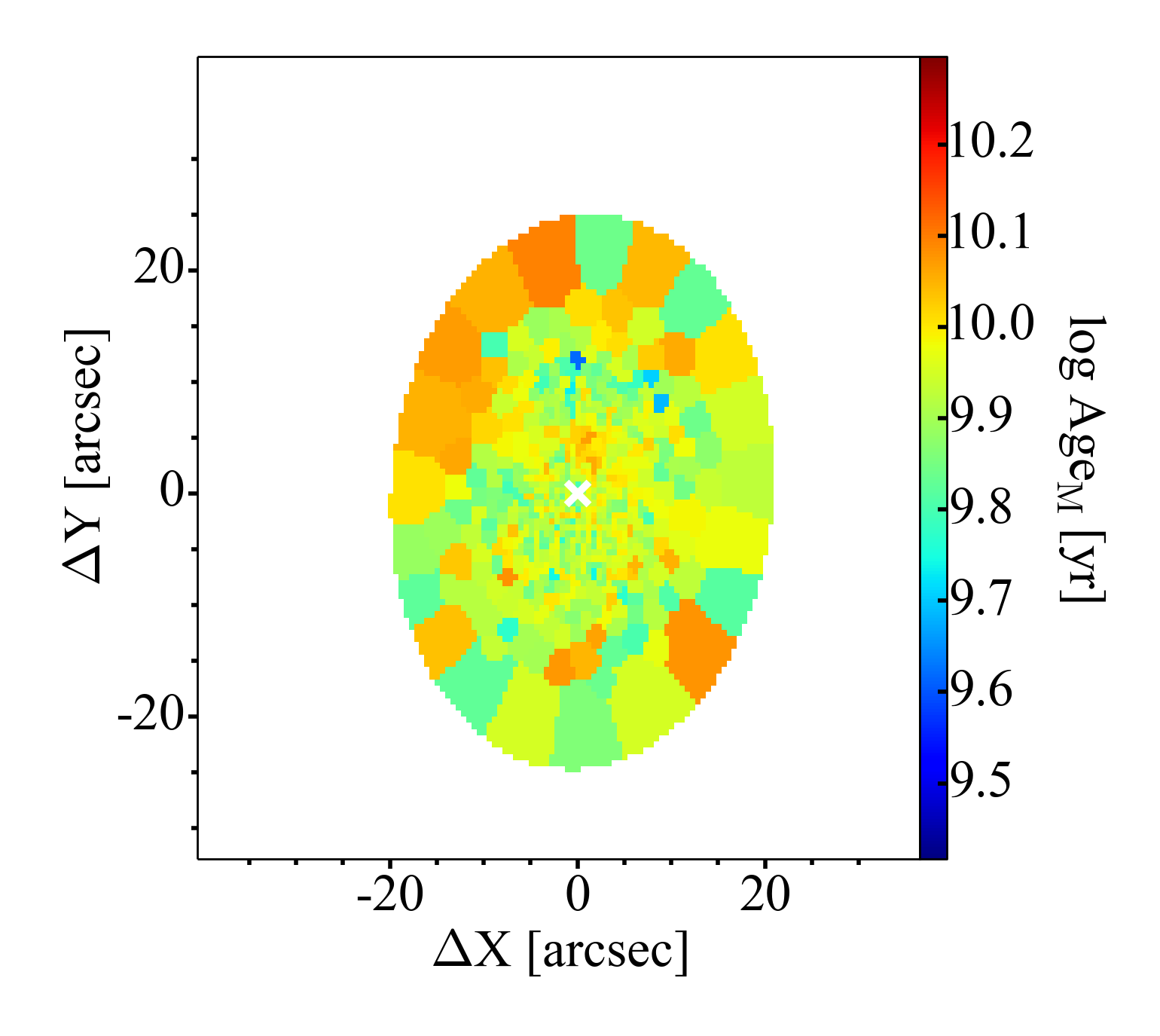}
\includegraphics[width=0.20\columnwidth]{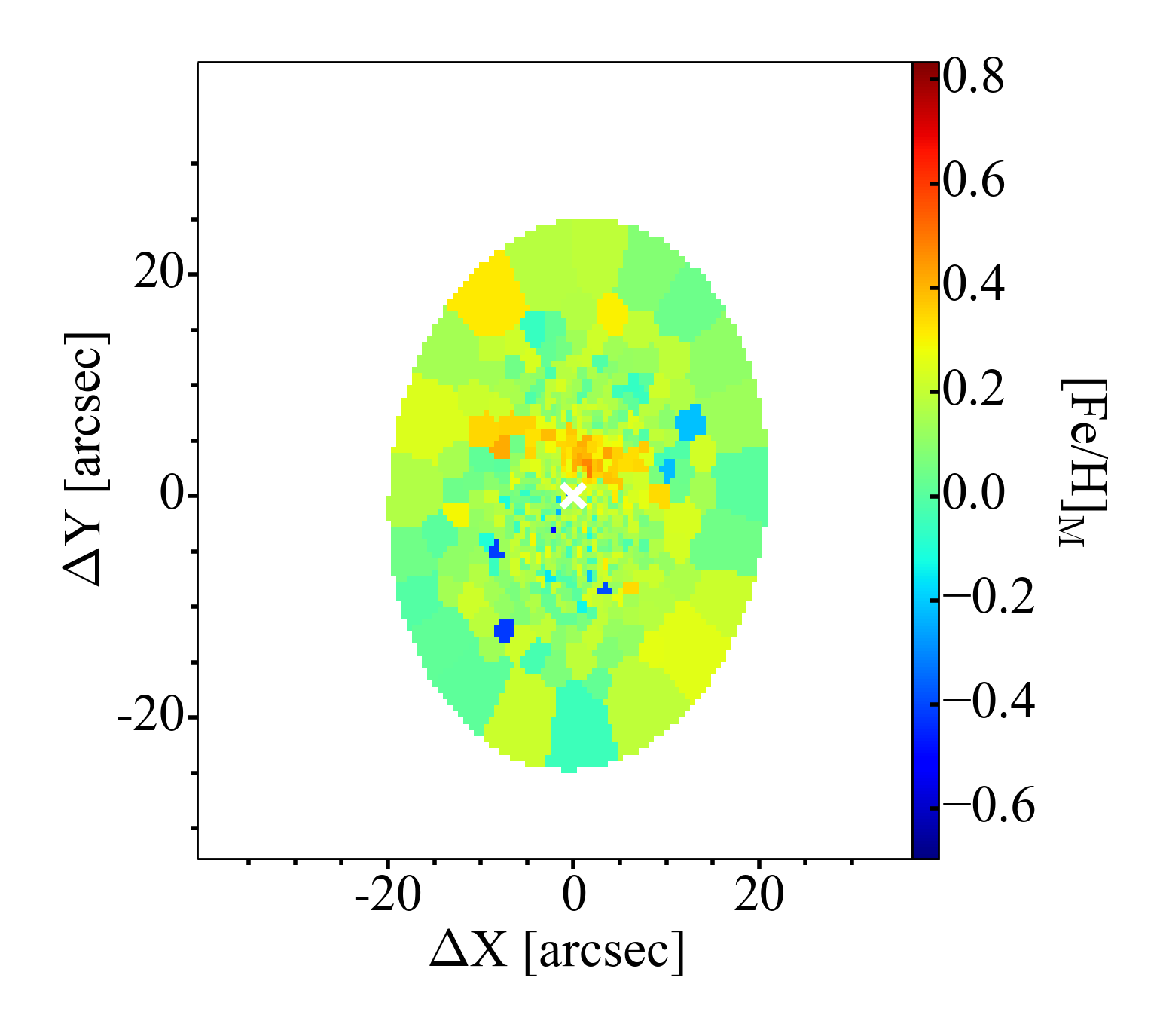}
\includegraphics[width=0.20\columnwidth]{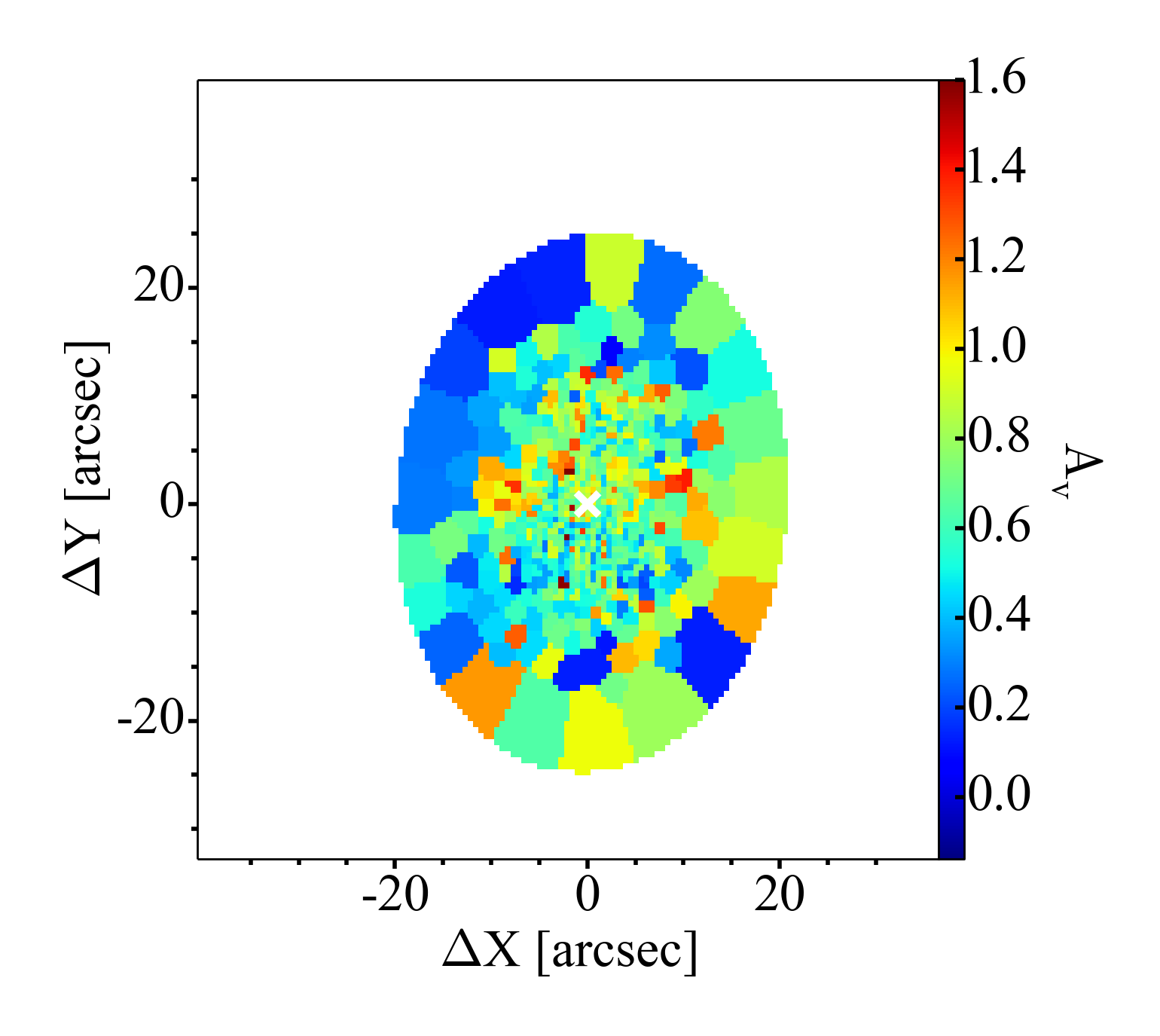}
\includegraphics[width=0.20\columnwidth]{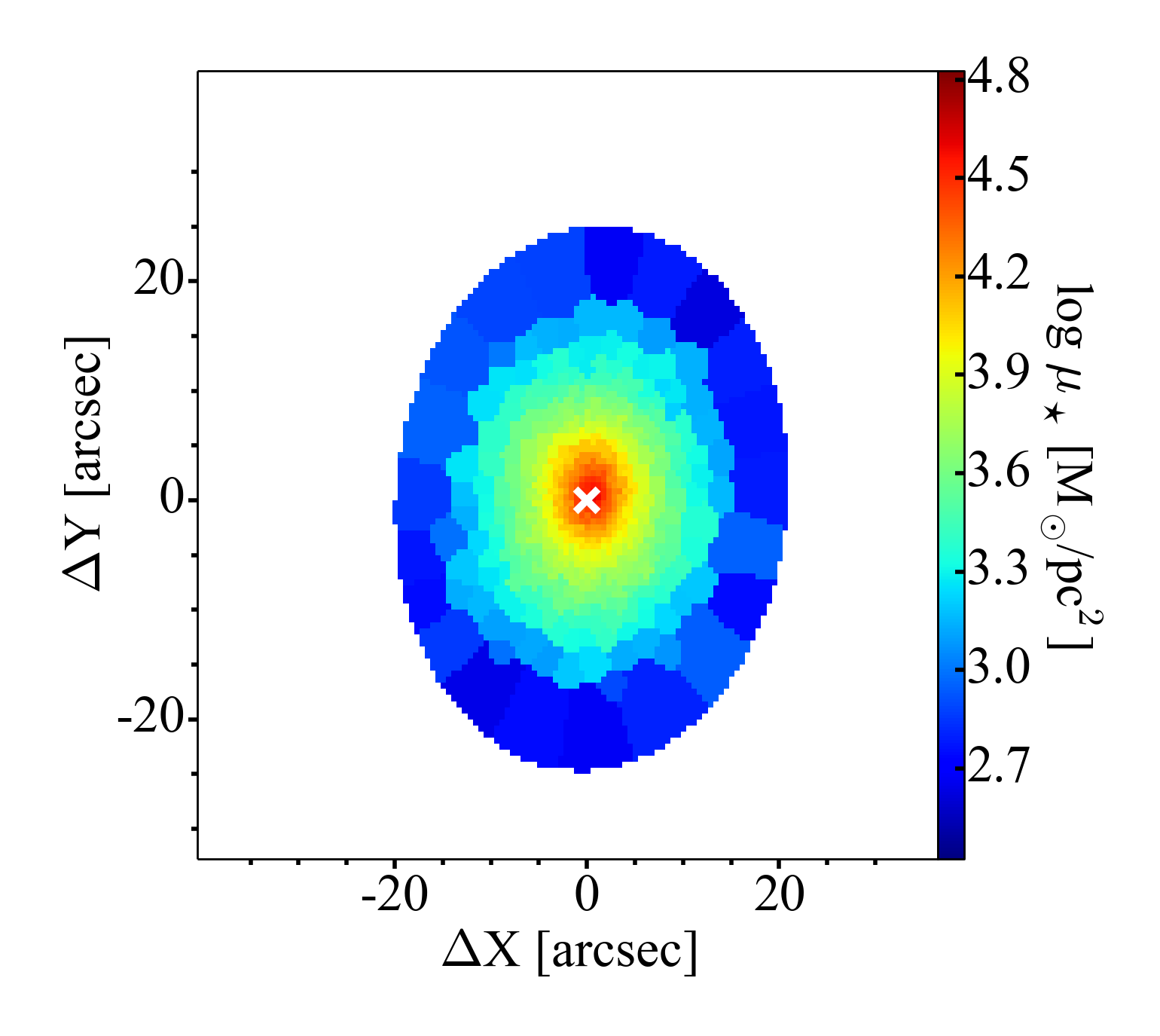}

\includegraphics[width=0.15\columnwidth, angle=90]{CALIFA_STARLIGHT.jpg}
\includegraphics[width=0.20\columnwidth]{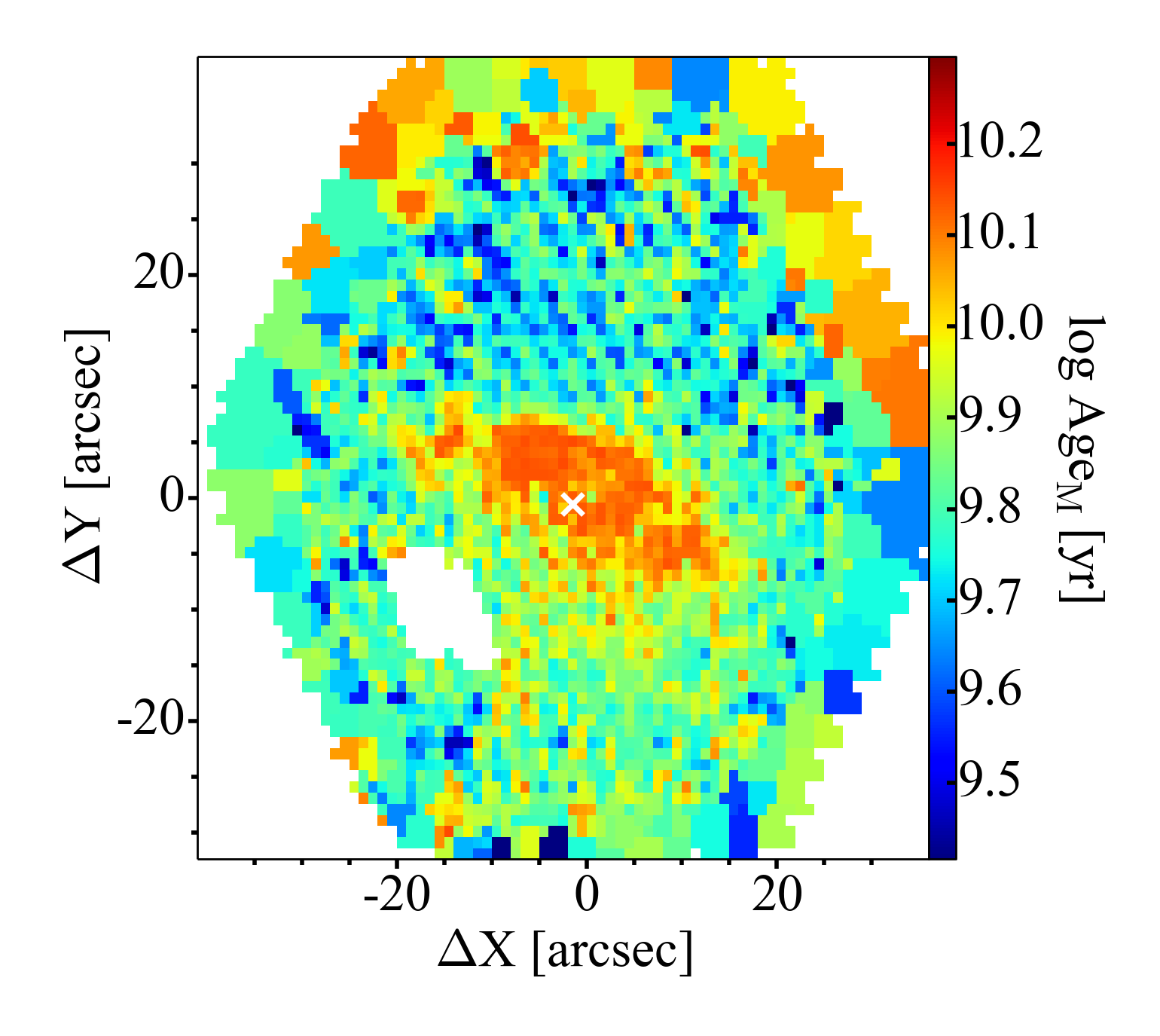}
\includegraphics[width=0.20\columnwidth]{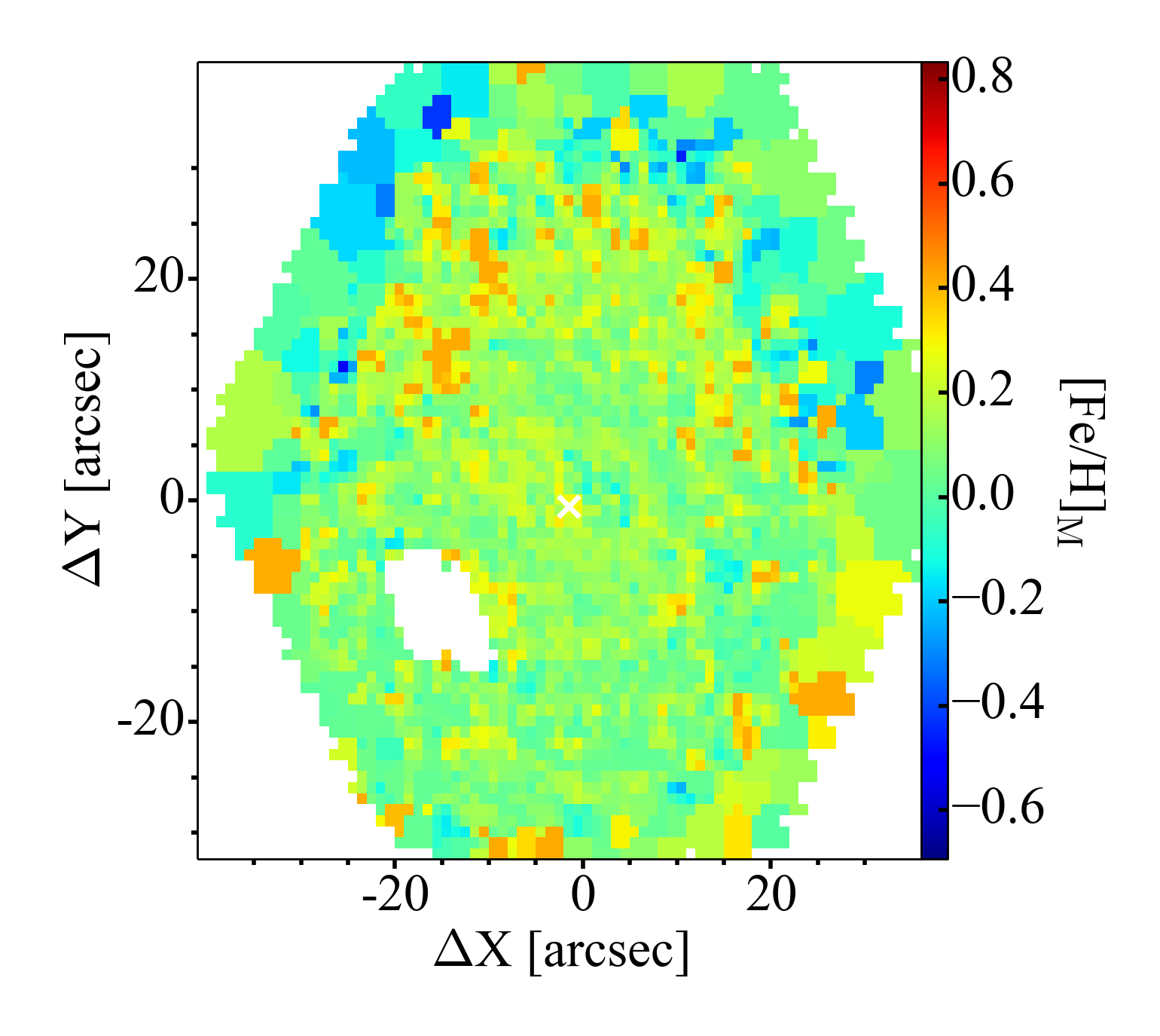}
\includegraphics[width=0.20\columnwidth]{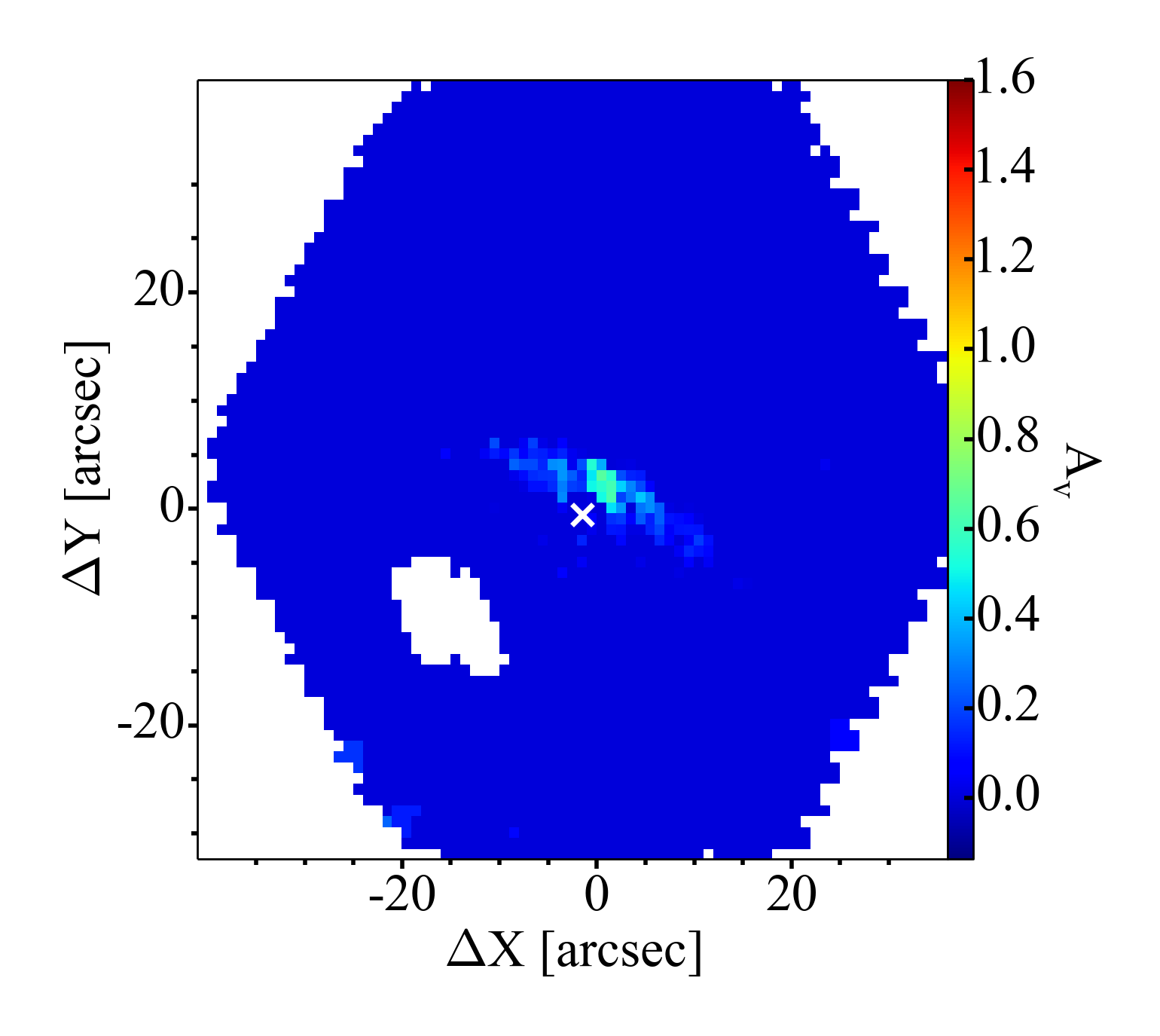}
\includegraphics[width=0.20\columnwidth]{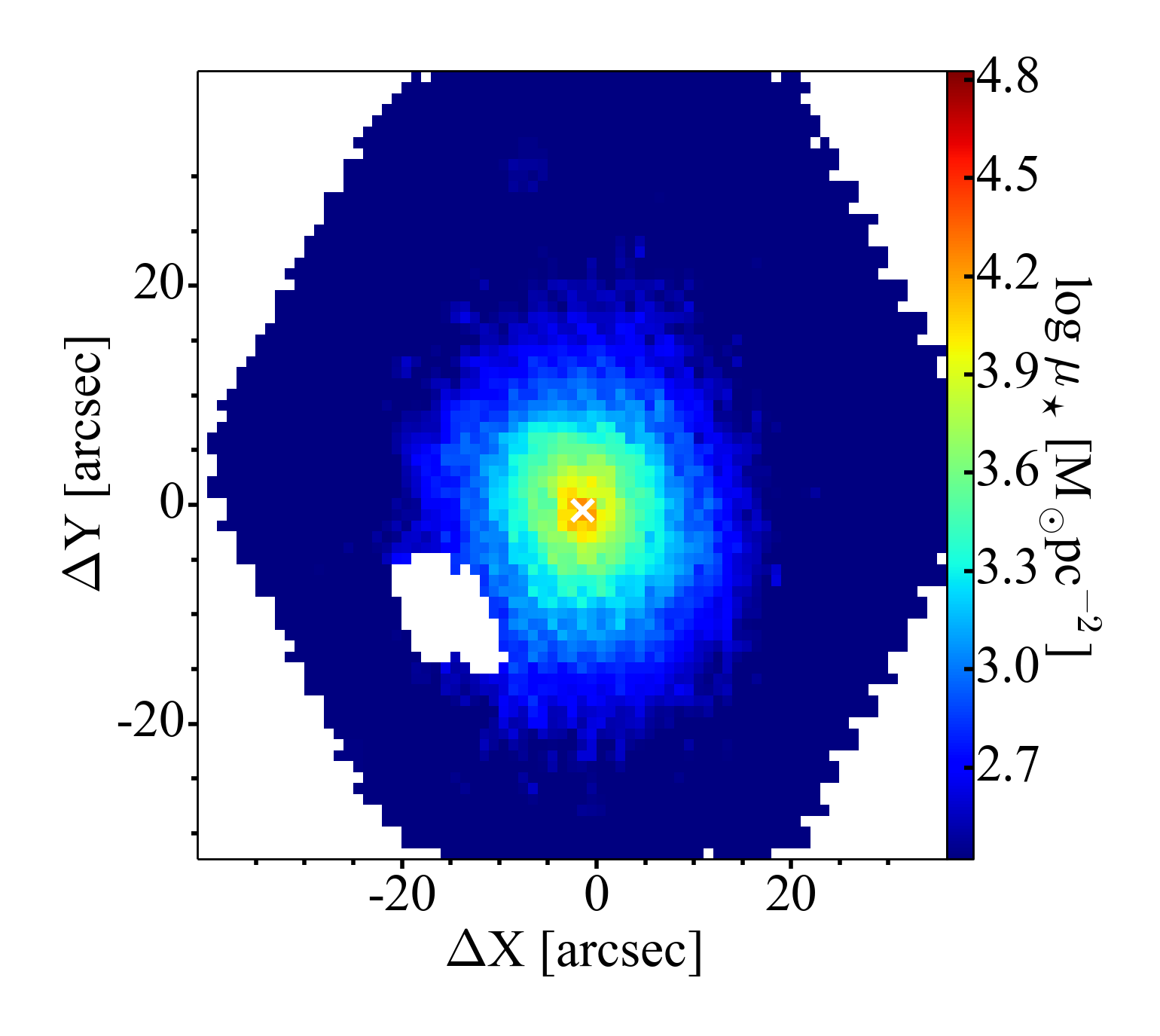}

\includegraphics[width=0.15\columnwidth, angle=90]{CALIFA_STECKMAP.jpg}
\includegraphics[width=0.20\columnwidth]{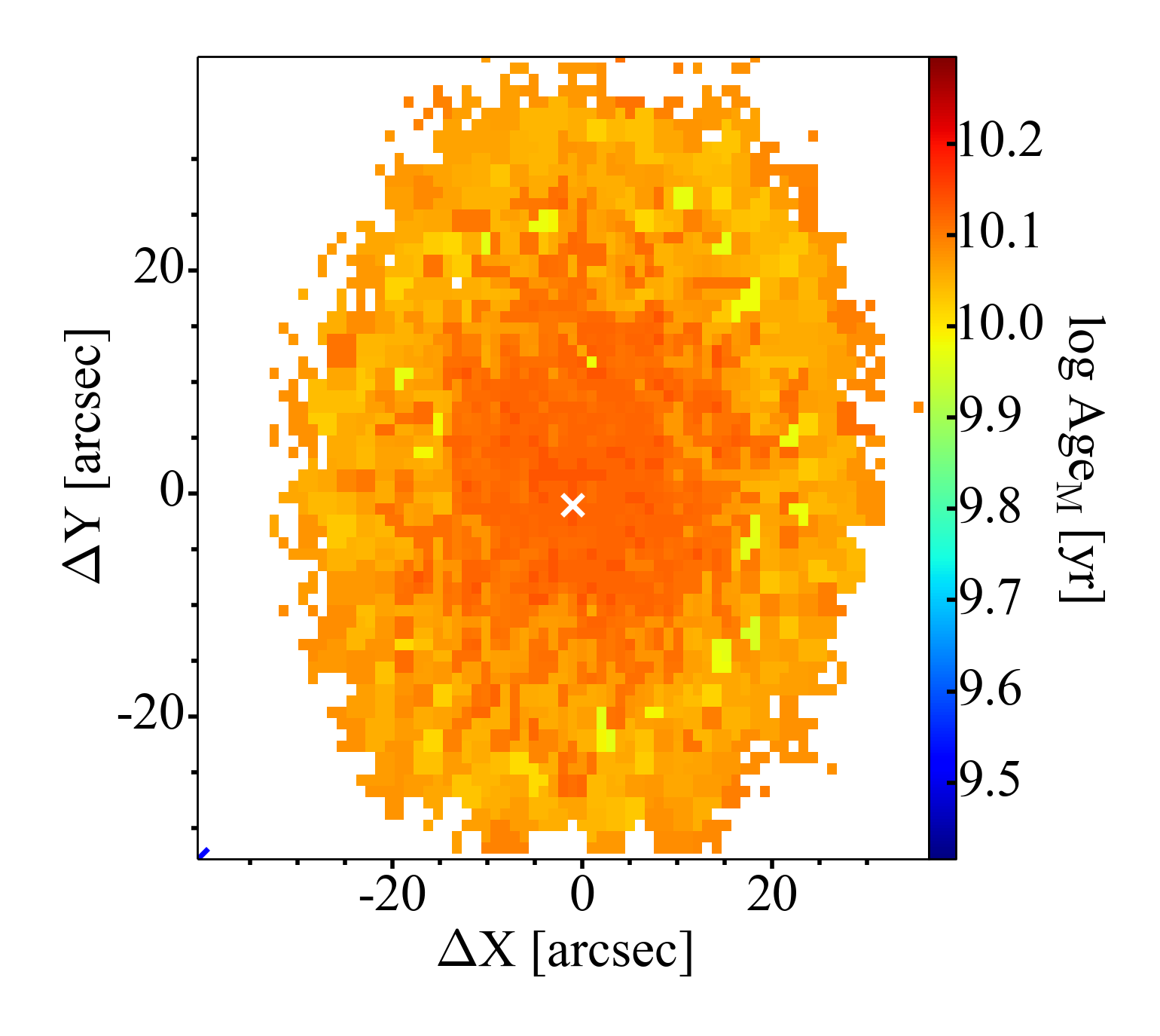}
\includegraphics[width=0.20\columnwidth]{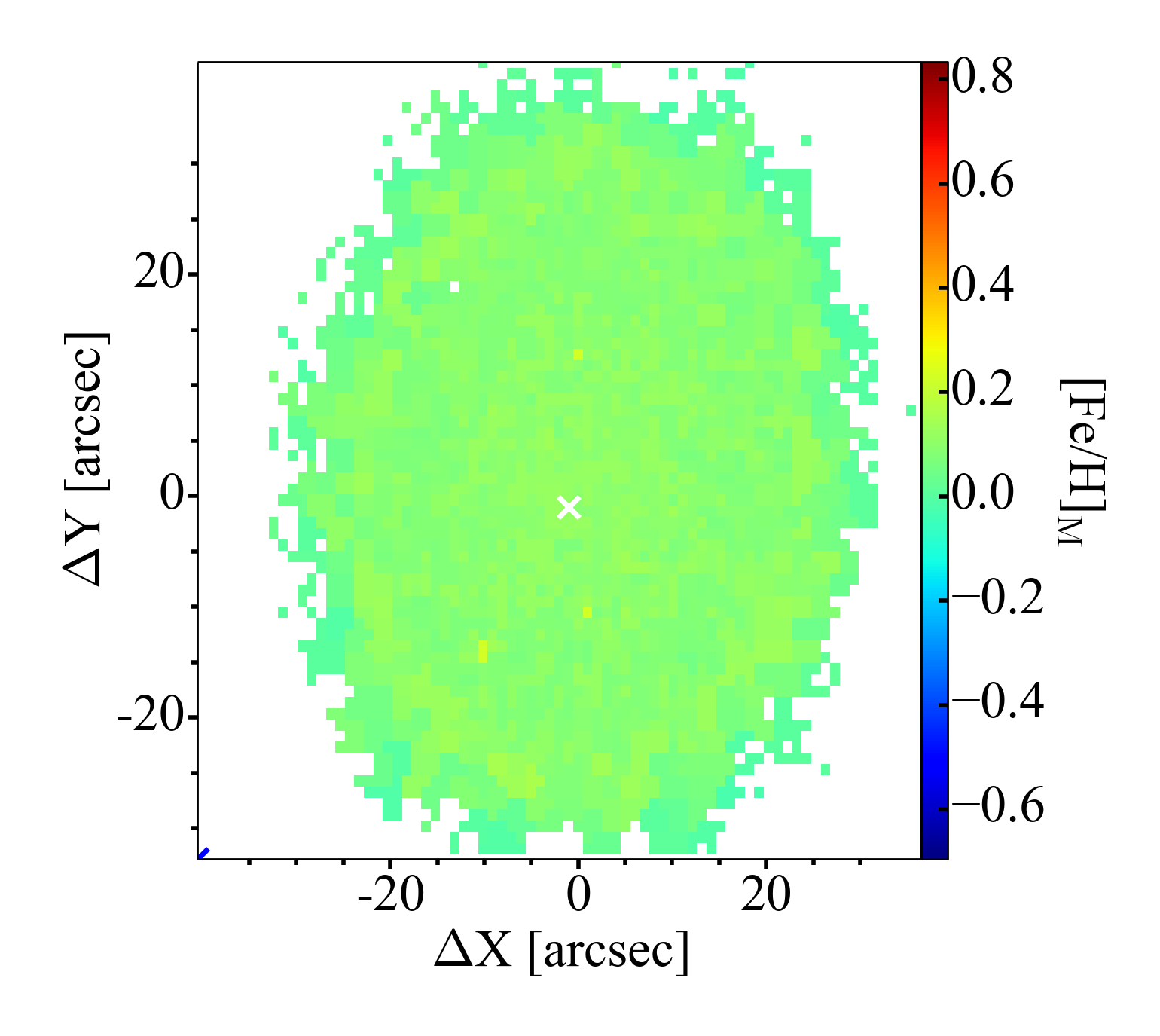}
\includegraphics[width=0.20\columnwidth]{blank.png}
\includegraphics[width=0.20\columnwidth]{blank.png}

\caption{Same as in Fig. \ref{fig:2Dmapsa}, but for NGC 5485.} 
\label{fig:2Dmapsb}
\end{center}
\end{sidewaysfigure*}

\subsection{CALIFA/STECKMAP} \label{sec:steckmap}

The second method for extracting stellar population information from the CALIFA data cubes is based on a spectral feature synthesis approach using STECKMAP code \citep[STEllar Content and Kinematics via Maximum A Posteriori likelihood; ][]{Ocvirketal2006a,Ocvirketal2006b}.  Previous to the spectral fitting, pre-processing steps include spatial masking of foreground/background sources, very low S/N spaxels, and bad pixels.  Although CALIFA/STECKMAP spatially bins the data cubes using also the centroidal Voronoi tessellation routine describe in \citet{Cappellarietal2003}, the minimum S/N required is more restrictive (40 per $\AA$ at 5800$\AA$) than for the CALIFA/STARLIGHT method. This conservative restriction produces a different Voronoi segmentation than the one used in CALIFA/STARLIGHT but ensures a reliable determination of the stellar population properties. STECKMAP is run on the emission lines cleaned spectra, where the emission line cleaning has been performed with the code GANDALF \citep{Sarzietal2006}.

STECKMAP is a Bayesian method that simultaneously recovers the kinematic and stellar population  properties via a maximum a posteriori algorithm. STECKMAP projects the observed spectrum onto a temporal sequence of models of SSPs to determine the linear combination that better fits the observed spectrum. The stellar content of the object is indicated by the weights of the various components of this linear combination, thus the method does not assume the shape of the star formation history. STECKMAP uses a penalized $\chi$$^{2}$ that imposes high penalization values for solutions with strong oscillations (i.e., a rapid variation of the metallicity with age or a noisy broadening function) and small penalization values for smoothly varying solutions. This initial condition avoids extreme oscillating solutions that are not robust and most likely unphysical.  We note that this method does not use the continuum in the derivation of the stellar population parameters. The model is multiplied by a smooth non parametric transmission curve. This curve extends uniformly along the wavelength range. By using this curve to remove the continuum, no extinction correction needs to be applied as dust extinction does not change the equivalent width of the absorption lines. Therefore, an extinction law is not assumed. This technique avoids spurious results due to possible flux calibration errors or extinction. For details about the analysis method see \citet{SanchezBlazquezetal2011, SanchezBlazquezetal2014}, while the performance of STECKMAP is described in \citet{Ocvirketal2006a, Ocvirketal2006b}.  The typical STECKMAP outputs give the proportion of stars at each age that are contributing to the observed flux and to the stellar mass and the evolution of the metallicity with time. 

\begin{figure}
\begin{center}
\includegraphics[width=0.90\columnwidth]{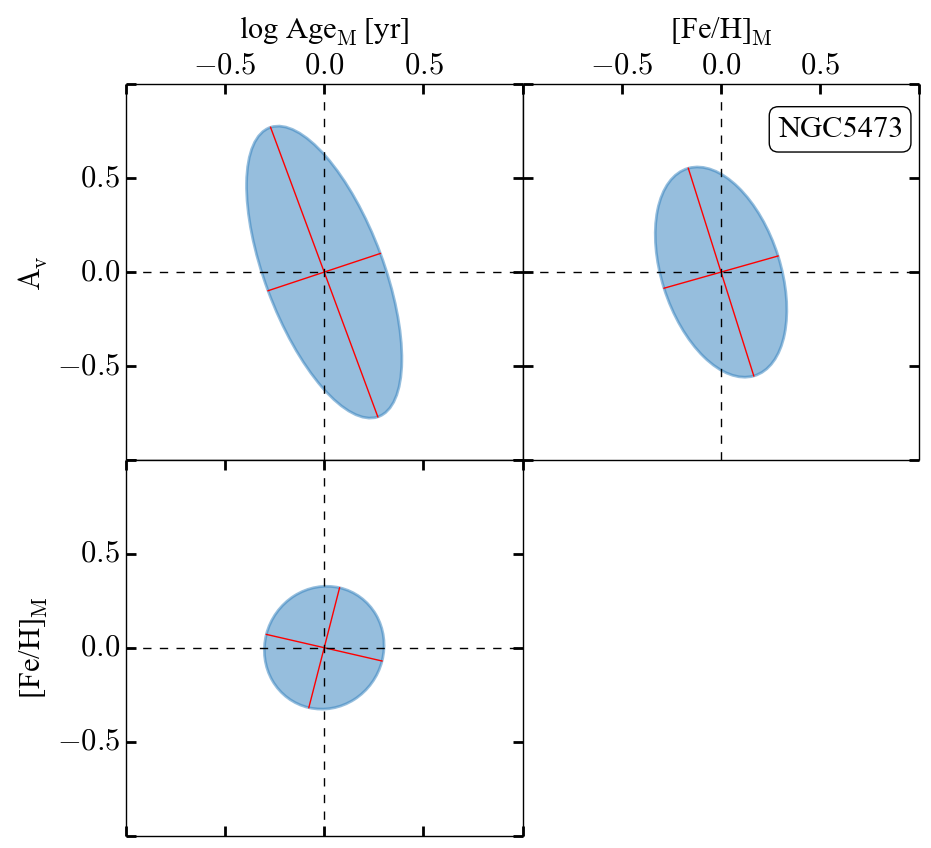}
\includegraphics[width=0.90\columnwidth]{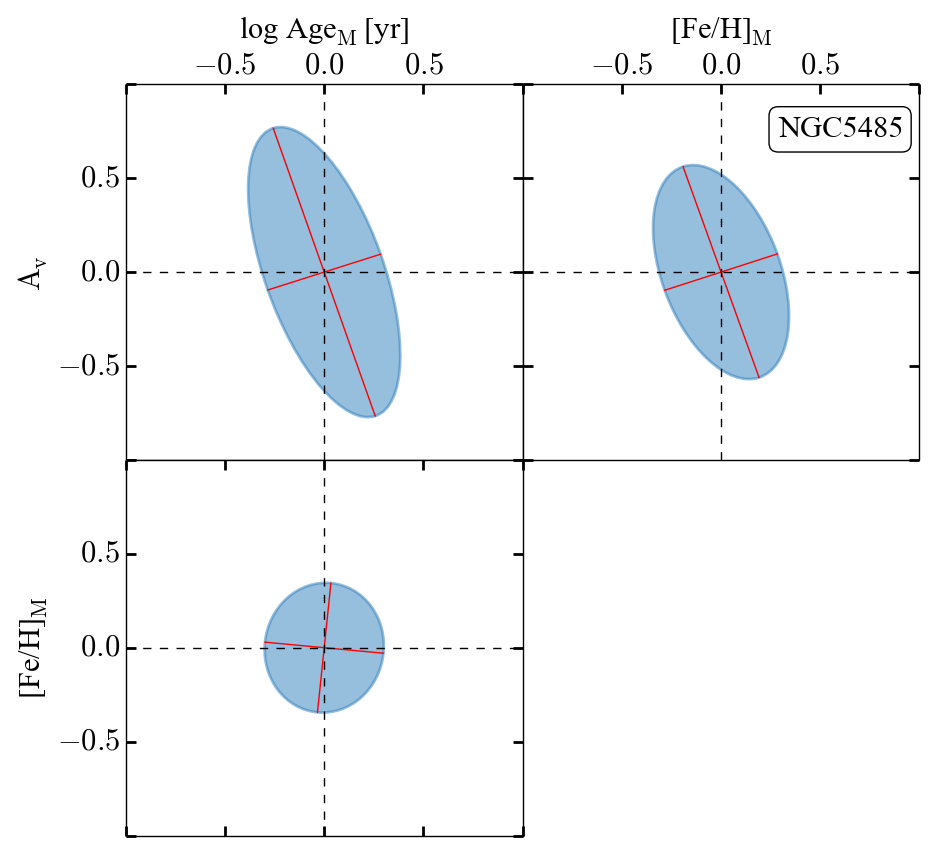}
\caption{Covariance error ellipses of the stellar population parameters for NGC 5473 (top panel) and NGC 5485 (bottom panel) using J-PLUS/MUFFIT.} 
\label{fig:degeneracies}
\end{center}
\end{figure}

\begin{figure*}
\begin{center}
\includegraphics[width=0.45\columnwidth, angle=90]{JPLUS_MUFFIT.jpg}
\includegraphics[width=0.60\columnwidth]{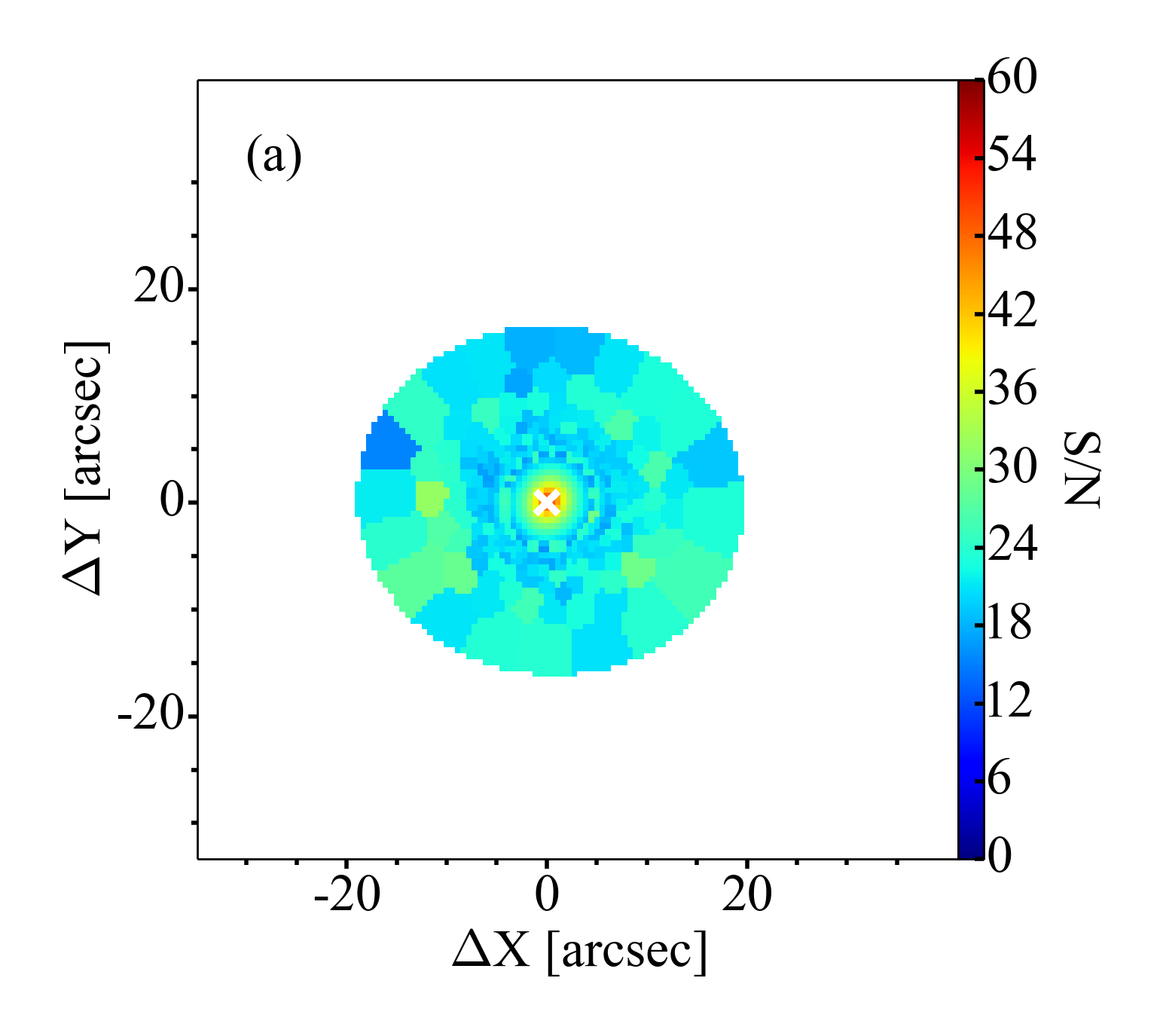}
\includegraphics[width=0.60\columnwidth]{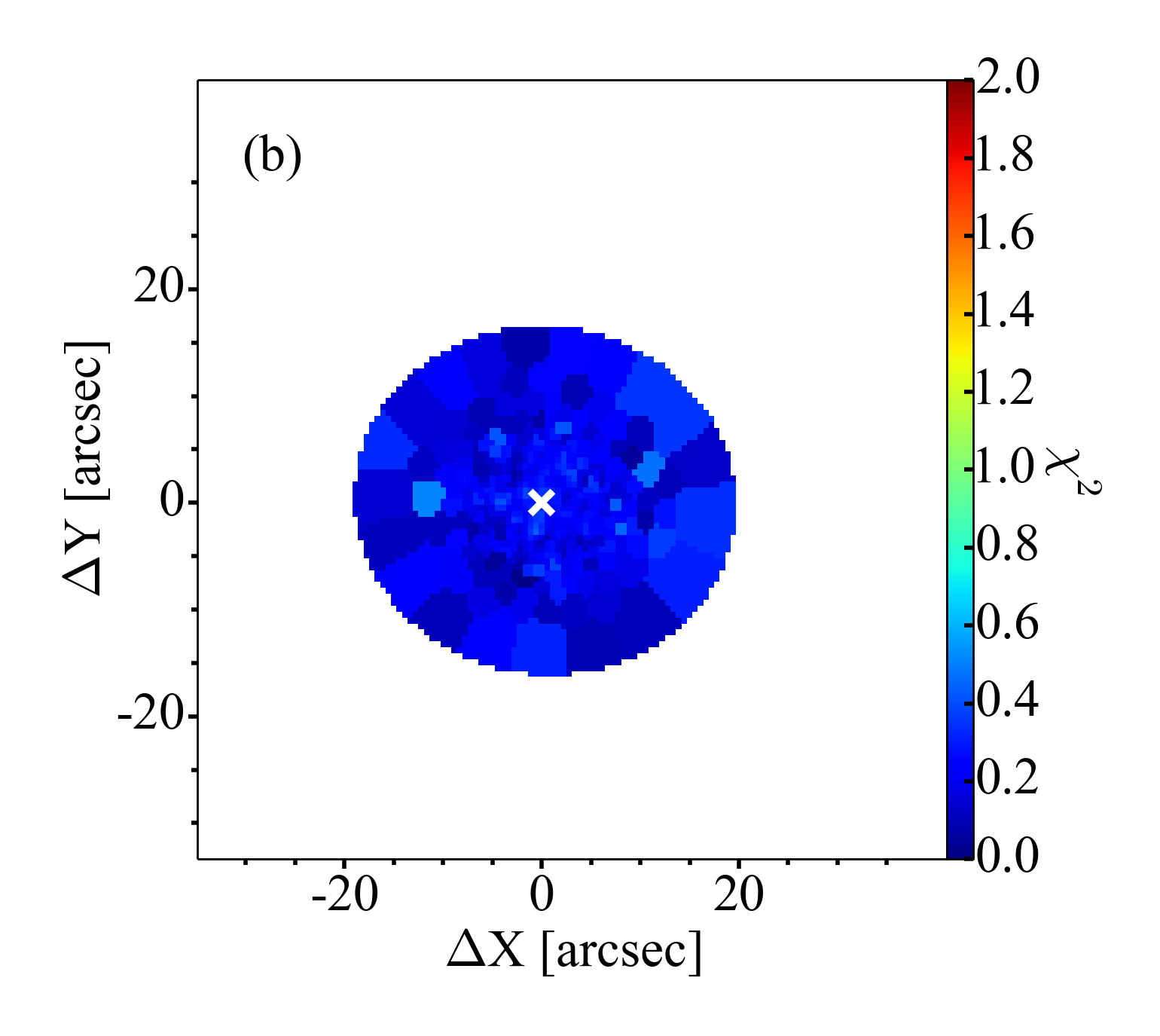}
\includegraphics[width=0.60\columnwidth]{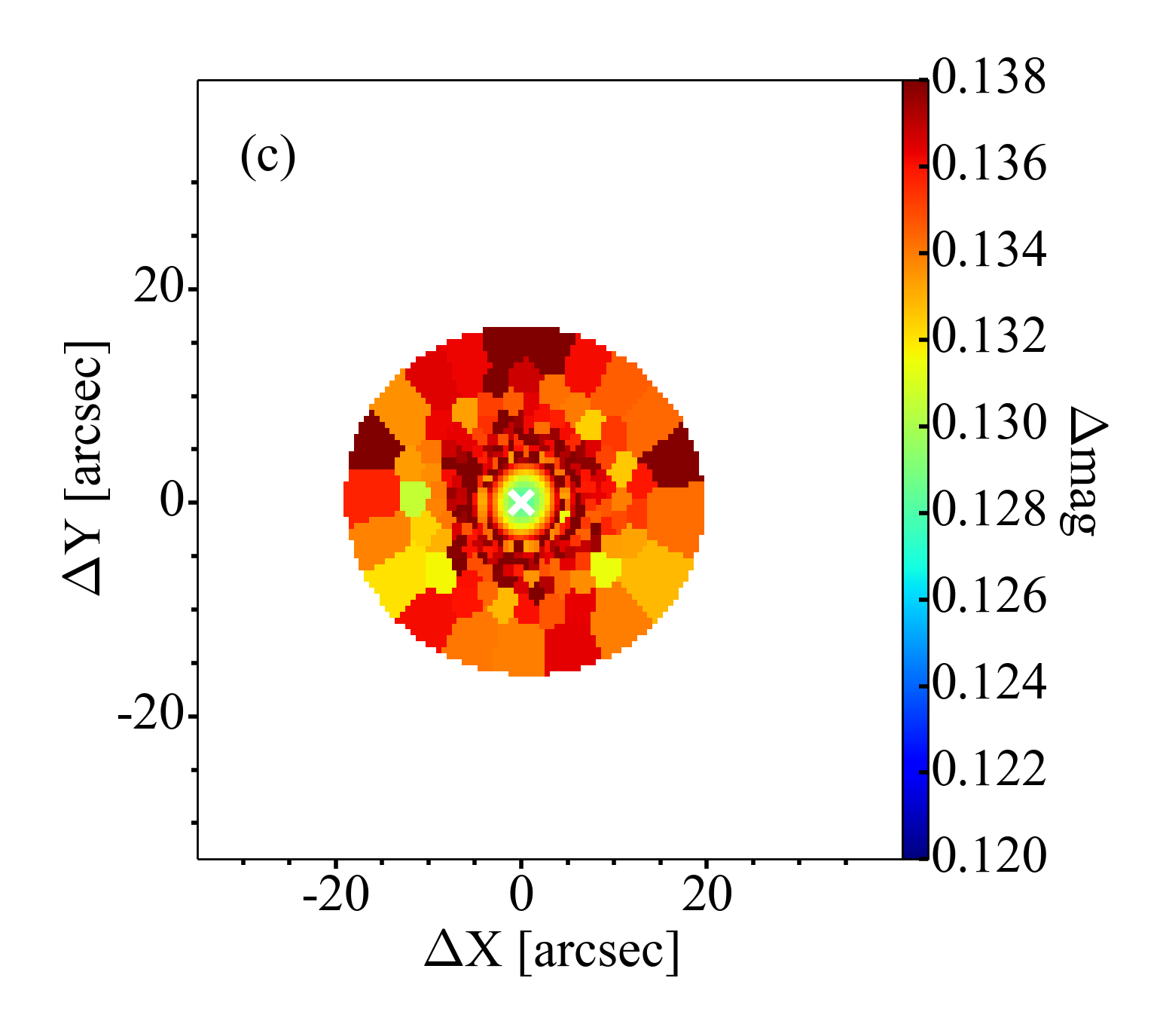}

\includegraphics[width=0.45\columnwidth, angle=90]{CALIFA_STARLIGHT.jpg}
\includegraphics[width=0.60\columnwidth]{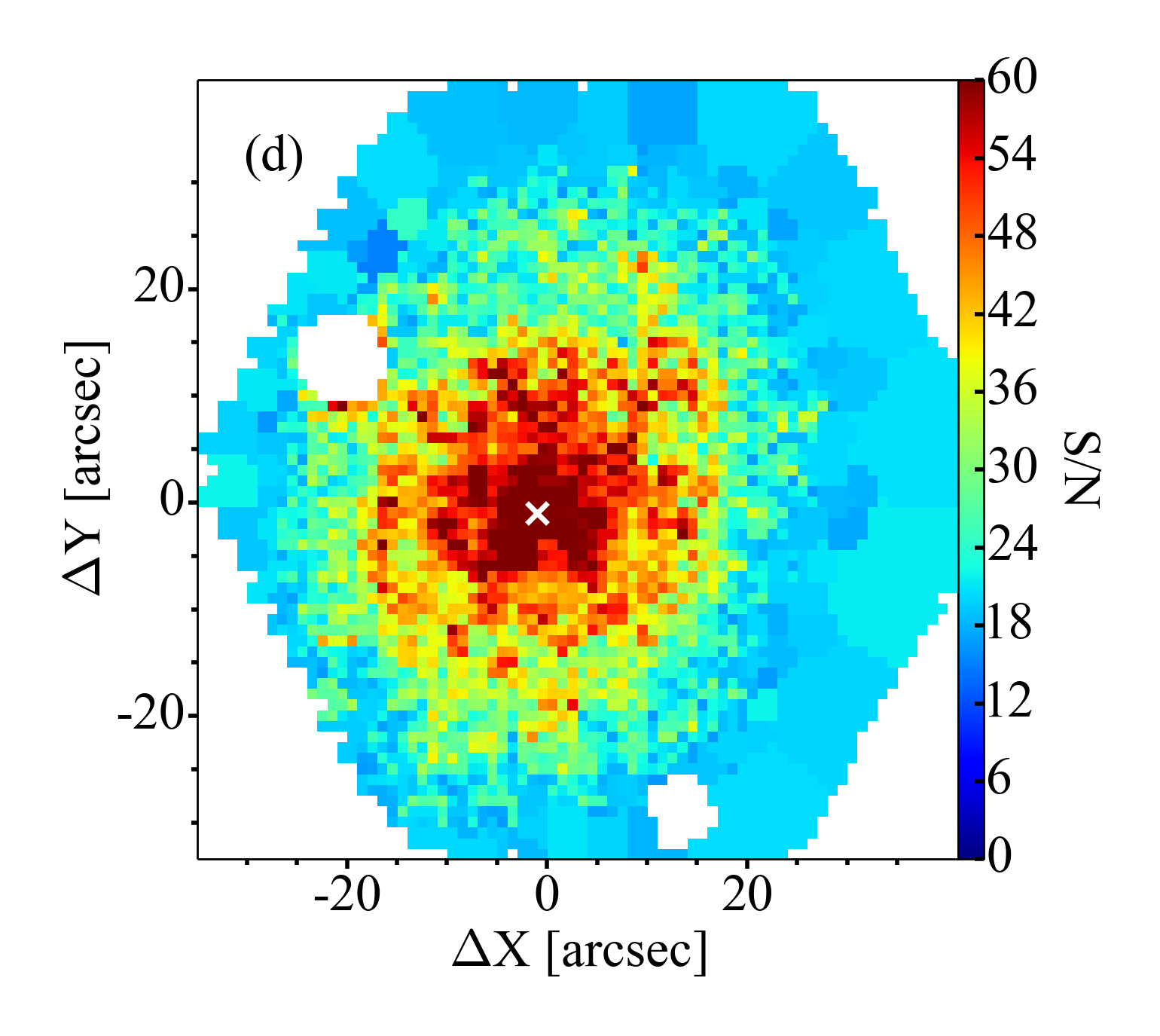}
\includegraphics[width=0.60\columnwidth]{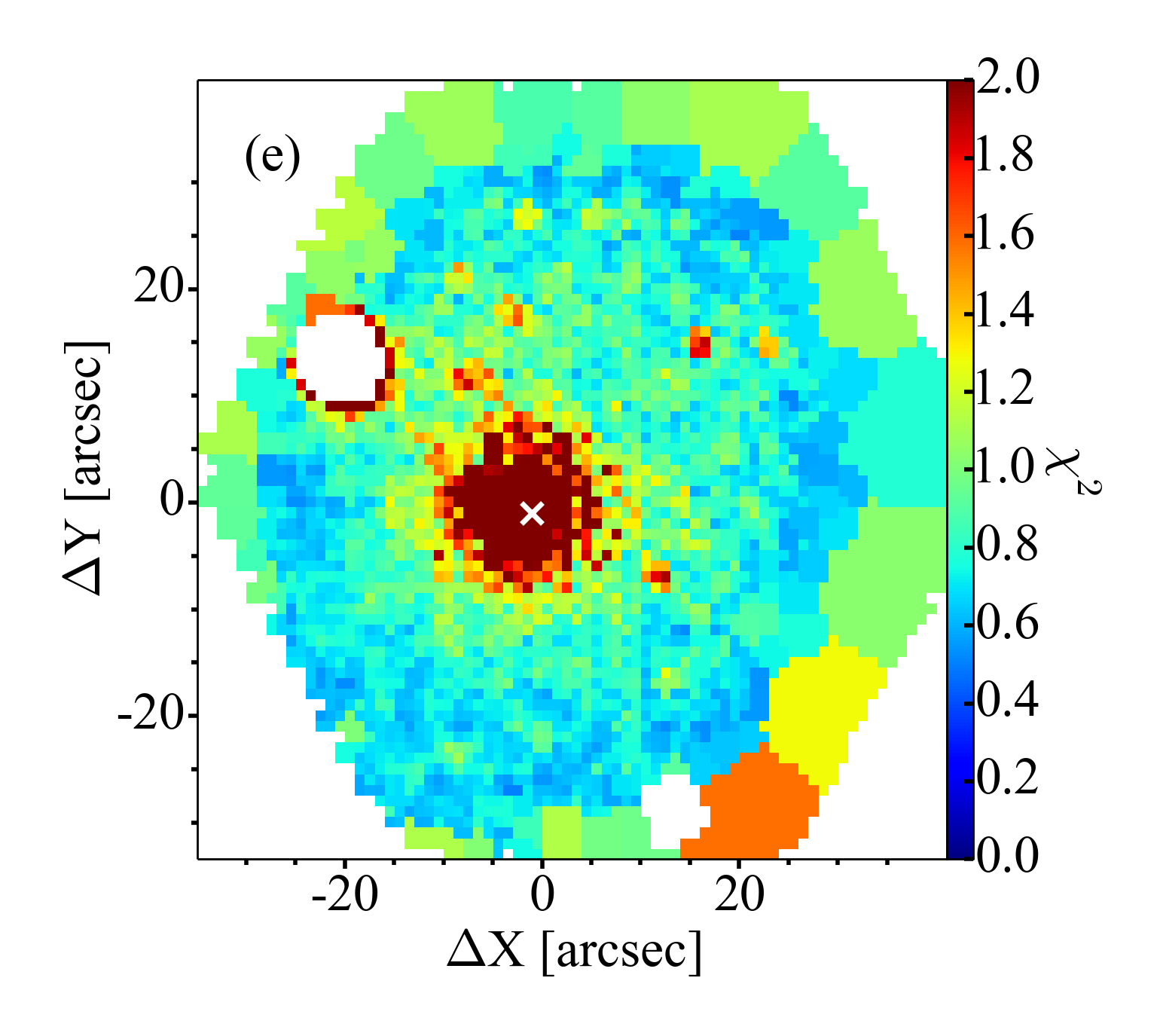}
\includegraphics[width=0.60\columnwidth]{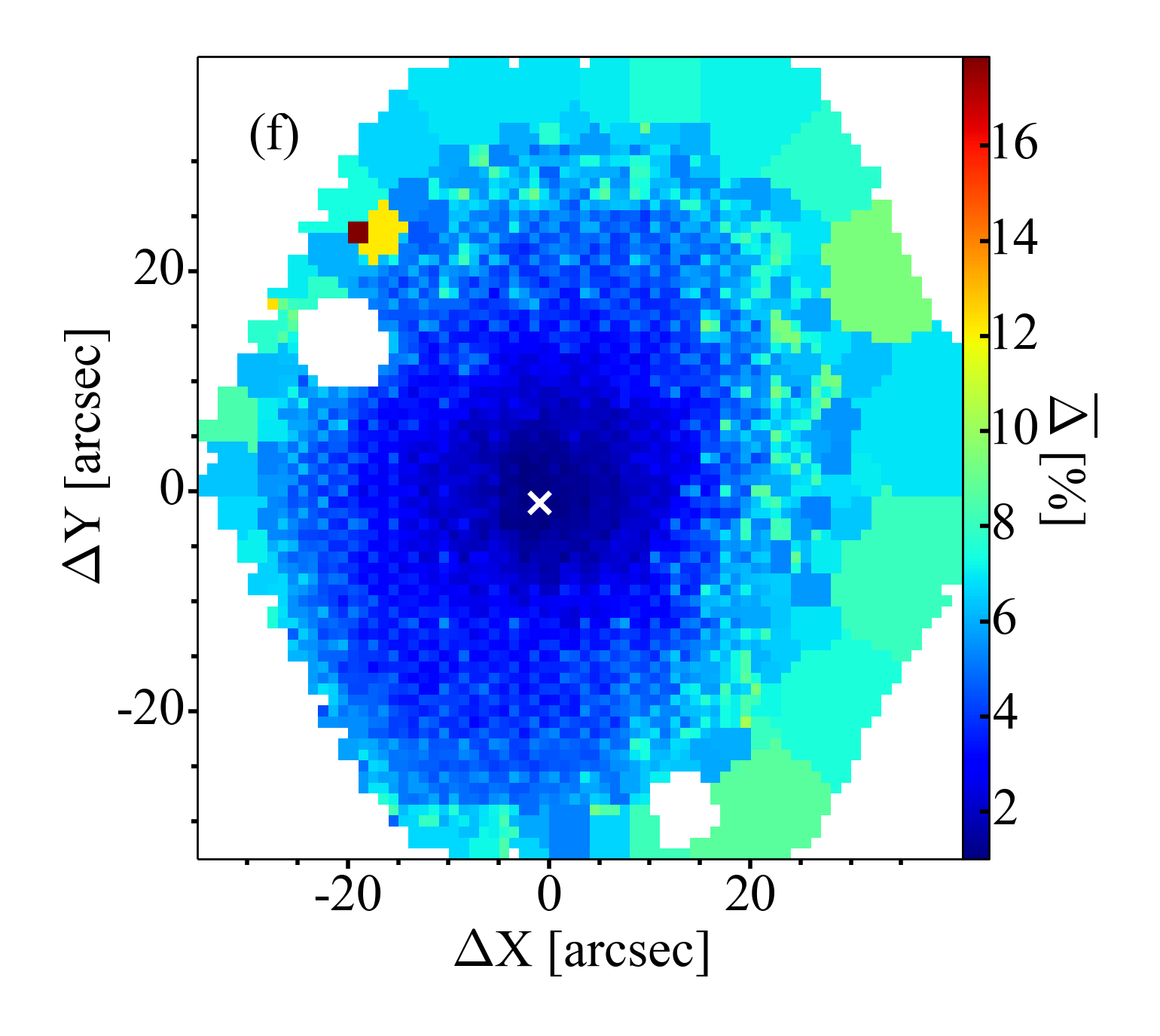}

\includegraphics[width=0.45\columnwidth, angle=90]{CALIFA_STECKMAP.jpg}
\includegraphics[width=0.60\columnwidth]{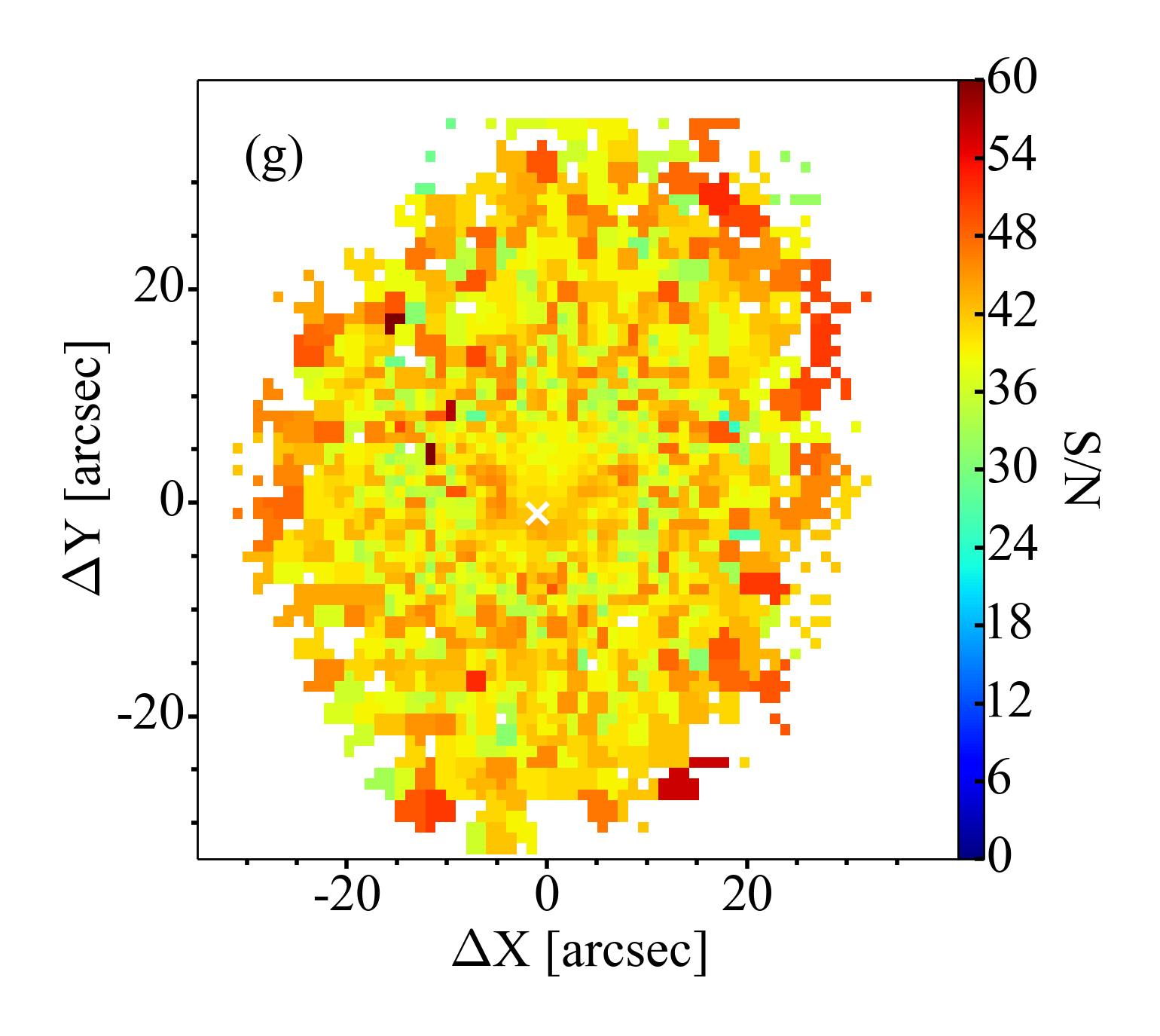}
\includegraphics[width=0.60\columnwidth]{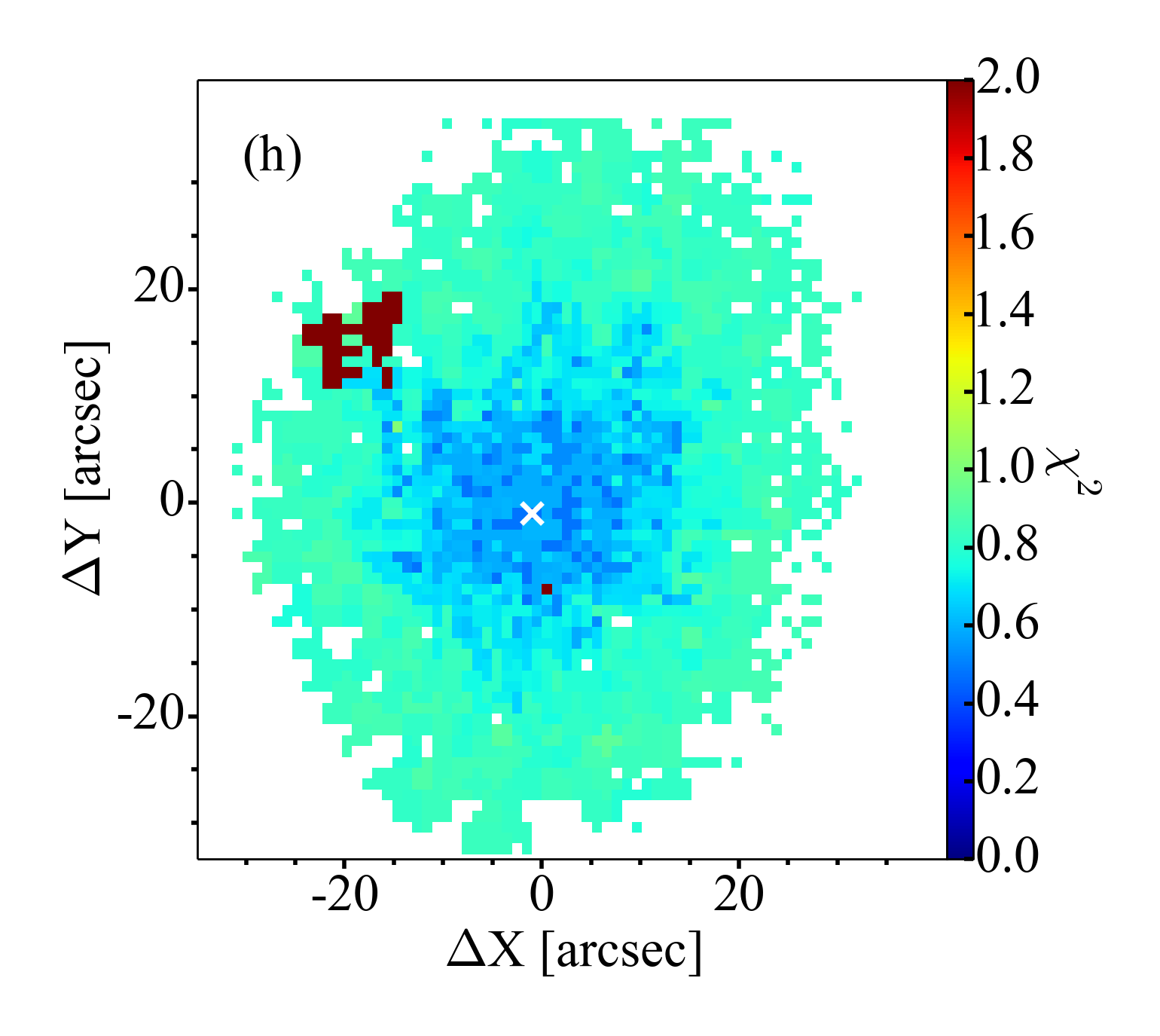}
\includegraphics[width=0.60\columnwidth]{blank.png}

\caption{Maps with data and fit quality indicators for NGC 5473 determined by J-PLUS/MUFFIT (top panels), CALIFA/STARLIGHT (middle panels) and CALIFA/STECKMAP (bottom panels) . The center of the galaxy is marked with a white cross in each panel.} 
\label{fig:fitsa}
\end{center}
\end{figure*}

\begin{figure*}
\begin{center}
\includegraphics[width=0.45\columnwidth, angle=90]{JPLUS_MUFFIT.jpg}
\includegraphics[width=0.60\columnwidth]{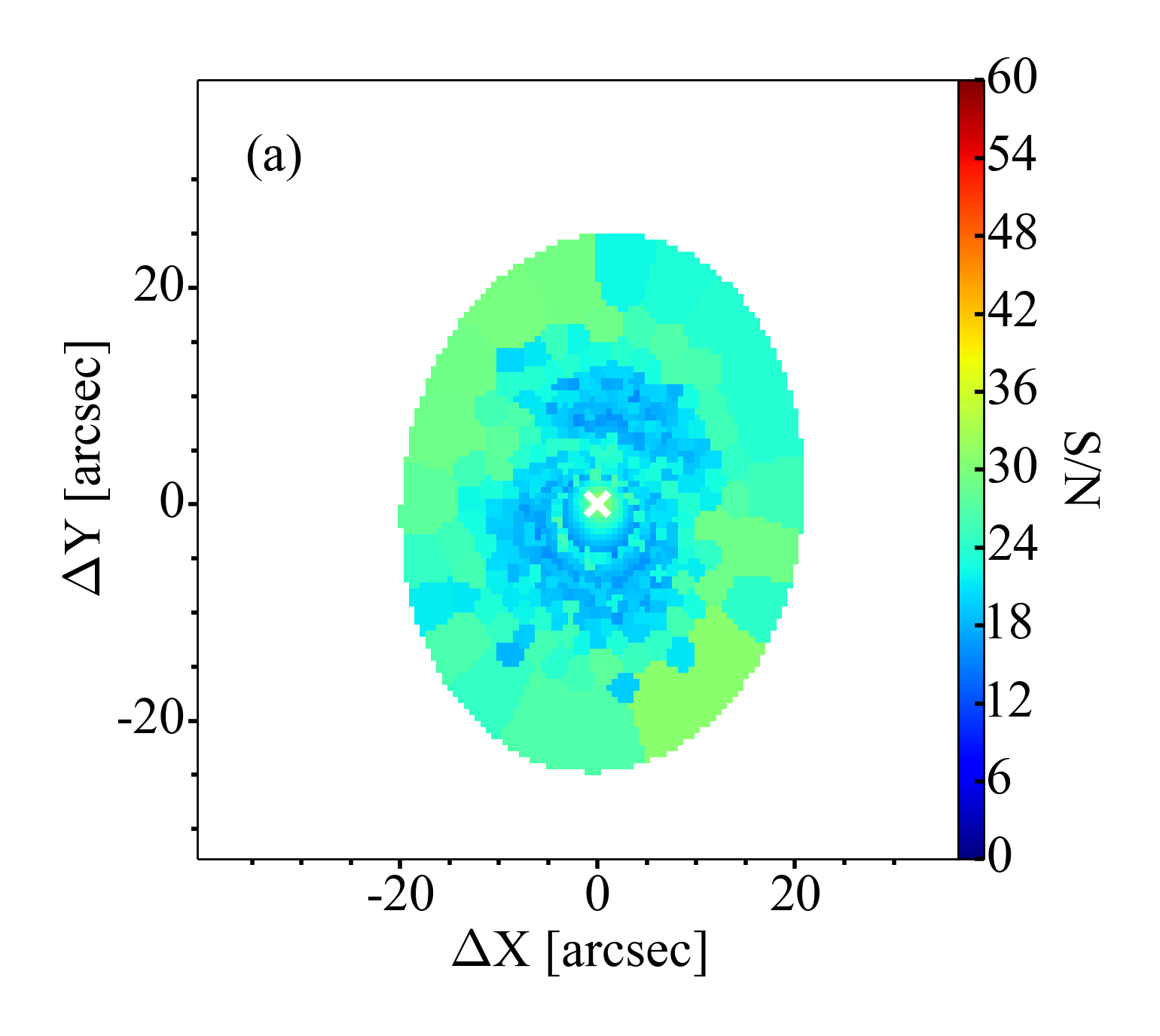}
\includegraphics[width=0.60\columnwidth]{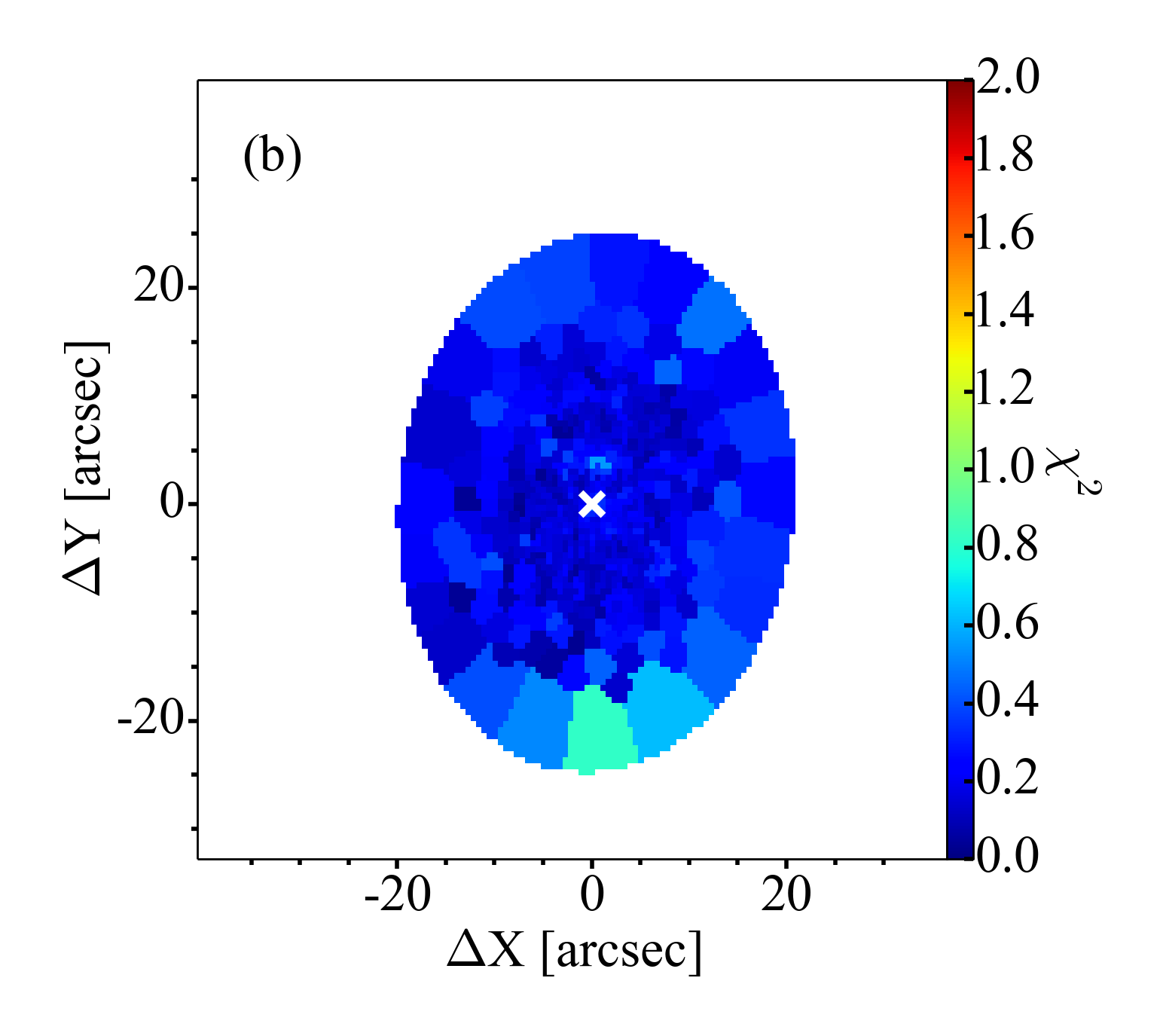}
\includegraphics[width=0.60\columnwidth]{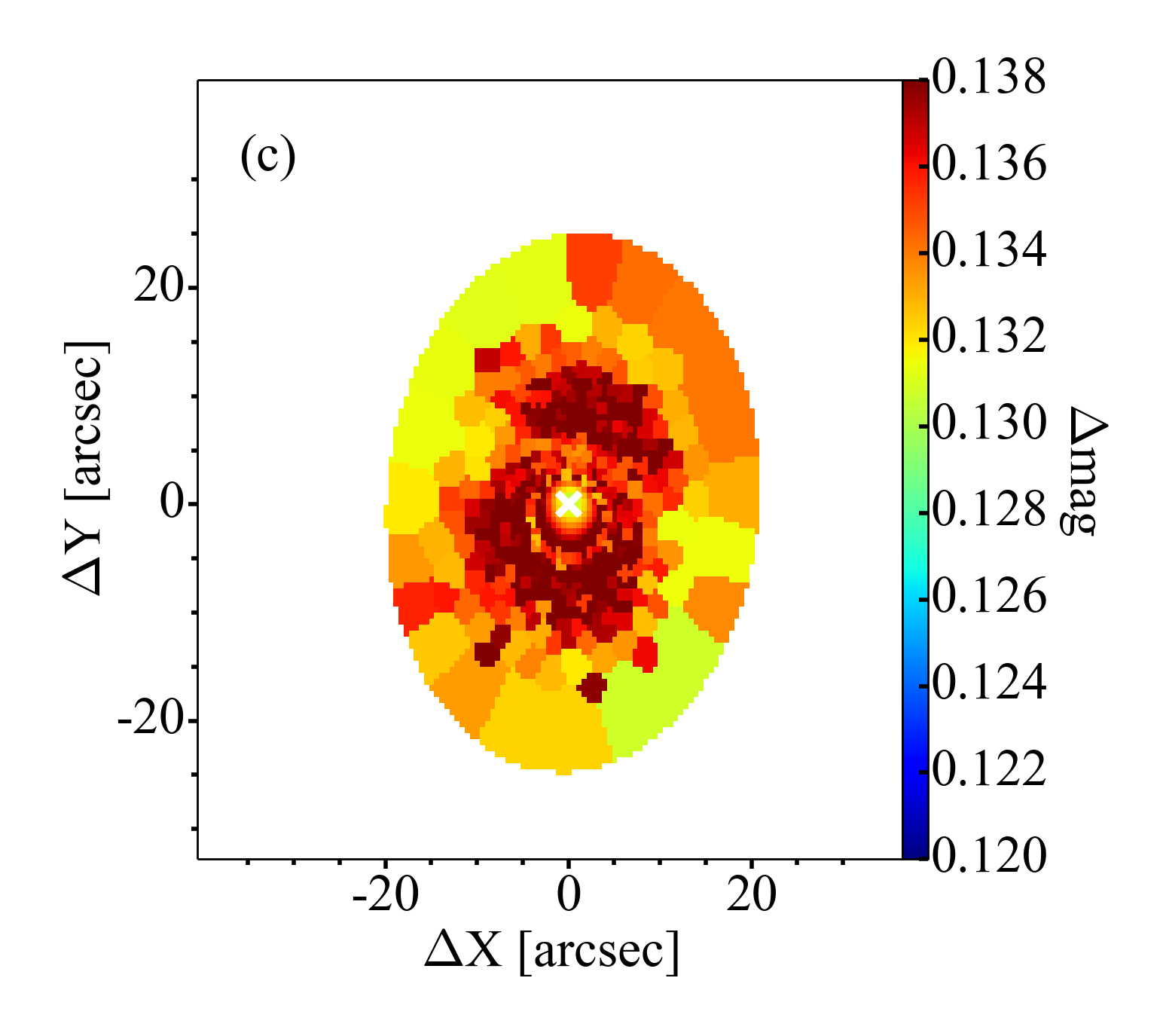}

\includegraphics[width=0.45\columnwidth, angle=90]{CALIFA_STARLIGHT.jpg}
\includegraphics[width=0.60\columnwidth]{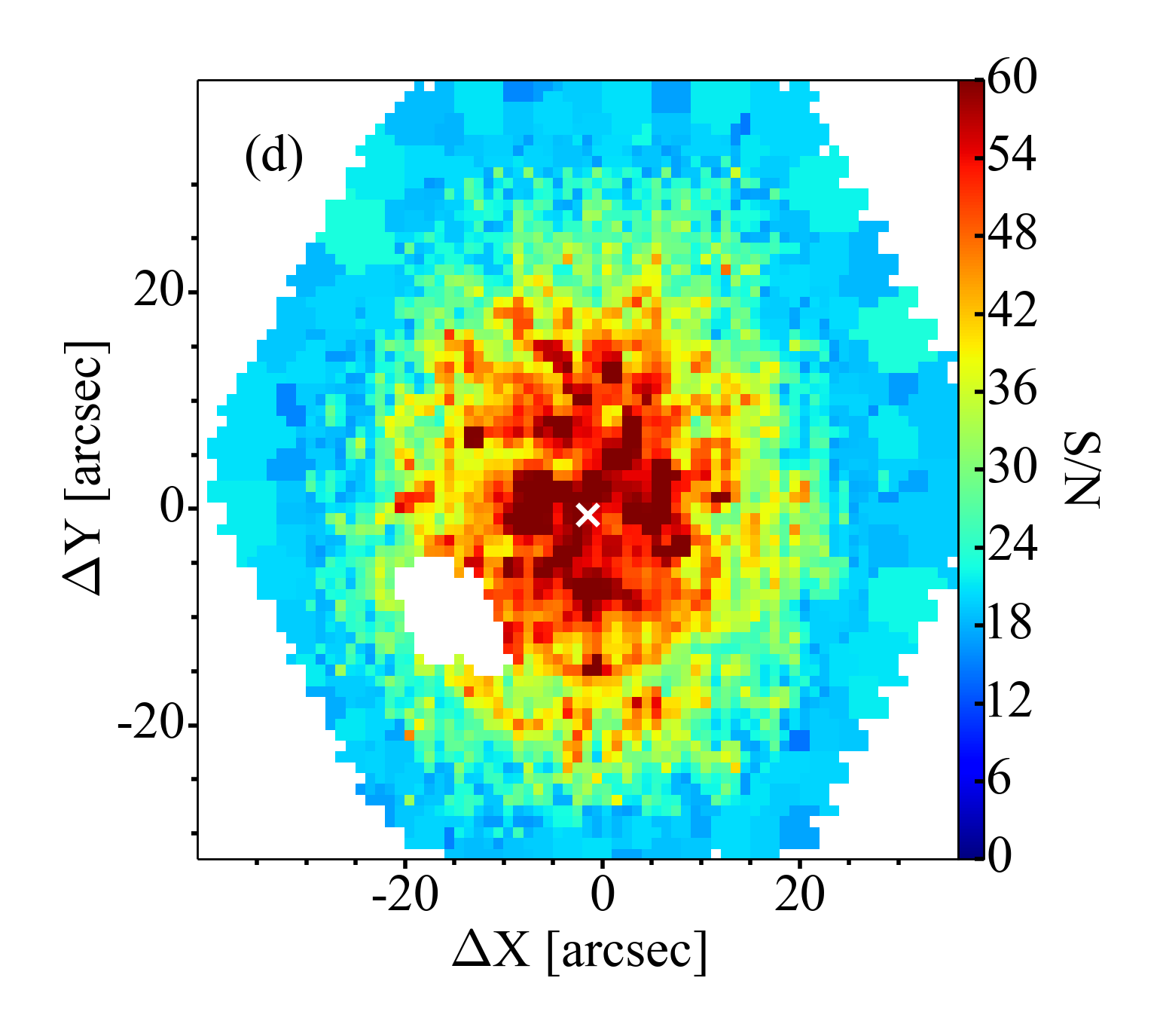}
\includegraphics[width=0.60\columnwidth]{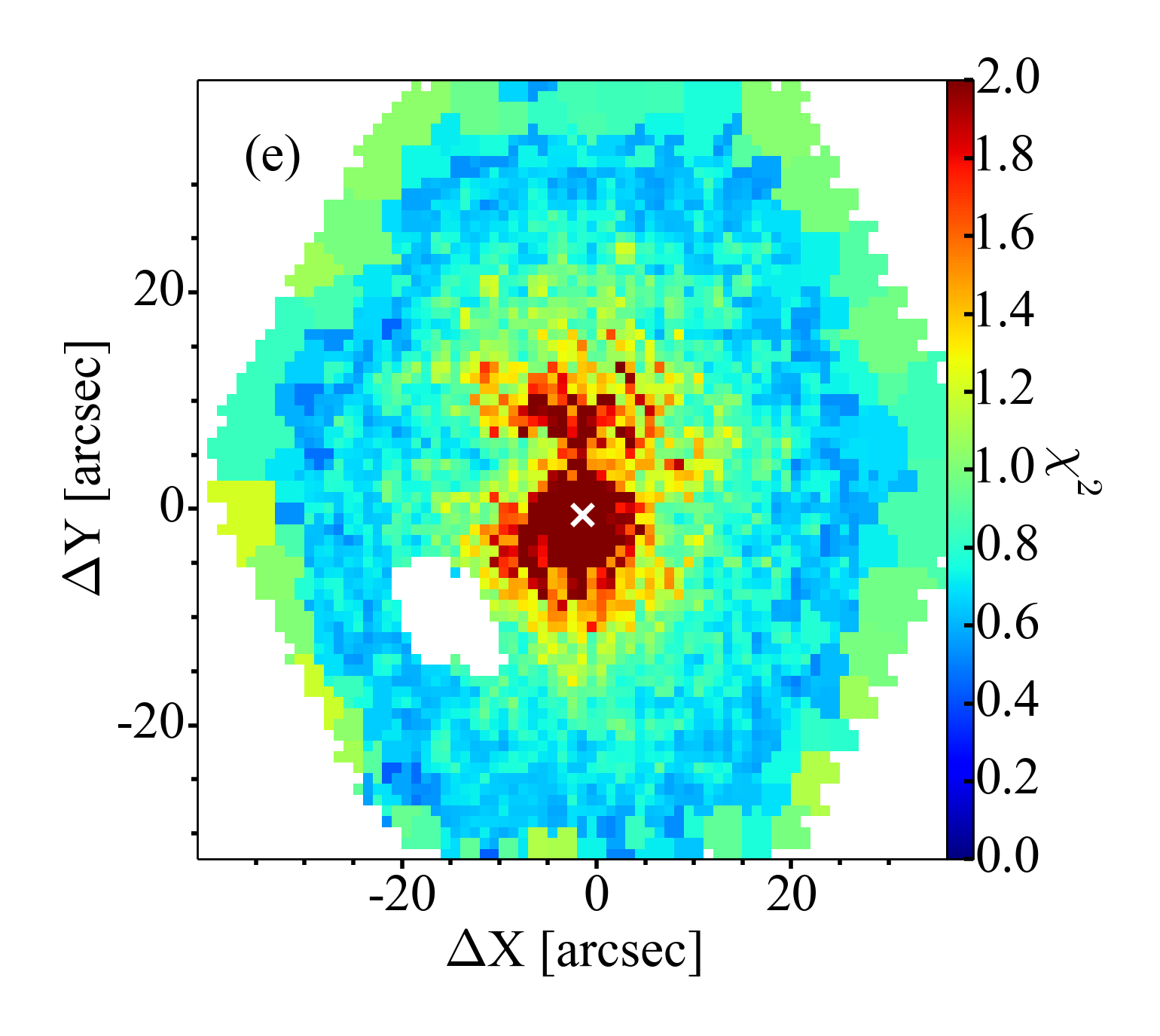}
\includegraphics[width=0.60\columnwidth]{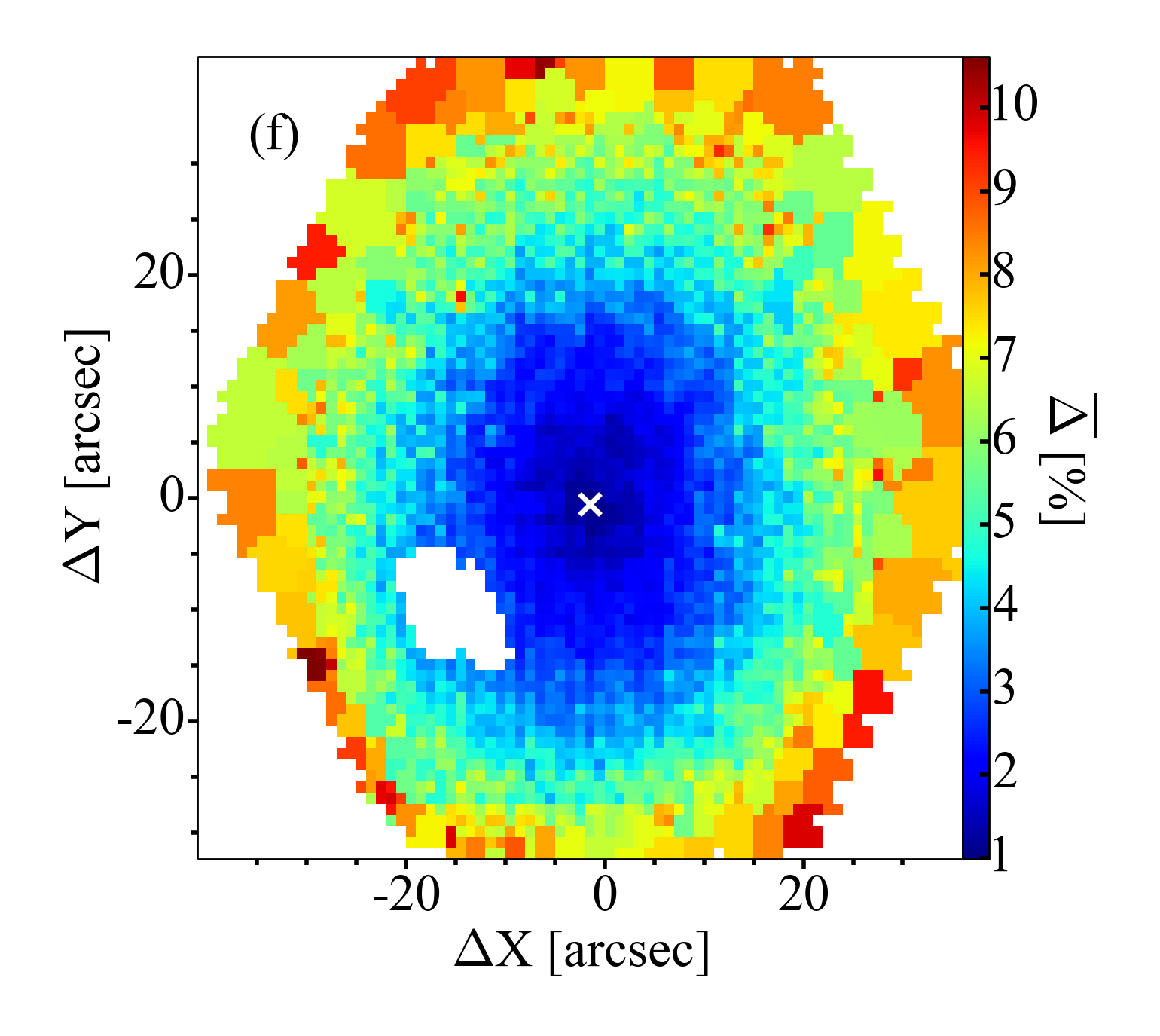}

\includegraphics[width=0.45\columnwidth, angle=90]{CALIFA_STECKMAP.jpg}
\includegraphics[width=0.60\columnwidth]{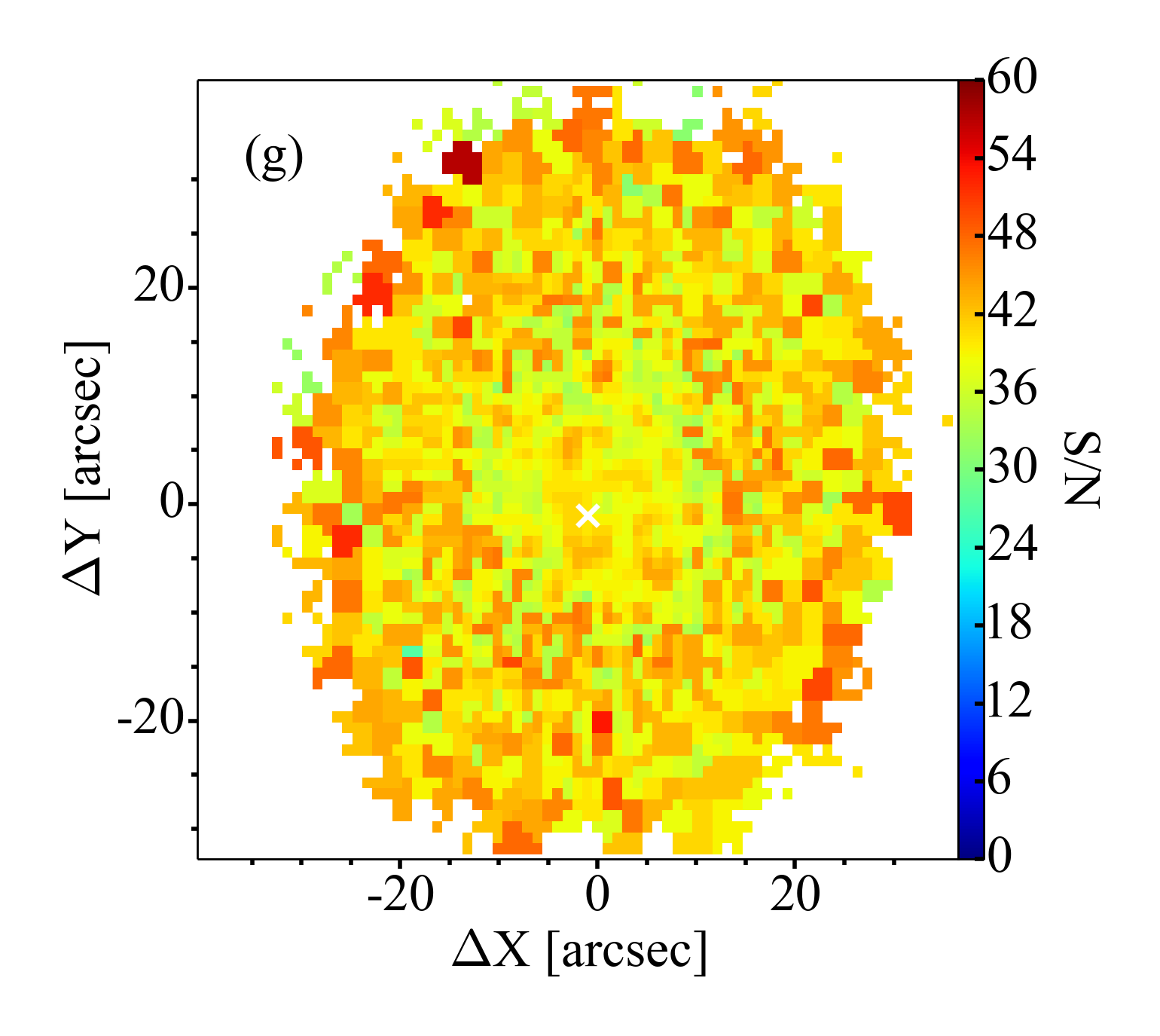}
\includegraphics[width=0.60\columnwidth]{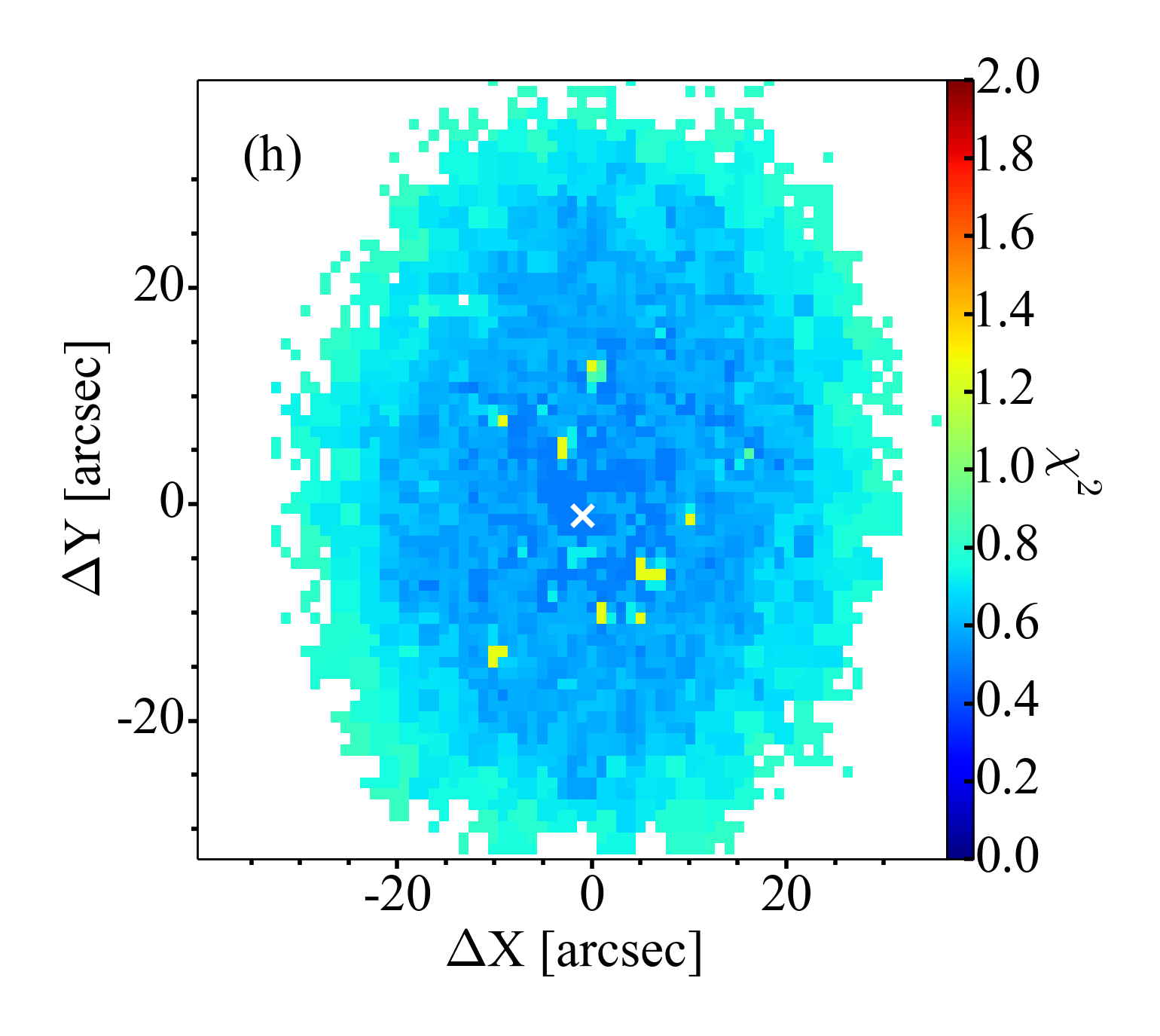}
\includegraphics[width=0.60\columnwidth]{blank.png}

\caption{Same as in Fig. \ref{fig:fitsa}, but for NGC 5485.} 
\label{fig:fitsb}
\end{center}
\end{figure*}

\section{2D Maps}\label{sec:map}

As explained in the previous section, J-PLUS/MUFFIT and CALIFA/STARLIGHT provide luminosity- and mass-weighted log Age, [Fe/H] and A$_\mathrm{v}$ maps while from the  CALIFA/STECKMAP outputs we can obtain luminosity- and mass-weighted log Age and [Fe/H] maps. Mass-weighted properties are more representative of the whole evolutionary history of the galaxy since they give insight into its mass assembly history. On the other hand, luminosity-weighted properties are better constrained and more sensitive to the fingerprints of the most recent periods of star formation in the galaxy. Throughout this study, we analyze both mass- and luminosity-weighted properties. We present in this section the mass-weighted properties maps. For completeness, Appendix \ref{appex} includes the luminosity-weighted maps. 

\subsection{Age, [Fe/H], A$_\mathrm{v}$ and stellar mass surface density}
Figures \ref{fig:2Dmapsa} and \ref{fig:2Dmapsb} show the 2D maps of the stellar populations for NGC 5473 and NGC 5485 derived with the different methods. The maps were shifted to center the galaxies and facilitate the comparison. The center of each galaxies (white crosses in each 2D maps) has been derived using the IRAF task ELLIPSE to fit elliptical isophotes to the stellar mass surface density maps.  Isophotes were fitted between 0.1 arcsec and the largest measurable semi-major axis. The overall center position was determined as the average of ELLIPSE output between 0.5 and 1.5 arcsec along the semi-major axis where the measurements are more reliable.

Figure \ref{fig:2Dmapsa} shows some differences between the values derived by each method.  NGC 5473 shows in the J-PLUS/MUFFIT maps a smooth behavior in log Age$_\mathrm{M}$, [Fe/H]$_\mathrm{M}$, and A$_\mathrm{v}$ suggesting flat age and metallicity gradients. These results are in agreement with the relatively flat age and metallicity maps derived by CALIFA/STECKMAP.  In contrast, CALIFA/STARLIGHT map suggests a mild negative age gradient. In addition, the upper part of the galaxy seems to be more metal-rich than the lower part in the CALIFA/STARLIGHT metallicity map. Significant differences in the extinction parameter, A$_\mathrm{v}$, are found between J-PLUS/MUFFIT and CALIFA/STARLIGHT. While J-PLUS/MUFFIT obtained a significant dust component, smoothly distributed within the galaxy, the CALIFA/STARLIGHT analysis finds A$_\mathrm{v}$ = 0 across the whole galaxy except in the central region.  We note that the older and metal-rich area ($\Delta$X, $\Delta$Y = --20", 15") visible in the CALIFA/STECKMAP corresponds to a background galaxy not masked during the pre-processing steps.

NGC 5485 (Fig. \ref{fig:2Dmapsb}) also presents a relative smooth log Age$_\mathrm{M}$ and [Fe/H]$_\mathrm{M}$ maps in J-PLUS/MUFFIT, but a higher extinction area (A$_\mathrm{v}$ $\sim$ 1.2) is present in the upper part of the galaxy in the A$_\mathrm{v}$ map. This high extinction seems to be associated with the prominent minor-axis dust lane visible in the colored images (Fig. \ref{fig:1}). We note that the [Fe/H] map shows a slightly more metal-rich population in that specific area that could be produce by a potential metallicity-extinction degeneracy.  CALIFA/STARLIGHT is also able to detect the prominent dust lane although A$_\mathrm{v}$ values are significantly lower than the values obtained by J-PLUS/MUFFIT. CALIFA/STECKMAP shows smooth log Age$_\mathrm{M}$ and [Fe/H]$_\mathrm{M}$ maps, although obtains an older population. The log Age$_\mathrm{M}$ map determined by CALIFA/STARLIGHT exhibits an older component in the center of the galaxy not present in J-PLUS/MUFFIT or CALIFA/STECKMAP maps. This old component seems to have the same position, size and orientation than the dust line crossing the galaxy. We checked that the general results do not significantly vary for luminosity-weighted parameters (see the luminosity-weighted maps in Appendix \ref{appex}). In a recent study, \citet{MartinNavarroetal2018} present a spatially-resolved stellar populations analysis of a sample of 45 elliptical galaxies using the CALIFA survey.  They measure the stellar population properties (age, metallicity,  and [Mg/Fe]) via standard line-strength analysis of the indices H$\beta_{o}$, Fe4383, Fe5015, Fe5270, and Mgb \citep{Wortheyetal1994, Bursteinetal1984}. Overall, they find flat age gradients and negative metallicity gradients. We note that their galaxy sample includes NGC 5485. Visual inspection of the NGC 5485 age map does not show any evidence for the old stellar component present in the CALIFA/STARLIGHT map.

Although some degeneracies are unavoidable, analysis of their extension and potential effects are crucial in order to avoid any misinterpretation. To address the degeneracy problem, we use the stellar population values recovered by MUFFIT during the Monte Carlo approach for both objects in every bin of the tessellation. This approach assumes an independent Gaussian distribution in each filter, centered on the band flux or magnitude, with a standard deviation equal to its photometric error.  Figure \ref{fig:degeneracies} presents the 2D confidence intervals. The ellipses are obtained from the covariance matrix of each distribution and following the method used in \citet{DiazGarciaetal2015}.  A value of the ellipticity  close to zero implies no degeneracy between the two parameters. Furthermore, when the position angle lies on any of the two axes (position angle multiple of $\pi$/2), the two parameters are not correlated and no degeneracy is found. Figure \ref{fig:degeneracies} shows no age-metallicity degeneracy but presents a degeneracy between A$_\mathrm{v}$ and the other two parameters. This means that a stellar population reddened by extinction can mimic a metal-rich population or an old one.  J-PLUS/MUFFIT  provides typical uncertainties of $\Delta$log Age$_\mathrm{M}$ =  0.18 dex, $\Delta$[Fe/H]$_\mathrm{M}$ =  0.3 dex, and $\Delta$A$_\mathrm{v}$ = 0.5 for our specific target galaxies. These parameter errors are determined as the best solution space based on the Monte Carlo method. A comprehensive discussion about the intrinsic uncertainties and degeneracies of MUFFIT using J-PLUS data (both simulated and real galaxies) is out of the scope of this paper and will be presented in a future work. CALIFA/STARLIGHT does not provide direct error estimates in its output. Based on simulations by \citet{CidFernandesetal2014} and \citet{deAmorimetal2017}, estimated uncertainties of physical quantities obtained by  STARLIGHT are $\Delta$Age$_\mathrm{M}$ = 1.4 Gyrs, $\Delta$[Fe/H]$_\mathrm{M}$ = 0.12 dex, and $\Delta$A$_\mathrm{v}$ = 0.05. These uncertainty estimates must be interpreted as approximate, as they are based on experiments with a single galaxy, and do not take into consideration sources of error other than random noise.

\begin{figure}
\begin{center}
\includegraphics[width=0.90\columnwidth]{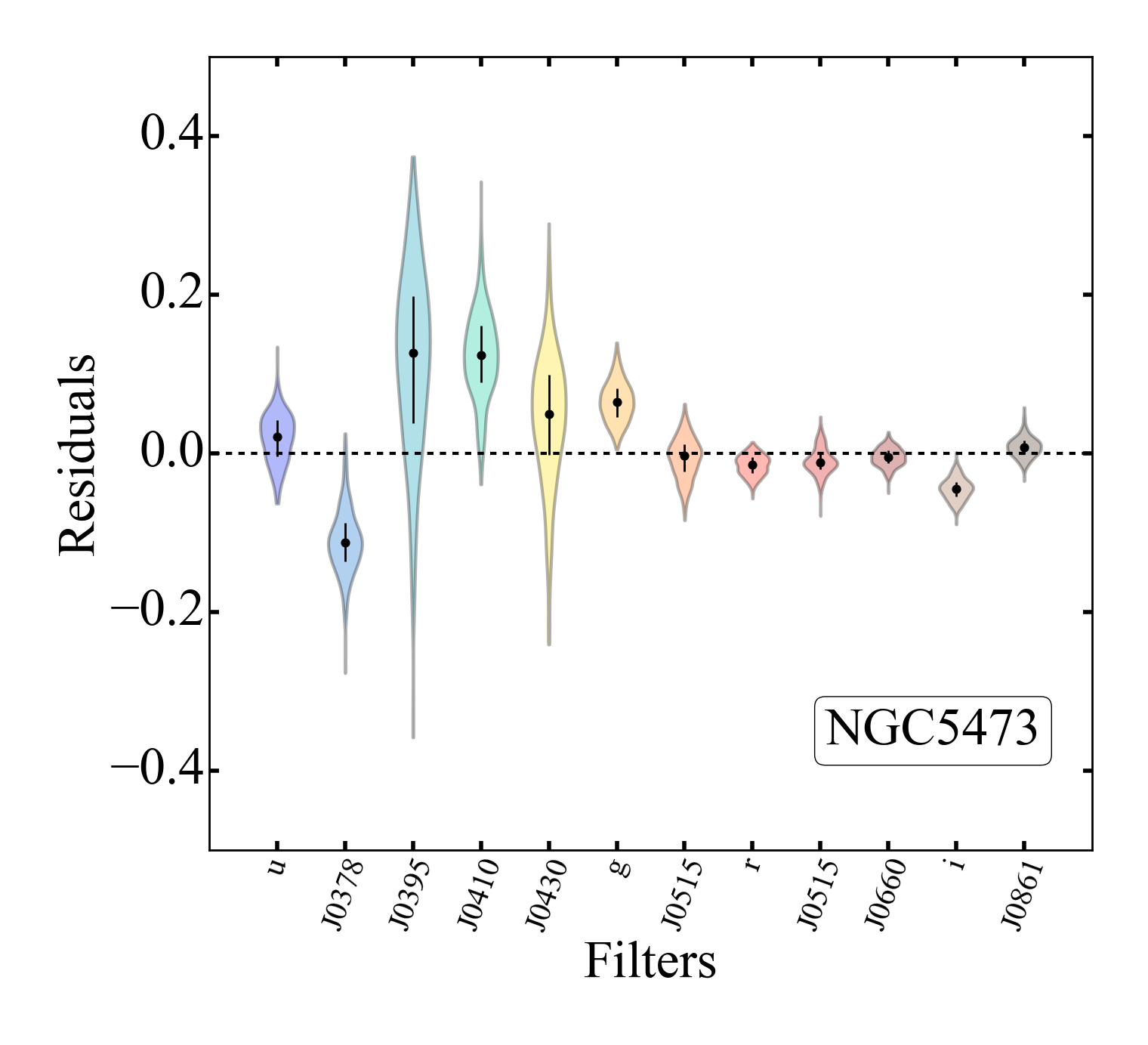}
\includegraphics[width=0.90\columnwidth]{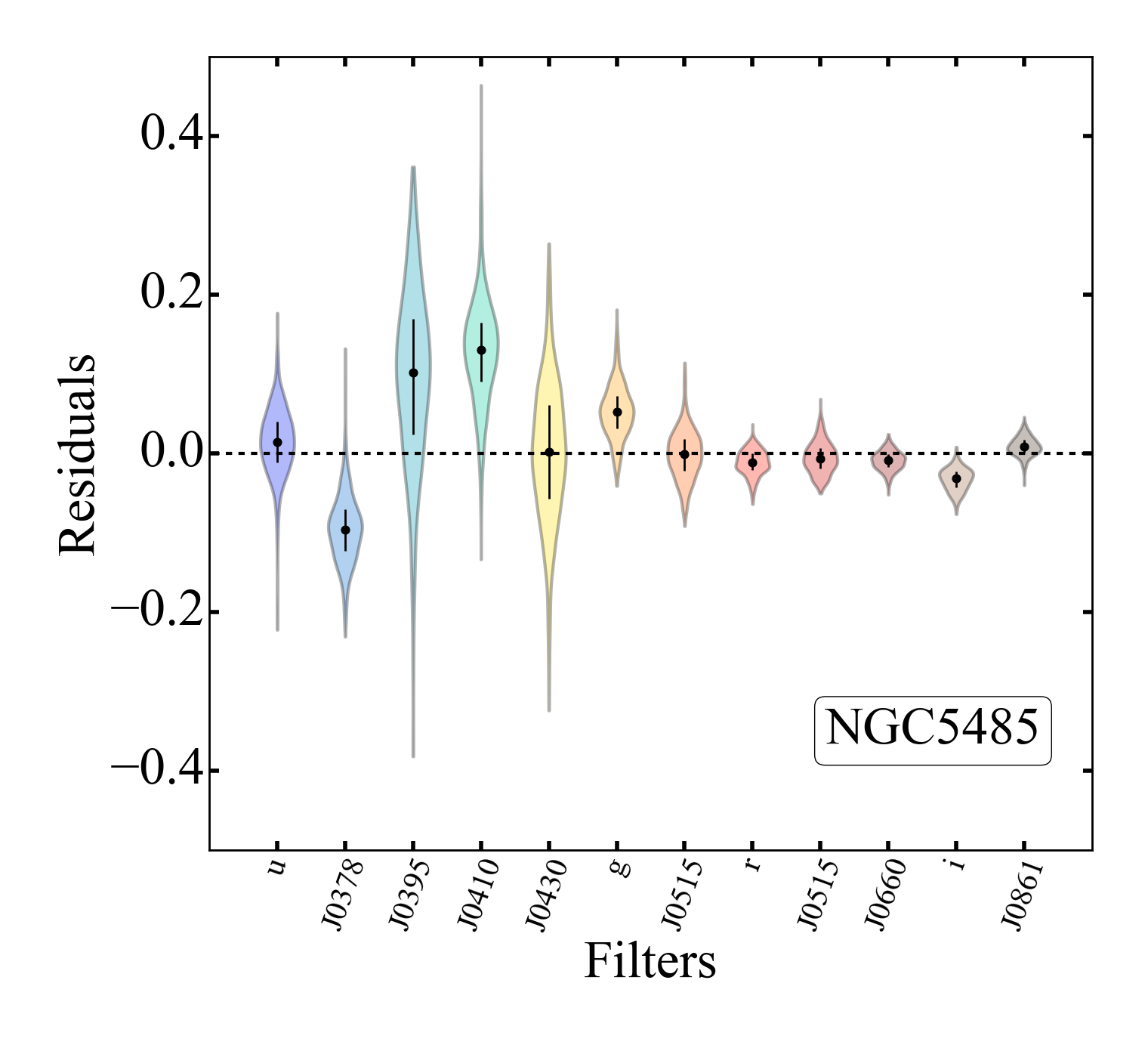}
\caption{Distribution of the residuals of the best fitting for each J-PLUS filter (enclosed colored regions).  The color scheme correspond to Fig. \ref{fig:transmission}. The black symbols and bars correspond to the medians and the interquartile range for each distribution.} 
\label{fig:residuals}
\end{center}
\end{figure}

\subsection{Goodness-of-fit maps}

Despite the general high quality data of CALIFA and J-PLUS, variations in data quality across the image of a given galaxy or from galaxy-to-galaxy can produce biased results. In addition, a poor fit can be produced even from good quality data (e.g., an unmasked emission line). Therefore, it is important to perform a quality control check of the data and the fit. Figs. \ref{fig:fitsa} and \ref{fig:fitsb} show the different data and fit quality maps reported by J-PLUS/MUFFIT, CALIFA/STARLIGHT,  and CALIFA/STECKMAP for NGC 5473 and NGC 5485, respectively. 

J-PLUS/MUFFIT shows as quality indicators the bin-by-bin S/N in the filter $J0515$, the magnitude error of the $J0515$ filter, and  the reduced $\chi$$^\mathrm{2}$ map of the SED fitting (first row of Figs. \ref{fig:fitsa} and \ref{fig:fitsb}). The reference filter $J0515$ was chosen exclusively for comparison purposes with the CALIFA methods. Figures \ref{fig:fitsa}a and \ref{fig:fitsb}a show the CVT zones for the two galaxies color-coded by the S/N in the $J0515$ filter.  As expected, the outer parts of the galaxies correspond to larger bins, while the bins associated to the central region are composed by single pixels. All the SEDs have S/N > 20 in the $J0515$ reference image. The $\chi$$^\mathrm{2}$ map of every object is inspected as a goodness-of-fit quality check (Figs. \ref{fig:fitsa}b and \ref{fig:fitsb}b). A detailed definition of the error-weighted $\chi$$^\mathrm{2}$ minimization process can be found in Sec. 3.2.1 of \citet{DiazGarciaetal2015}. Although generally speaking, a value of $\chi$$^\mathrm{2}$ $\sim$ 1 represents a good fit, the value of $\chi$$^\mathrm{2}$ should be considered only as an indicator because strongly depends on the photometric errors estimate.  $\chi$$^\mathrm{2}$ maps show small values in both cases. Visual inspection of the error maps (Figs. \ref{fig:fitsa}c and \ref{fig:fitsb}c) does not show evidence of significantly higher photometric errors that could suggest any artificial feature. Finally the distribution of the residuals of the best fitting are also examined. Figure \ref{fig:residuals} shows the distribution of the residuals for each filter.  We note that while the red filters are always well fitted producing a small median residual and a small interquartile range (black symbols), the blue filters show a larger residual distribution. In particular, median residuals for filters J0395 and J0410 are $\sim$ 0.1. This effect could be a consequence of the calibration technique performed since the zero point uncertainties of those filters are larger than in the rest of the filters.  J-PLUS calibration applies a series of calibration procedures rather than relying on a single calibration technique. While the photometric calibration in some filters is performed based on SDSS spectroscopic observations, photometric SDSS observations are used to calibrate the bands uncovered by SDSS spectra. The spectrophotometric standard star technique is critical in the calibration of the J0378 filter, since neither SDSS photometry nor SDSS spectroscopy cover this bandpass. Although this procedure has the advantage of providing an independent calibration for each filter, by combining the information from different bands, it is also possible to apply methods that enable to anchor the calibration across the spectral range. One particular promising approach is the use of the stellar locus method \citep[e.g.][]{Highetal2009}. The stellar locus approach for the calibration of J-PLUS is currently under development but preliminary results suggest consistent zero point calibrations over the full J-PLUS spectral range with $\sigma_{\mathrm{zp}}$ $\lesssim$ 0.02. The details of this procedure and its application to J-PLUS data will be presented in a future work.

The second row in Figs. \ref{fig:fitsa} and \ref{fig:fitsb} shows quality indicator maps for CALIFA/STARLIGHT method. The spaxels with artifacts, foreground objects, and very low S/N (< 3) are masked and appear as white regions in the maps. The spaxels with S/N lower than 20 in the 5635 $\pm$ 45 $\AA$ band are binned into Voronoi zones (e.g., two or more spaxels are contained in a given zone). As shown by Figs. \ref{fig:fitsa}d and \ref{fig:fitsb}d, only the very outer parts of each galaxy are affected by low S/N spaxels. After the Voronoi binning, all the spectra have S/N > 20 at 5635 $\AA$.  Figs \ref{fig:fitsa}e and \ref{fig:fitsb}e present the reduced $\chi$$^\mathrm{2}$ for the CALIFA/STARLIGHT analysis. As discussed previously, $\chi$$^\mathrm{2}$ is closely tied to the uncertainty of the spectra, meaning that inspection of Figs \ref{fig:fitsa}e and \ref{fig:fitsb}e may lead to the wrong conclusion that the fits are worse in the central regions than in the outskirts. Based on this argument, CALIFA/STARLIGHT provides also the mean absolute model deviation, $\overline{\Delta}$ maps (Figs. \ref{fig:fitsa}f and \ref{fig:fitsb}f). $\overline{\Delta}$ does not depend explicitly on the uncertainties so it is a more appropriate measure of the fit quality. A detailed definition of $\chi$$^\mathrm{2}$ and $\overline{\Delta}$ can be found in \citet{CidFernandesetal2013}. As noted by \citet{CidFernandesetal2013}, the inspection of the highest $\overline{\Delta}$  spectra often reveals non-masked emission lines or artifacts. The median $\overline{\Delta}$ value for the $\sim$ 10$^{5}$ CALIFA analyzed spectra was 4$\%$ (corresponding to an equivalent S/N of 25), and in less than 2$\%$ of the cases $\overline{\Delta}$  exceeds 10$\%$.

As explained in Sec. \ref{sec:method}, the Voronoi tessellation is different for each method. Although ideally the same binning segmentation should be used for a fair comparison (i.e., same areas/spectra of the object are compared), in practice this is not convenient. Different observing conditions between J-PLUS and CALIFA would require degrading J-PLUS quality data to match CALIFA point-spread function and spatial resolution (e.g., strong homogenization of the data). Even when considering the same observing data (CALIFA/STARLIGHT and CALIFA/STECKMAP),  the peculiarities of each method require a different treatment to ensure a reliable determination of the output parameters (e.g., different minimum S/N required). For these reasons, each method has been applied under the best possible conditions and produce different S/N and binning maps (Figs. \ref{fig:fitsa} and \ref{fig:fitsb}) from survey to survey (J-PLUS versus CALIFA) and also from technique to technique (STARLIGHT versus STECKMAP).

\begin{figure*}
\begin{center}
\includegraphics[width=0.70\columnwidth]{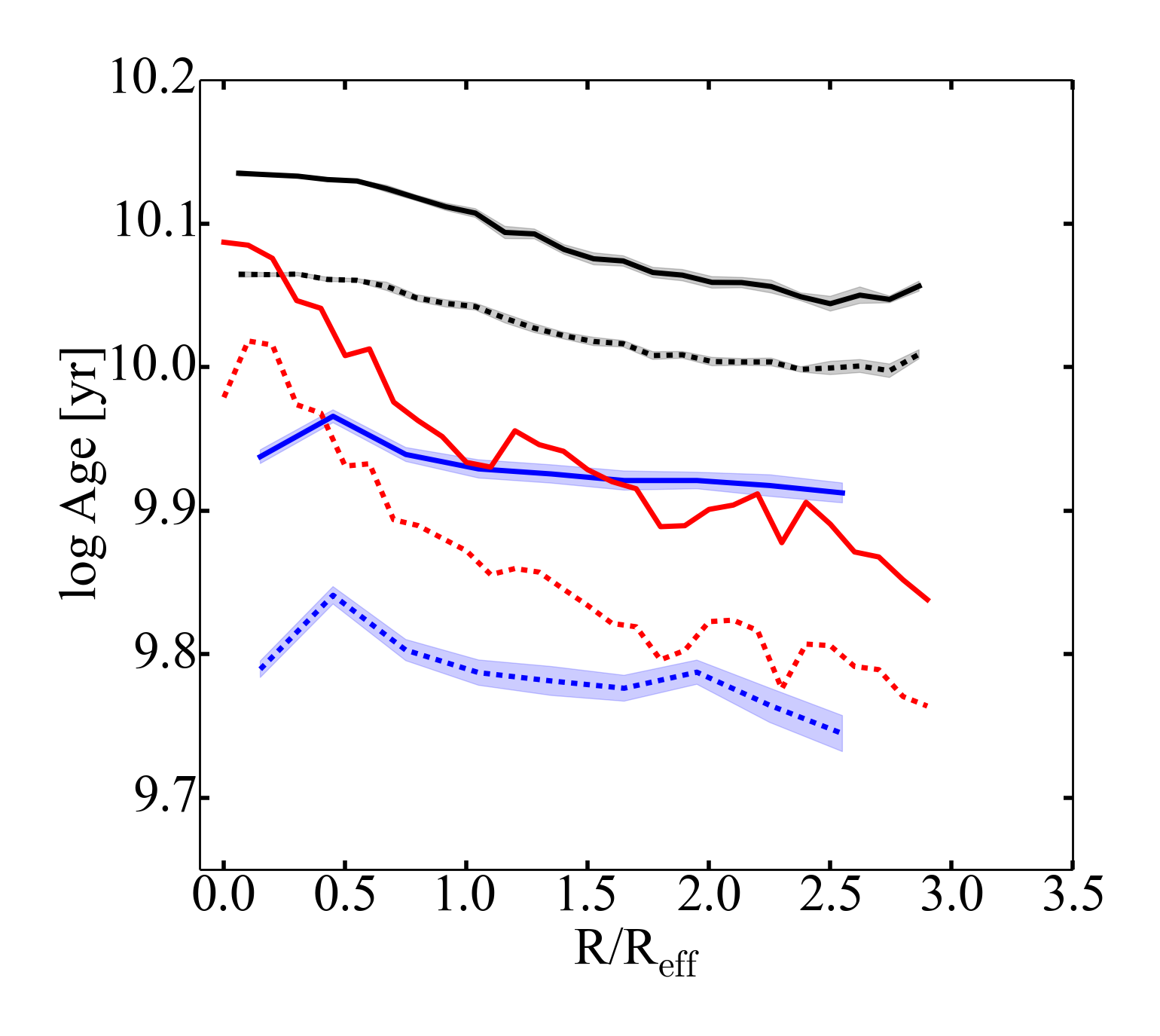}
\includegraphics[width=0.70\columnwidth]{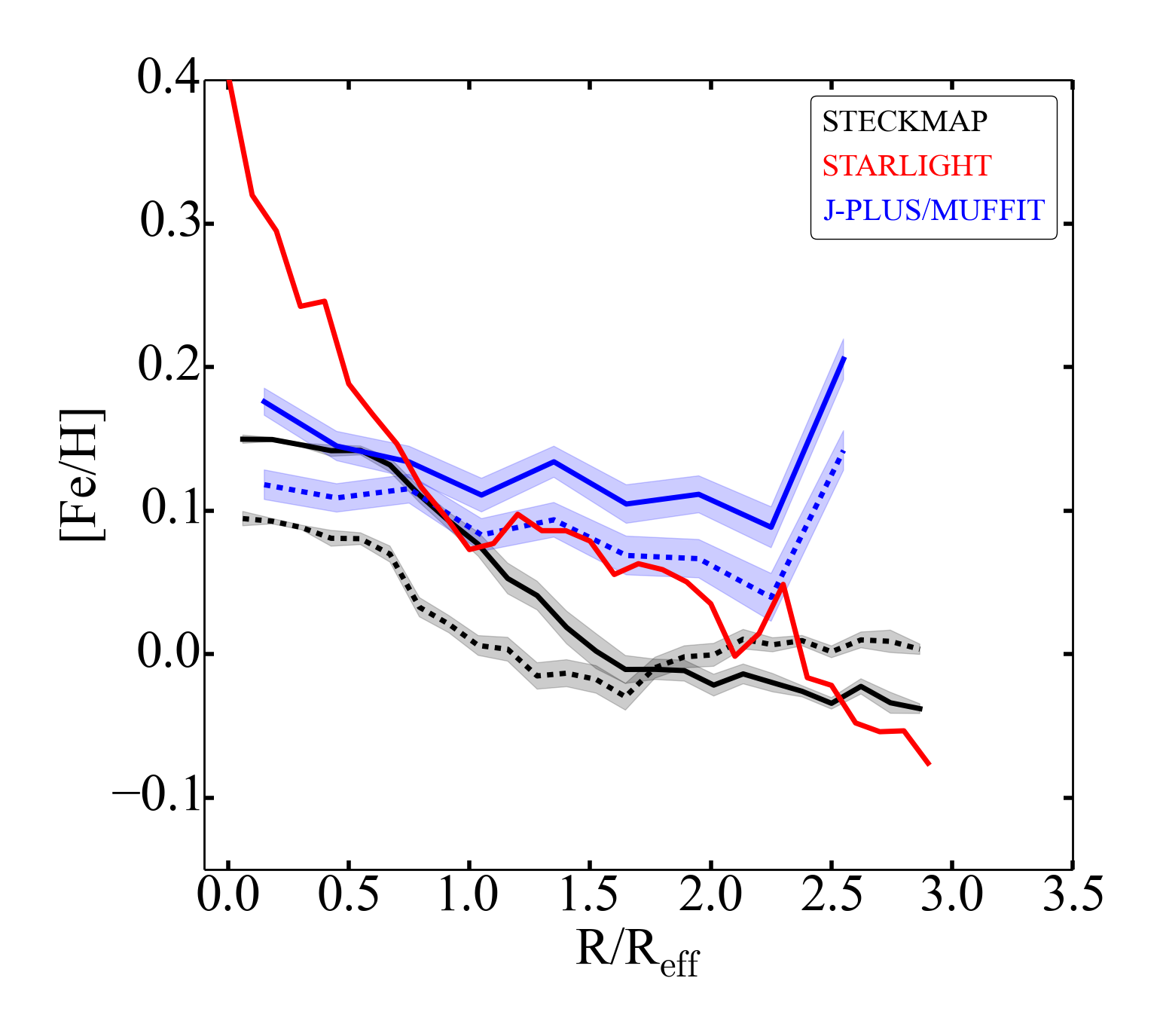}
\includegraphics[width=0.70\columnwidth]{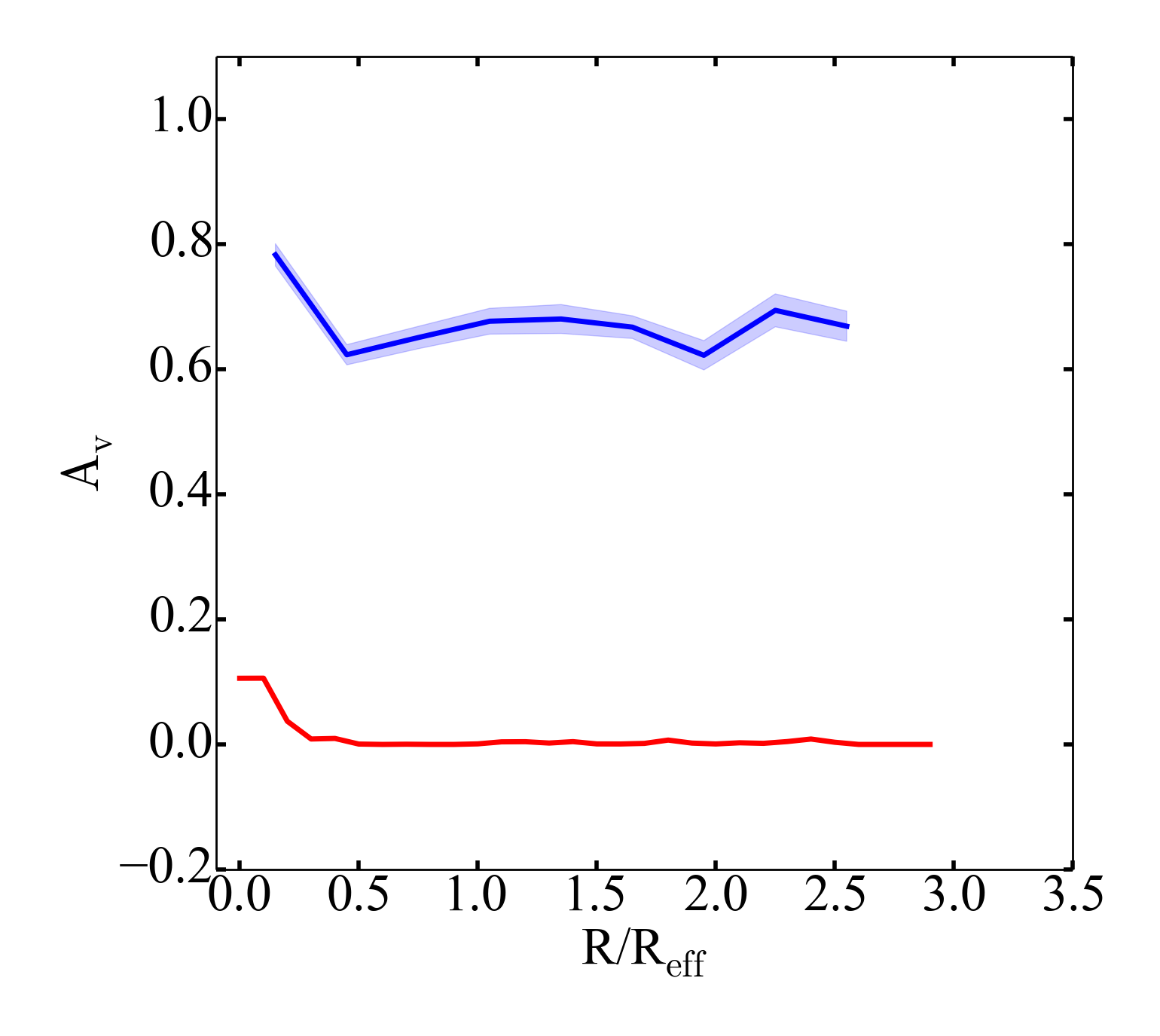}
\includegraphics[width=0.70\columnwidth]{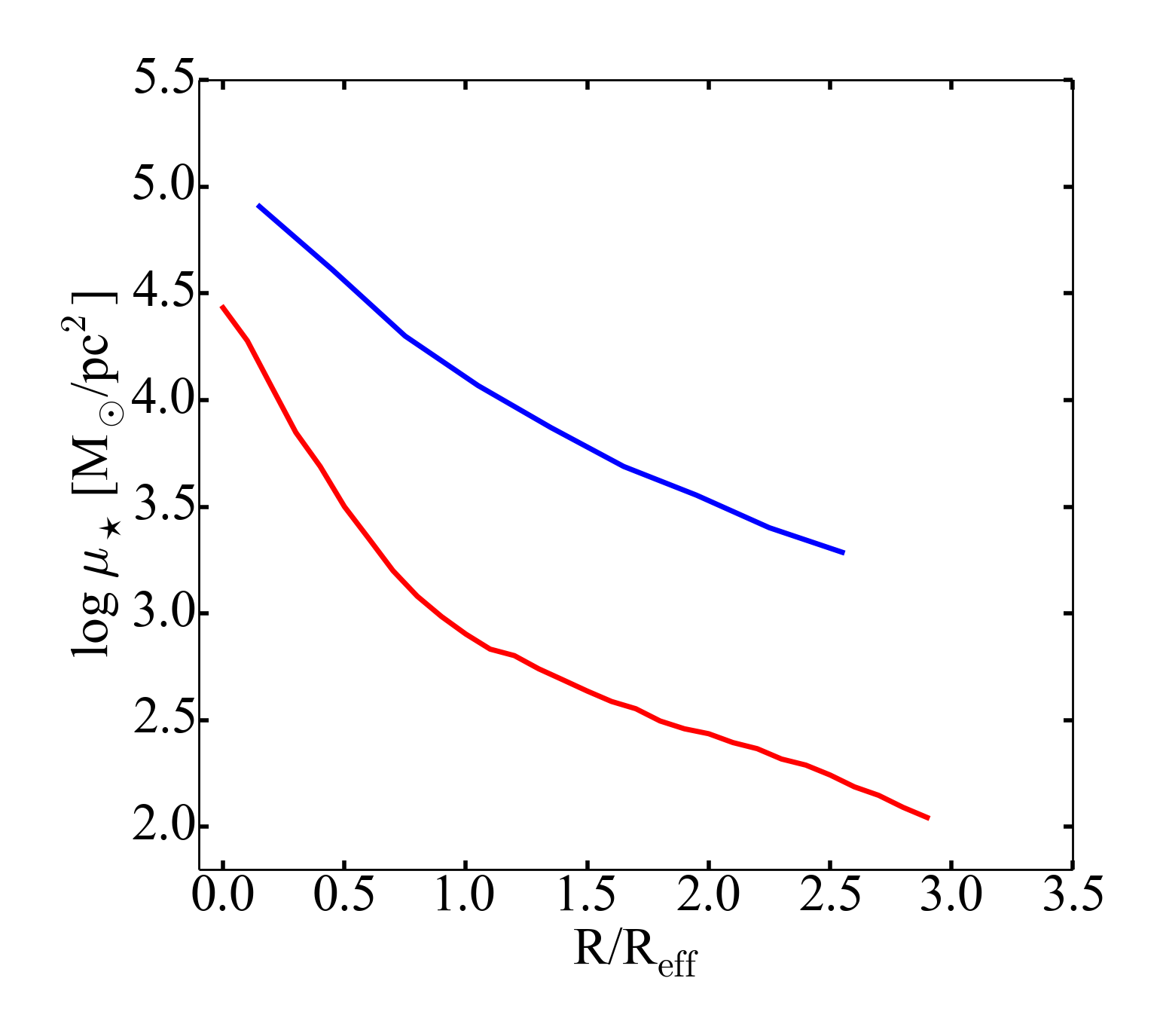}

\caption{Mass-weighted (solid lines) and luminosity-weighted (dashed lines) radial profiles for NGC 5473 of log Age, [Fe/H], A$_\mathrm{v}$ and log $\mu$$_{\star}$ for the three different methods. Enclosed shadowed regions correspond to the uncertainties of each profile.} 
\label{fig:profa}
\end{center}
\end{figure*}

\begin{figure*}
\begin{center}
\includegraphics[width=0.70\columnwidth]{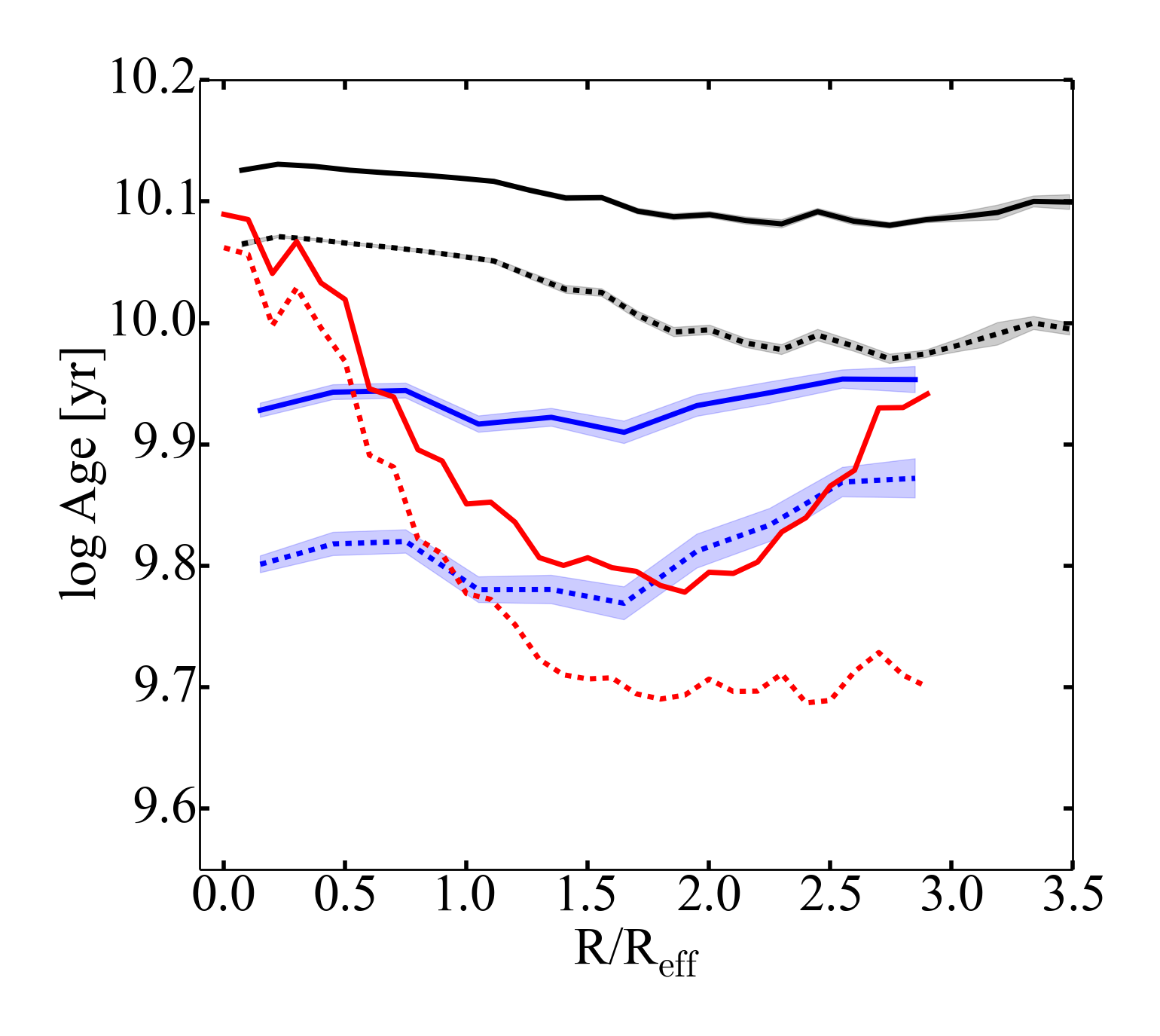}
\includegraphics[width=0.70\columnwidth]{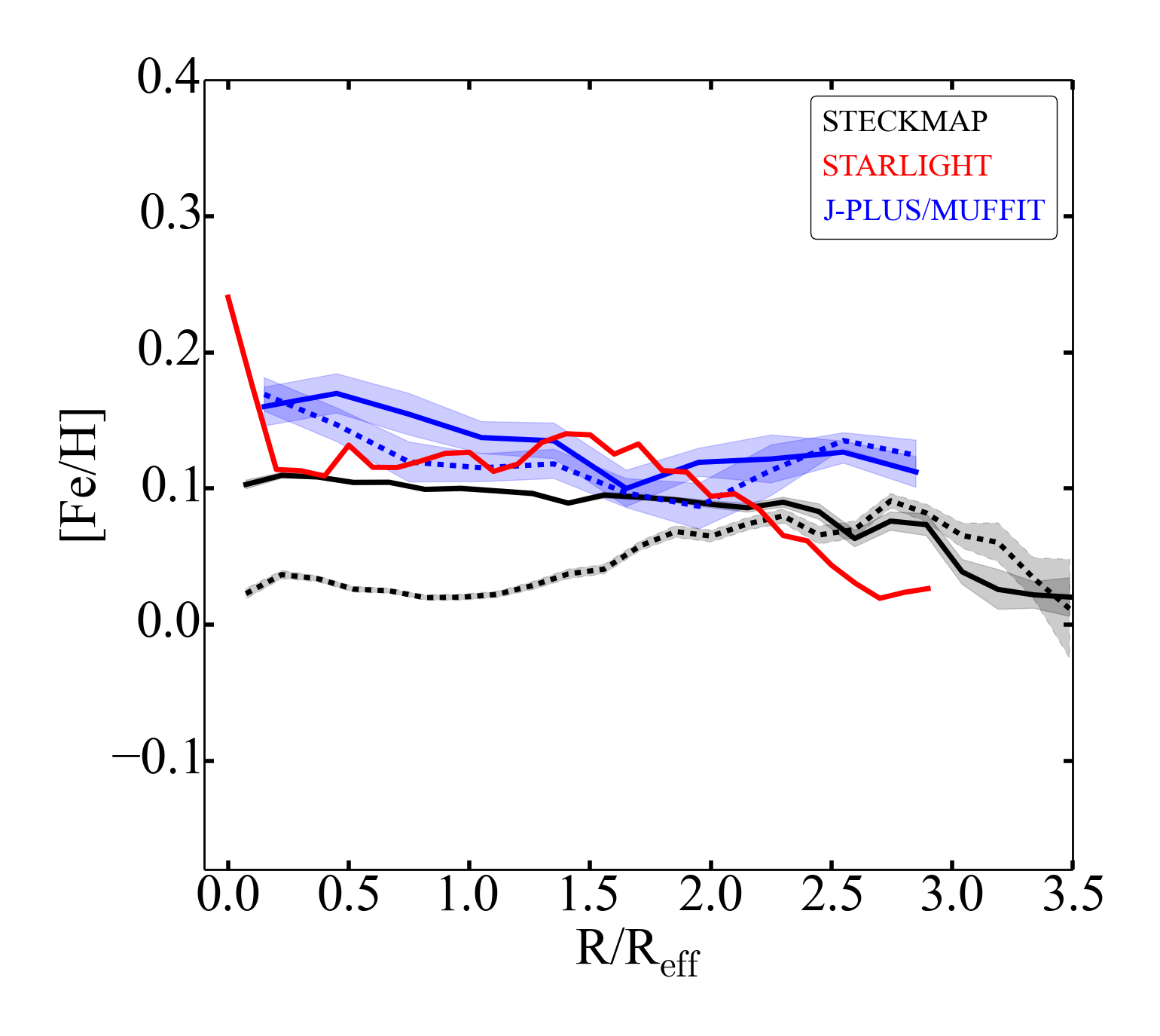}
\includegraphics[width=0.70\columnwidth]{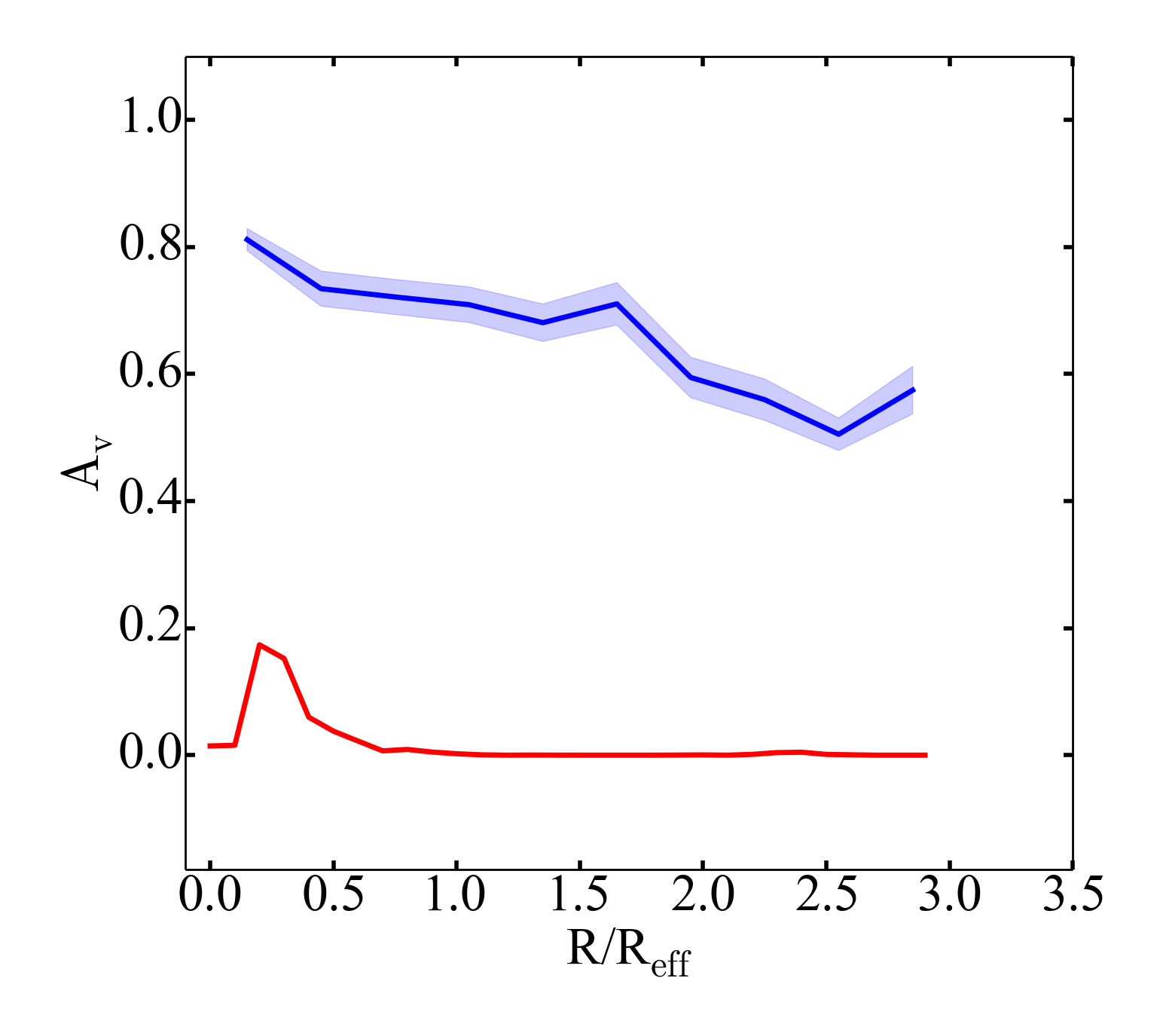}
\includegraphics[width=0.70\columnwidth]{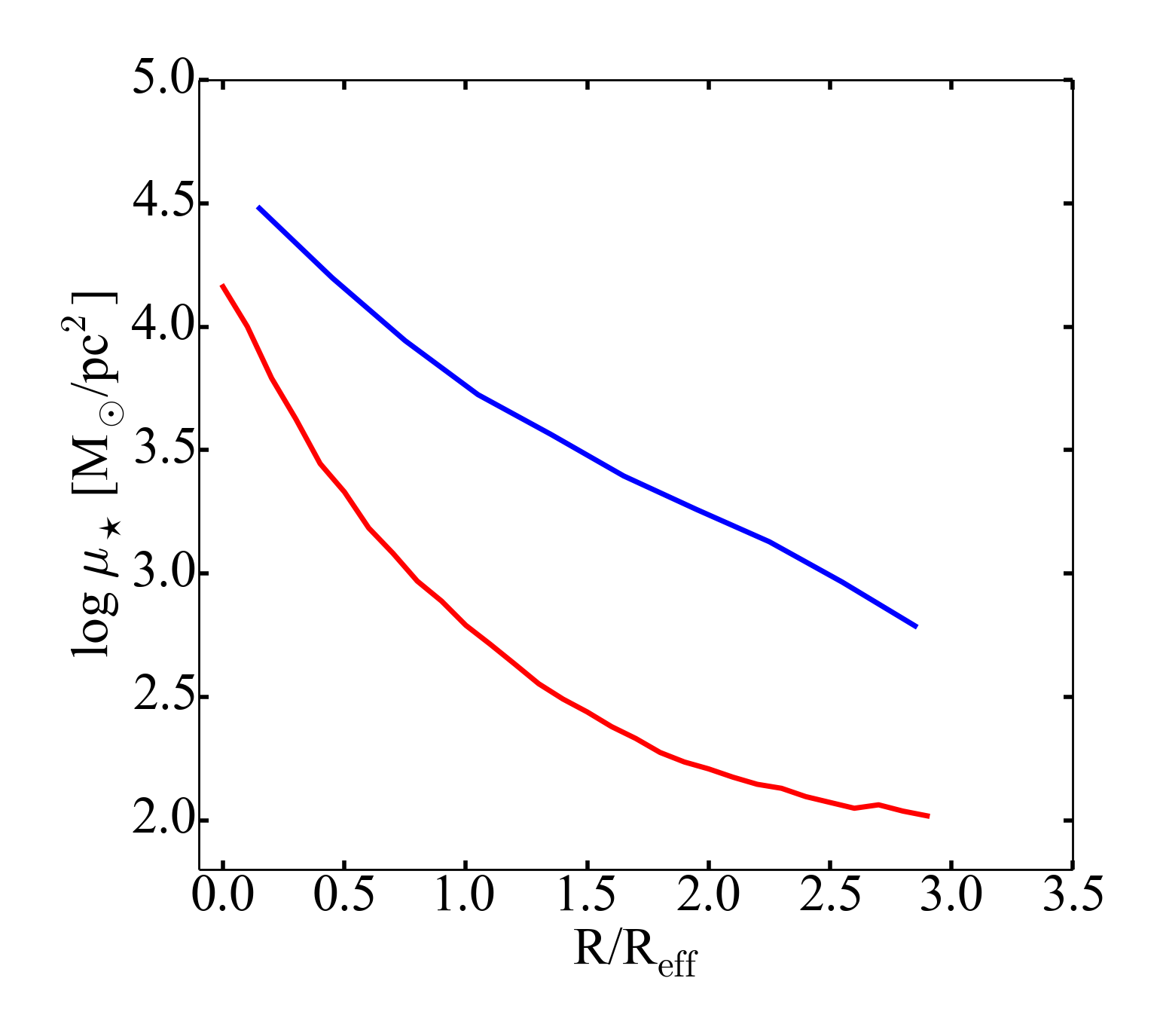}

\caption{Same as in Fig. \ref{fig:profa}, but for NGC 5485.} 
\label{fig:profb}
\end{center}
\end{figure*}

\section{Radial profiles}\label{sec:rad}
To quantify radial variations of the galaxies properties, we present in Figs. \ref{fig:profa} and \ref{fig:profb} the mass- and luminosity-weighted radial profiles of the stellar population parameters. J-PLUS/MUFFIT profiles were obtained following the technique described in \citet{Sanromanetal2018}. They plot the stellar properties values of each bin in each galaxy as a function of the circularized galactocentric distance, R$^{\prime}$ (see their eq. 3). The final profiles were obtained by averaging the stellar population properties of the sample in constant bins of 0.2 R$_\mathrm{eff}$ for 0 $\leq$ R $\leq$ 3.5 R$\textsubscript{eff}$. The errors correspond to the standard deviation of the mean in each bin. CALIFA/STARLIGHT and CALIFA/STECKMAP profiles were derived by binning the output values into elliptical annuli that are scaled in along the major axis such that the bins are constant in effective radius. Elliptical apertures of 0.1 R$_\mathrm{eff}$ are used to extract the radial profiles. These azimuthally averaged radial profiles assume a priori symmetry in the stellar population of the galaxies by directly collapsing the information to a 1D plot.  Same position angles, ellipticities and R$_\mathrm{eff}$ are used to obtain J-PLUS/MUFFIT and CALIFA/STECKMAP profiles. Along the semi-major axis R$^{\prime}$ = R, so the profiles derived by the different techniques are directly comparable. Enclosed shadowed regions correspond to the uncertainties of each profile.

Figures \ref{fig:profa} and \ref{fig:profb} show an offset between the different methodologies with differences up to $\Delta$log Age = 0.3 dex and $\Delta$[Fe/H] = 0.1 dex. The existence of intrinsic systematic differences between the three methods seems to be the most plausible reason for the different absolute values of the derived stellar parameters. The discrepancies between the analysis of spectral features versus colors, together with the assumptions of different star formation histories may be responsible for the quantitative discrepancies. For each individual method, the age and metallicity radial profiles are very similar (i.e., same gradient) when luminosity- and mass-weighted properties are used. This result agrees with previous studies \citep{Sanromanetal2018, GonzalezDelgadoetal2014}, and confirms that the contribution of the second SSP (the younger component) is small and the star formation  history of early-type galaxies is well represented by mainly an old SSP component. 

The radial profiles shapes, however, show clear differences between the three methodologies. NGC 5473 (Fig. \ref{fig:profa}) shows flat or slightly negative age  profiles in J-PLUS/MUFFIT  and CALIFA/STECKMAP analysis while the CALIFA/STARLIGHT age profile is significantly steeper. The metallicity profiles are negative in all the cases although the J-PLUS/MUFFIT metallicity gradient seems flatter than in the other two methods.  On the other side, NGC 5485 profiles present significant differences from method to method, more clearly evident in the age profiles. While J-PLUS/MUFFIT and CALIFA/STECKMAP show similar flat age gradients, CALIFA/STARLIGHT presents a u-shaped log Age$_\mathrm{M}$ profile with a strong negative age gradient inside 1.5 R$_\mathrm{eff}$ that becomes positive at larger radii. The luminosity-weighted age profile, log Age$_\mathrm{L}$, of STARLIGHT also presents significant differences with the other methods showing a strong negative inner (< 1.5 R$_\mathrm{eff}$) gradient that flattens at larger radii. The slightly negative [Fe/H] profiles seem to be compatible between the different methods. Results of J-PLUS/MUFFIT of the stellar extinction behavior are consistent with  a flat or slightly negative A$_\mathrm{v}$ profile with a constant A$_\mathrm{v}$ $\sim$ 0.7 suggesting no significant changes in the dust content. In contrast, CALIFA/STARLIGHT results show a dust-free content (A$_\mathrm{v}$ = 0) at R > 0.5 R$_\mathrm{eff}$ with inner regions showing  A$_\mathrm{v}$  < 0.2 value for both galaxies. The stellar mass surface density profiles, log $\mu$$_{\star}$, also show differences in the structures where J-PLUS/MUFFIT presents a more linear decline in the profiles. These differences in the stellar mass surface density may be a consequence of the large differences in the extinction parameter.

Overall, Figs. \ref{fig:profa} and \ref{fig:profb} show that the profiles obtained by J-PLUS/MUFFIT and CALIFA/STECKMAP present a linear behavior with the galactocentric distance (i.e., flat age gradient and negative metallicity gradient). On the contrary, CALIFA/STARLIGHT  presents non-linear profiles (i.e., negative gradients in the inner part of the galaxies (< R$_\mathrm{eff}$) that flatten at larger galactocentric distances) producing different inner and outer gradients. 

\begin{table}
\caption{Ages and metallicities values determined within different circular aperture using Lick index measurements (SSP) and mass-weighted parameters from spectral fitting (SFH) by the ATLAS$^{3D}$ survey.}
\label{tab:atlas}
\centering
\small
\begin{tabular}{lrrrr}
\hline 
\hline \\[-1ex]

 & \multicolumn{2}{c}{NGC 5473} & \multicolumn{2}{c}{NGC 5485} \\[0.5ex]
R$_\mathrm{circ}$& Age$_\mathrm{SSP}$  &   [Fe/H]$_\mathrm{SSP}$ & Age$_\mathrm{SSP}$  &   [Fe/H]$_\mathrm{SSP}$ \\[0.5ex]
&  (Gyrs) &    & (Gyrs) &   \\[0.5ex]
 
\hline \\[-1ex]

R$_\mathrm{eff}$/8  &    6.87 $\pm$ 1.25  &     0.21 $\pm$ 0.05 &  9.69  $\pm$ 1.68    &   0.05 $\pm$ 0.05  \\   
R$_\mathrm{eff}$/2  &    9.69  $\pm$ 1.68   &  --0.01 $\pm$ 0.05 &  11.51 $\pm$  2.14   &  --0.12 $\pm$ 0.05 \\    
R$_\mathrm{eff}$     &   11.51 $\pm$ 1.99  &   --0.14  $\pm$ 0.06 &  12.55 $\pm$ 2.28   &  --0.20 $\pm$ 0.05  \\ 

\hline\\[0.5ex]
   & Age$_\mathrm{SFH}$  &   [Fe/H]$_\mathrm{SFH}$ & Age$_\mathrm{SHF}$  &   [Fe/H]$_\mathrm{SHF}$ \\[0.5ex]
&  (Gyrs) &    & (Gyrs) &   \\[0.5ex]
 
\hline \\[-1ex]

R$_\mathrm{eff}$ &    11.63 $\pm$ 0.65  &     --0.10 $\pm$ 0.02 &  12.78 $\pm$ 0.75    &   --0.12 $\pm$ 0.02  \\   

\hline
\hline \\[-1ex]

\end{tabular}
\end{table}

As mentioned previously, ATLAS$^{3D}$ survey observed our two target objects using SAURON spectrograph. These IFU observations are limited by a small wavelength range (480 -- 538 nm) and focused on the very center of the galaxies. They determined the stellar population content applying two methods: one based on measuring line-strength indices and applying SSP models to derive SSP-equivalent values; and another one based on spectral fitting to derive non-parametric star formation histories, mass-weighted average values of age, metallicity and half-mass formation timescales. Using spectra integrated within three apertures covering up to one effective radius (R$_\mathrm{eff}$/8, R$_\mathrm{eff}$/2 and 1 R$_\mathrm{eff}$), \citet{McDermidetal2015} obtain average values of age and metallicity based on measuring the Lick indices H$\beta$, Fe5015, Mgb, and Fe5270 \citep{Wortheyetal1994} and using SSP models. Age values inferred at different apertures show that the young stars are more centrally concentrated implying positive age gradients. The derived metallicity becomes lower at larger apertures, due to the inclusion of the metal-poor outer regions. To obtain the mass-weighted parameters from spectral fitting, they use the penalized pixel fitting code pPXF \citep{Cappellarietal2004} to fit a linear combination of SSP model spectra from the MIUSCAT model library \citep{Vazdekisetal2012}. They fit the integrated spectra within one effective radius. We note that the ages and metallicities obtained by \citet{McDermidetal2015} are integrated aperture measurements so are not directly comparable with the radial profiles presented in Figs. \ref{fig:profa} and \ref{fig:profb}. The ages and values are presented in Table \ref{tab:atlas}.  The results of ATLAS$^{3D}$ for NGC 5485 would agree with the flat or slightly positive age gradient found by J-PLUS/MUFFIT and CALIFA/STECKMAP up to one effective radius, however they would contrast with the strong negative age gradient observed in CALIFA/STARLIGHT profile for the same area.

\begin{figure}
\begin{center}
\includegraphics[width=0.90\columnwidth]{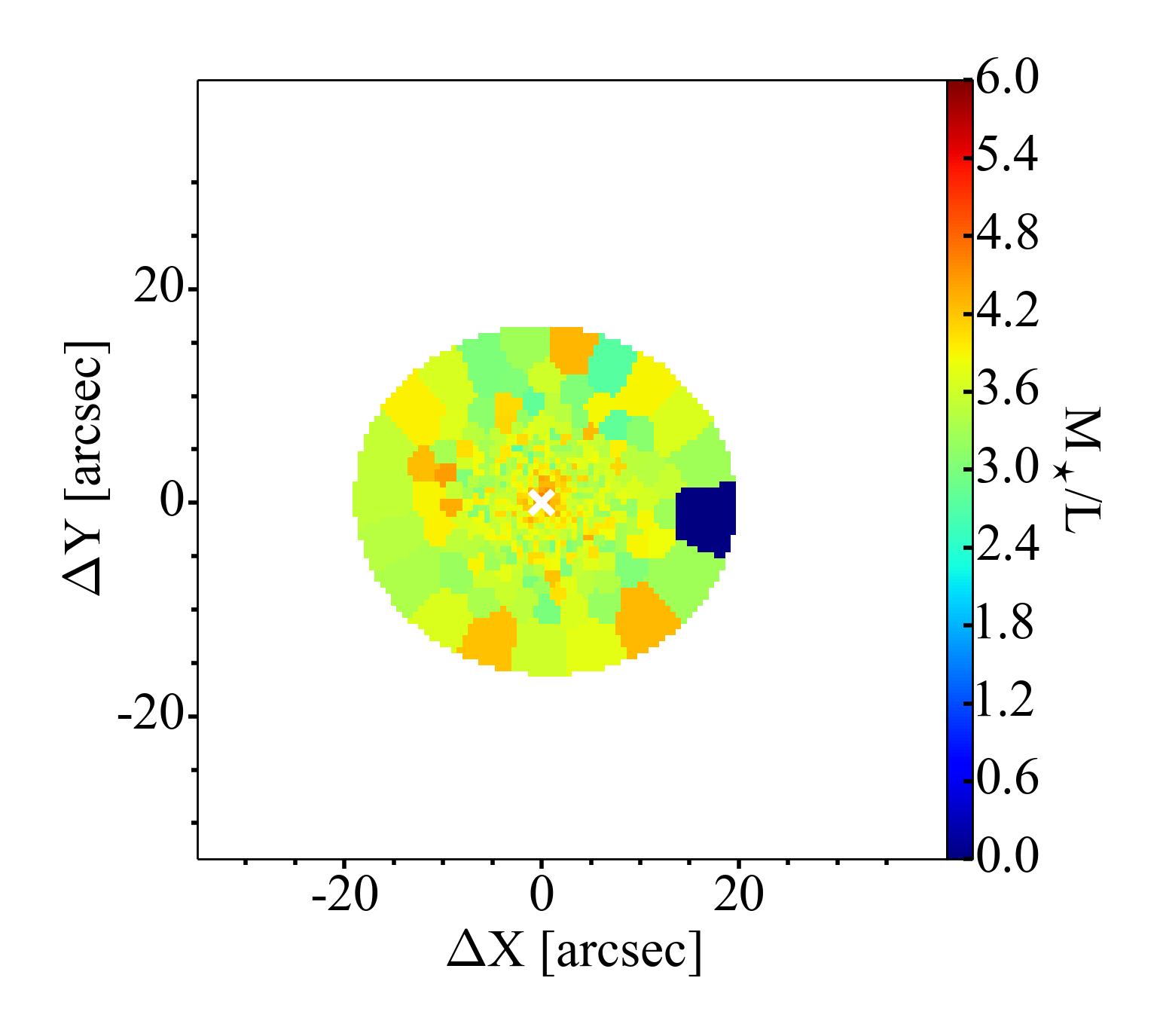}

\includegraphics[width=0.90\columnwidth]{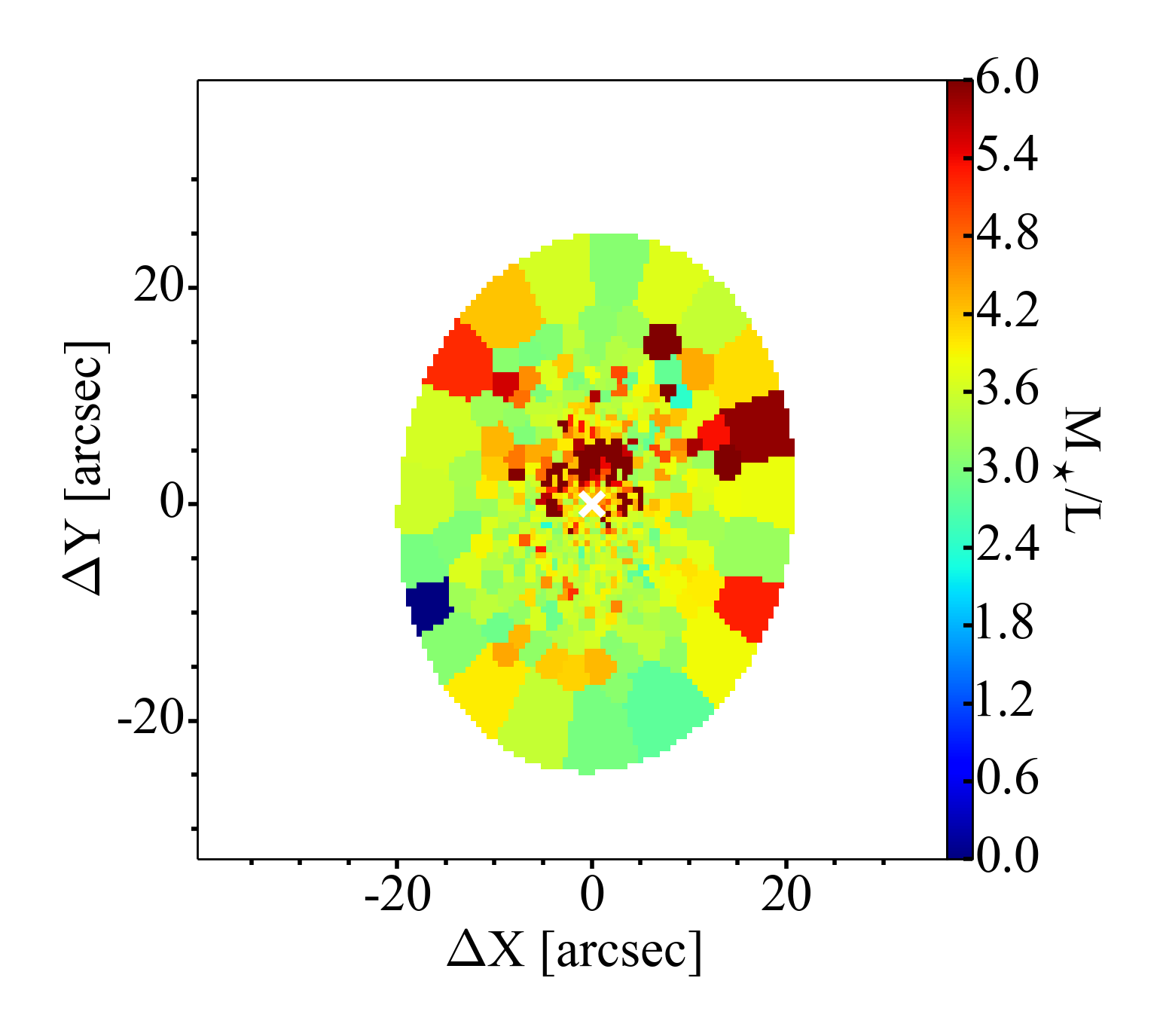}

\caption{Stellar mass-to-light ratio maps determined by J-PLUS/MUFFIT for NGC 5473 (top panel) and NGC 5485 (bottom panel). The center of the galaxy is marked with a white cross in each panel.} 
\label{fig:ML}
\end{center}
\end{figure}

\section{Stellar mass-to-light ratio}\label{sec:ML}

Stellar masses play a crucial role in the study of galaxy properties and the evolution of the galaxy population. Even though it is generally accepted that the analysis of galaxies by their  estimated  stellar  masses  rather  than  observed luminosities provides a more physical insight, it is also recognized that we are limited by a number of statistical and systematic uncertainties when translating observational quantities into physical parameters. Besides the accuracy of the population synthesis models used to interpret observations (e.g., different stellar libraries or particular stellar evolutionary phases), dust attenuation is a key uncertainty in stellar mass, M$_{\star}$, and mass-to-light ratio,  M$_{\star}/$L, values. Although dust can affect also absorption features like the 4000$\AA$-break \citep{MacArthur2005}, this uncertainty is especially relevant when using color information in the analysis \citep[e.g.,][]{Zibettietal2009, Sorbaetal2015}. If the attenuation is patchy, such as the case of NGC 5485, using spatially resolved M$_{\star}$ and M$_{\star}$/L and then integrating the results galaxy-wide, reduces this systematic uncertainty \citep[e.g.,][]{Zibettietal2009, Sorbaetal2015}.

Figure \ref{fig:ML} shows the M$_{\star}$/L maps derived with J-PLUS/MUFFIT for NGC 5473 and NGC 5485. M$_{\star}$/L ratios are derived considering the $J0515$ filter as the reference band. For the absolute magnitude of the Sun in the reference filter, we convolved J-PLUS $J0515$ filter with the solar spectrum. It can be seen that M$_{\star}$/L is almost constant across both galaxies, although NGC 5485 presents a significant increase in M$_{\star}$/L clearly associated with the minor-axis dust lane. Table \ref{tab:int} presents the log M$_{\star}$$^\mathrm{resolved}$ and (M$_{\star}$/L$)^\mathrm{resolved}$, obtained integrating the spatially resolved maps. 

\begin{figure}
\begin{center}
\includegraphics[width=0.90\columnwidth]{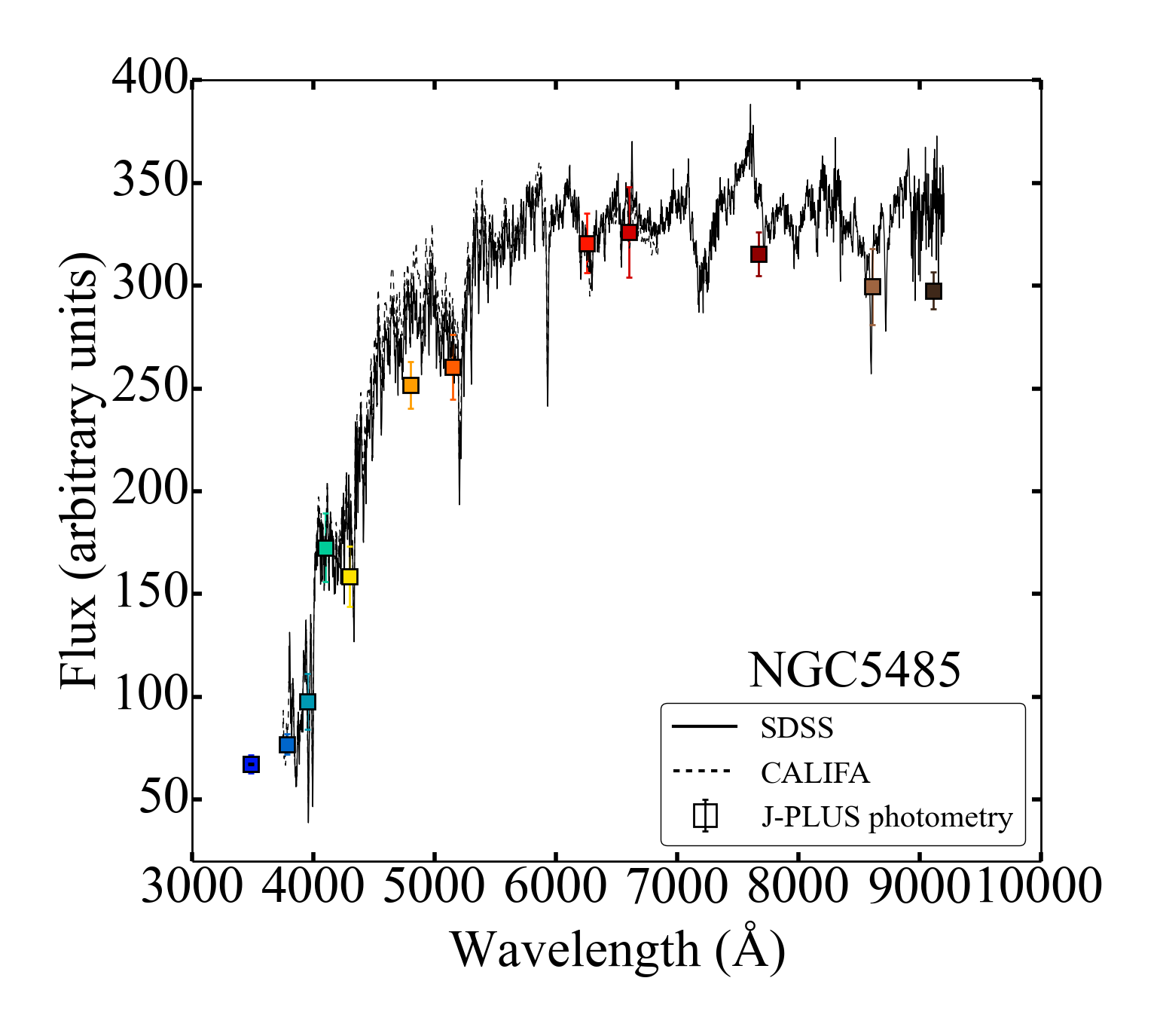}
\caption{Comparison between the integrated spectra in a 3" diameter fiber of SDSS, CALIFA and the photo-spectrum of J-PLUS for NGC 5485. The color scheme correspond to Fig. \ref{fig:transmission}. Error bars correspond to the photometric errors.}
\label{fig:data3arsec}
\end{center}
\end{figure}

\section{Integrated properties}\label{sec:int}
In addition to the  spatially resolved stellar properties of each galaxy, we also determined the global stellar properties of the two galaxies. Table \ref{tab:int} summarizes the global properties of the two galaxies determined using the integrated photometry for J-PLUS/MUFFIT, and the integrated spectra for CALIFA/STARLIGHT and CALIFA/STECKMAP. 

Table \ref{tab:int} shows some discrepancies between the global properties of the galaxies analyzed by  J-PLUS/MUFFIT, CALIFA/STARLIGHT, and CALIFA/STECKMAP. Differences can reach up to $\Delta$log Age = 0.1 dex ($\Delta$Age $\sim$ 2 Gyrs) and up to $\Delta$Fe/H = 0.1 dex. \citet{DiazGarciaetal2015}, using results from MUFFIT, analyzed a subsample of red sequence galaxies shared by ALHAMBRA and SDSS. The one-to-one comparison between metallicities and ages derived spectroscopically for the SDSS data \citep{Gallazzietal2005} and those determined from MUFFIT and ALHAMBRA reveal good qualitative agreement but a systematic difference of $\sim$ 2 Gyr between the two methods.  After an exhaustive investigation of the potential origin of the offset, they conclude that the existence of intrinsic systematic differences between the two methods seems to be the most plausible reason for the difference in the absolute values of the derived ages. More significant are the differences between the extinction parameters obtained by the different methods. 

To closely  inspect any potential systematic effect between J-PLUS and CALIFA data, we compare directly the photo-spectra analyzed by J-PLUS and the integrated spectra of CALIFA. Figure \ref{fig:data3arsec} shows the comparison between the integrated spectra in a 3" diameter fiber of SDSS, CALIFA and the photo-spectrum of J-PLUS for NGC 5485. The spectra are normalized to the $r$ band. We note that SDSS does not provide the NGC 5473 spectrum so the analogous comparison for this object can not be shown. We also note that the apertures used are not exactly equivalent. The CALIFA extraction is made in a 3" x 3" area centered in the continuum peak of the V500 spectral setup while the integrated J-PLUS photo-spectrum is obtained using a circular aperture of 3" diameter. In addition, the precise position of the SDSS fiber is unknown producing potential differences in the aperture centering. In spite of these differences both observations closely follow the SDSS spectrum. This is not surprising since the calibration of J-PLUS and CALIFA observations are anchored to SDSS. In order to ultimately compare the three methodologies, we have performed the analysis on this spectra/pseudospectra where the spectra and the photometry concur. Table \ref{tab:int3} shows the results of J-PLUS/MUFFIT on the 3" integrated photometry, and the results of CALIFA/STARLIGHT and CALIFA/STECKMAP on the same 3"x 3" integrated spectra. Overall, differences reach up to $\Delta$log Age = 0.2 dex ($\Delta$Age $\sim$ 3.5 Gyrs) and up to $\Delta$Fe/H = 0.2 dex. Once again, the differences between the extinction parameters obtained by J-PLUS/MUFFIT and CALIFA/STARLIGHT are very large. The assumption of different star formation histories or the different spectral range covered by CALIFA and J-PLUS may be responsible for the quantitative discrepancies.  More remarkable are the differences between CALIFA/STARLIGHT and CALIFA/STECKMAP. Even when the same spectra and very similar stellar population models are used (see Sect. \ref{sec:method}) discrepancies are significant reaching  $\Delta$log Age$_\mathrm{M}$ = 0.12 dex (i.e., $\Delta$Age$_\mathrm{M}$ = 3 Gyrs) and $\Delta$Fe/H$_\mathrm{M}$ = 0.27 dex. This comparison must be interpreted with caution as it is based on one single spectra. Future work is required to evaluate in detail the origin of these differences and to clearly quantify them.

\begin{table*}
\caption{Global stellar population properties of the galaxies using  the integrated photometry/spectra}
\label{tab:int}
\centering
\small
\begin{tabular}{ccccccc}

\hline 
\hline \\[-1ex]
\multirow{2}{*}{NGC 5473}  &  log Age$_\mathrm{L}$  &  log Age$_\mathrm{M}$   &   [Fe/H]$_\mathrm{L}$  &  [Fe/H]$_\mathrm{M}$  &  A$_\mathrm{v}$  &  log M$_{\star}$  \\[0.5ex]
 &  (dex) &  (dex)  & (dex) & (dex) & (mag) & (M$_{\sun}$)   \\[0.5ex] 
 
\hline \\[-1ex]

J-PLUS/MUFFIT  & 9.92 $\pm$ 0.15 &  10.06 $\pm$ 0.08   & 0.17 $\pm$ 0.23   &  0.12 $ \pm$ 0.26 &   0.43 $\pm$ 0.37 & 10.81 \\
CALIFA/STARLIGHT & 9.89  & 9.97  & 0.08  & 0.16  &  0.0  & 10.62 \\
CALIFA/STECKMAP & 9.83 $\pm$ 0.01   & 9.99 $\pm$ 0.01  &  --0.01 $\pm$ 0.01 & --0.08 $\pm$ 0.01 & ...  & ... \\

\hline \\[-1ex]
 \multirow{2}{*}{NGC 5485}   & log Age$_\mathrm{L}$  &  log Age$_\mathrm{M}$   &   [Fe/H]$_\mathrm{L}$  &  [Fe/H]$_\mathrm{M}$  & A$_\mathrm{v}$ & log M$_{\star}$ \\[0.5ex] 
&  (dex) &  (dex)  & (dex) & (dex) & (mag)  & (M$_{\sun}$) \\[0.5ex] 
\hline \\[-1ex]

J-PLUS/MUFFIT & 9.93 $\pm$ 0.18 &  10.02 $\pm$ 0.14 &  0.25 $\pm$ 0.24     &  0.28 $\pm$ 0.21 & 0.34 $\pm$ 0.43 & 10.80 \\
CALIFA/STARLIGHT & 9.85 & 9.92  & 0.05 & 0.12  & 0.0 & 10.49 \\
CALIFA/STECKMAP  & 9.88 $\pm$ 0.01   & 9.92 $\pm$ 0.01  &  0.04 $\pm$ 0.01 & --0.01 $\pm$ 0.01 & ...  & ... \\

\hline
\hline \\[-1ex]
\end{tabular}
\end{table*}

\begin{table*}
\caption{Global stellar population properties of the galaxies using the integrated photometry/spectra in a 3" aperture.}
\label{tab:int3}
\centering
\small
\begin{tabular}{cccccc}

\hline 
\hline \\[-1ex]
 \multirow{2}{*}{NGC 5485}   & log Age$_\mathrm{L}$  &  log Age$_\mathrm{M}$   &   [Fe/H]$_\mathrm{L}$  &  [Fe/H]$_\mathrm{M}$  & A$_\mathrm{v}$  \\[0.5ex]
&  (Gyrs) &  (Gyrs)  & (dex) & (dex) & (mag)  \\[0.5ex]
\hline \\[-1ex]

J-PLUS/MUFFIT  & 9.76 $\pm$ 0.26  & 9.89 $\pm$ 0.18  & 0.12 $\pm$ 0.37  &  0.16 $\pm$ 0.37 & 0.91 $\pm$ 0.70 \\
CALIFA/STARLIGHT & 9.96 & 10.09  & 0.40 & 0.30  & 0.0 \\
CALIFA/STECKMAP  & 9.87 $\pm$ 0.05  & 9.97 $\pm$ 0.05  & 0.13 $\pm$ 0.02  &  0.21 $\pm$ 0.03 & ... \\

\hline
\hline \\[-1ex]
\end{tabular}
\end{table*}

\section{Discussion}\label{sec:discussion} 
Early-type galaxies were once considered uniform stellar systems with little gas, dust and nuclear activity.  We now know that early-type galaxies commonly contain large amount of dust in either organized or complex structures  \citep[e.g.,][]{Sadleretal1985, vanDokkumetal1995, Tranetal2001}. In a fraction of the early-type galaxy population, the dust is organized in prominent and large-scale dust lanes. These so-called dust-lane early-type galaxies are considered to be the remnants of recent gas-rich minor mergers with a low star-formation efficiency \citep{Hawardenetal1981, Kavirajetal2012, Shabalaetal2012, Davisetal2015}. If we assume that dust and gas settle in the principal planes of a galaxy, the existence of a minor axis dust lane in NGC 5485 is a visual evidence for its triaxiality. This triaxiality is also supported by its rather exceptional kinematical structure, which shows strong minor-axis rotation -- prolate rotation -- \citep{Wagneretal1988, Krajnovicetal2011, Emsellemetal2011, Tsatsietal2017}. 

\citet{Baesetal2014} point out that despite a noticeable amount of dust, neither neutral nor molecular hydrogen has been detected in NGC 5485. This anomaly produces an extremely low gas-to-dust ratio, almost an order of magnitude lower than the canonical value for the Milky Way. \citet{Baesetal2014} propose a potential scenario where NGC 5485 would be recently merged with an SMC-type metal-poor galaxy where a substantial fraction of the HI could have been lost during the interactions. Using IFU CALIFA data and studying N-body merger simulations,  \citet{Tsatsietal2017} propose a different formation scenario. They find that a prolate early-type galaxy, such as NGC 5485, may have been formed by gas-poor, polar major merger that happened 10 Gyr ago. The galaxy was imaged in H$\alpha$ by \citet{Finkelmanetal2010} who detected an ionized gas disk that closely follow the dust structure. They find that the H$\alpha$ emission and color of NGC 5485 is consistent with the presence of an old stellar population ($\sim$ 4.5 Gyr) and a small fraction of a young population ($\sim$10-350 Myr).  The nature of the ionized gas emission found in early-type galaxies is still under debate since non-stellar ionization mechanisms (e.g., induced shocks), stellar ionization mechanisms (e.g., low-level of star formation, post-AGB) or even AGN effects may contribute to the excitation of the warm ionized medium \citep[e.g.,][]{Dopitaetal1995, Stasinskaetal2008, Papaderosetal2013}.

As shown in Fig. \ref{fig:2Dmapsb}, the age maps of NGC 5485 present significant differences between the 3 methodologies. If the old stellar component, aligned with the dust lane, present in the CALIFA/STARLIGHT maps is a real feature, the stellar component could be associated with the NGC 5485 kinematic structure and would favor the polar major merger scenario at 10 Gyrs.  This interpretation would not however explain the absence of the aligned old stellar component in the CALIFA/STECKMAP and J-PLUS/MUFFIT maps. It would also contrast the centrally concentrated young stars found by ATLAS$^{3D}$. On the other hand, if the position, size and orientation of the old component in the CALIFA/STARLIGHT is an artificial feature, this will suggest that a potential age-extinction degeneracy could be affecting the results.  This means that a stellar population reddened by the old content can mimic the behavior of a population that has been reddened by extinction. Overall, MaNGA early-type galaxies exhibit relatively flat radial profiles with reddening values between E(B--V) $\sim$ 0.06, that corresponds\footnote{Considering R$_\mathrm{v}$=A$_\mathrm{v}$/E(B--V) and assuming  a value of R$_\mathrm{v}$=3.1} to A$_\mathrm{v}$ $\sim$ 0.2. However, higher values of E(B--V) = 0.25  corresponding to A$_\mathrm{v}$=0.8 are not unusual in dusty early-type galaxies \citep{Goddardetal2016, Wilkinsonetal2015}. Unfortunately, the lack of previous extinction values studies of NGC 5485 does not allow for further comparison. Although the peculiarities of NGC 5485 imposes an additional challenge to the comparison between the methodologies, the more standard case of NGC 5473\footnote{We note that although visual inspection reveals no photometric peculiarities suggesting that NGC 5473 is an elliptical galaxy, the morpho-kinematic study of \citet{MendezAbreuetal2017} suggests that NGC 5473 is a S0 galaxy.} (i.e., absence of a dusty structure) reveals that the main source of discrepancies are due to the actual methodologies and not the singularities of the objects.  

IFU MaNGA studies also show discrepant conclusions when analyzing the same galaxy sample but using different spectral fitting techniques. \citet{Goddardetal2016}, using the full spectral fitting code FIREFLY \citep{Wilkinsonetal2015, Wilkinsonetal2017} and the spectral population models of \citet{Marastonetal2011}, found flat luminosity-weighted age gradients inside R < R$_\mathrm{eff}$. Contrary, styding the same data set but using the full spectral fitting code STARLIGHT  and BC03 models, \citet{Zhengetal2016} found a slightly negative gradient (-- 0.05 dex/R$_\mathrm{eff}$) at the same galactocentric distances. \citet{Goddardetal2016} perform a comparison between fitting codes and stellar population models. They conclude that overall, the luminosity-weighted ages are affected by systematic offsets between the various codes and underlying stellar population models of the order of --0.13 dex with a large scatter of 0.37 dex. The comparison for the luminosity-weighted metallicity is even more complex showing an overall difference of -- 0.24 dex with a large scatter of 0.34 dex. When the dependence of the stellar population models is isolated (i.e., same stellar population models are used), the choice of fitting technique also yield significant effects producing age offsets of --0.04 $\pm$ 0.45 dex and metallicity offsets of --0.11 $\pm$ 0.37 dex. A fundamental difference between both techniques is the treatment of dust: while STARLIGHT assumes a dust reddening law, FIREFLY is parameter free, because it does not fit the continuum shape to constrain the stellar population properties. Differences in the extinction treatment method of our study are also present (Sect. \ref{sec:method}). 

The age and metallicity measurements are considerably affected by systematic differences, not only because of the stellar population models used, but also based on the fitting technique chosen. As a consequence, measurements of quantities such as age gradients are affected by uncertainties of similar magnitude as the signal itself. This problem clearly requires more investigation to include other spectral fitting codes. Future work is required to evaluate in detail the origin of these differences and explore possible paths to mitigate them.

\section{Summary and conclusions}\label{sec:summary}
We illustrate the scientific potential of J-PLUS data to explore the spatially resolved stellar populations of local galaxies using a method that combines a centroidal Voronoi tessellation and MUFFIT multi-filter SED fitting method. This technique allows us to analyze unresolved stellar populations of spatially resolved galaxies based on multi-filter photometry.  We present detailed 2D maps of stellar population properties (age, metallicity, extinction, and stellar mass surface density) for two early-type galaxies: NGC 5473 and NGC 5485. Radial structures were also obtained and luminosity- and mass-weighted profiles were derived out to R=3 R$_\mathrm{eff}$. J-PLUS/MUFFIT results were compared with analysis from IFU CALIFA data for the same galaxies. Two different techniques to analyze IFU CALIFA were used: STARLIGHT and STECKMAP.  We demonstrate that our alternative technique derives radial stellar population gradients in good agreement with IFU technique such as CALIFA/STECKMAP but differs when CALIFA/STARLIGHT methodology is used.

Comparison of the absolute values reveals the existence of intrinsic systematic differences between the three methods. Differences are also found in the 2D maps. While NGC 5473 shows flat age and slightly negative metallicity profiles, NGC 5485 age and extinction profiles are more challenging. CALIFA/STARLIGHT shows an older component in the center of the galaxy not present in the J-PLUS/MUFFIT and CALIFA/STECKMAP analysis. This older component has the same position, size and orientation that the prominent dust line visible along the minor-axis of the galaxy. Although CALIFA/STARLIGHT detects the dust feature in the A$_\mathrm{v}$ map, values are significantly lower than the ones obtained by J-PLUS/MUFFIT. Radial profile shape of NGC 5485 also presents a different behavior between different methodologies. CALIFA/STARLIGHT presents a u-shaped age profile with strong negative age gradient inside 1.5 R$_\mathrm{eff}$ that become positive at larger radii while a flat age gradient is present in the J-PLUS/MUFFIT and CALIFA/STECKMAP analysis.

For each methodology, the age and metallicity radial profiles are very similar when luminosity- or mass-weighted properties are used, suggesting that the mass assembly of the early-type galaxies NGC 5473 and NGC 5485 are followed by their luminosity components. 

Although discrepancies between the analysis of spectral features and colors together with different star formation histories assumptions and the different spectral range may be responsible for the discrepancies between J-PLUS/MUFFIT and CALIFA/STECKMAP, significant offsets are also present when similar analysis conditions are present (e.g., CALIFA/STARLIGHT versus CALIFA/STECKMAP). This result suggests that the specific characteristics of each methodology such as the extinction treatment used may cause important differences. We conclude that the ages, metallicities and extinction derived for individual galaxies not only depend on the chosen models but also depend on the methodology used. This problem clearly requires more investigation to evaluate in detail the origin of these differences. 

Finally, we remark that although detailed investigations will require larger data sets, it is clear that photometric surveys such as the current J-PLUS (Paper I) and the upcoming J-PAS \citep{Benitezetal2014} will extend 2D multi-filter studies such as the one presented here to scientific cases not available to current IFU techniques (e.g., larger galactocentric distances, effect of the environments on the 2D structures, ...)

\begin{acknowledgements} 
We thank Rosa M. Gonz\'{a}lez Delgado for the fruitful discussions and the helpful comments.

Based on observations made with the JAST/T80 telescope at the Observatorio Astrof\'{\i}sico de
Javalambre (OAJ), in Teruel, owned, managed and operated by the Centro de Estudios de F\'{\i}sica del
Cosmos de Arag\'on. We acknowledge the OAJ Data Processing and Archiving Unit
(UPAD) for reducing and calibrating the OAJ data used in this work. 

Funding for the J-PLUS Project has been provided by
the Governments of Spain and Arag\'on through the Fondo de Inversiones
de Teruel, the Arag\'on Government through the Reseach Groups E96 and E103,
the Spanish Ministry of Economy and Competitiveness (MINECO; under
grants AYA2015-66211-C2-1-P, AYA2015-66211-C2-2, AYA2012-30789 and ICTS-2009-14),
and European FEDER funding (FCDD10-4E-867, FCDD13-4E-2685). The Brazilian agencies FAPESP and the National Observatory of Brazil have also contributed to this project. R.L.O. was partially supported by the Brazilian agency CNPq (Universal Grants 459553/2014-3, PQ 302037/2015-2, and PDE 200289/2017-9).

This research made use of NASA's Astrophysics Data System Bibliographic Services, as well as the following software packages: Astropy \citep{Astropy2013}, Matplotlib \citep{Hunteretal2007}, IPython \citep{PerezGranger2007}, SciPy \citep{Jonesetal2001} and NumPy \citep{vanderWaltetal2011}.

\end{acknowledgements}

\nocite{*}


\appendix 
\section{Luminosity-weighted Maps} \label{appex}
\begin{sidewaysfigure*}[htbp]
\begin{center}
\hspace{0.5cm}
\includegraphics[width=0.07\columnwidth]{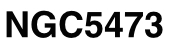}
\hspace{8.2cm}
\includegraphics[width=0.07\columnwidth]{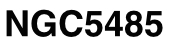}

\includegraphics[width=0.15\columnwidth, angle=90]{JPLUS_MUFFIT.jpg}
\includegraphics[width=0.20\columnwidth]{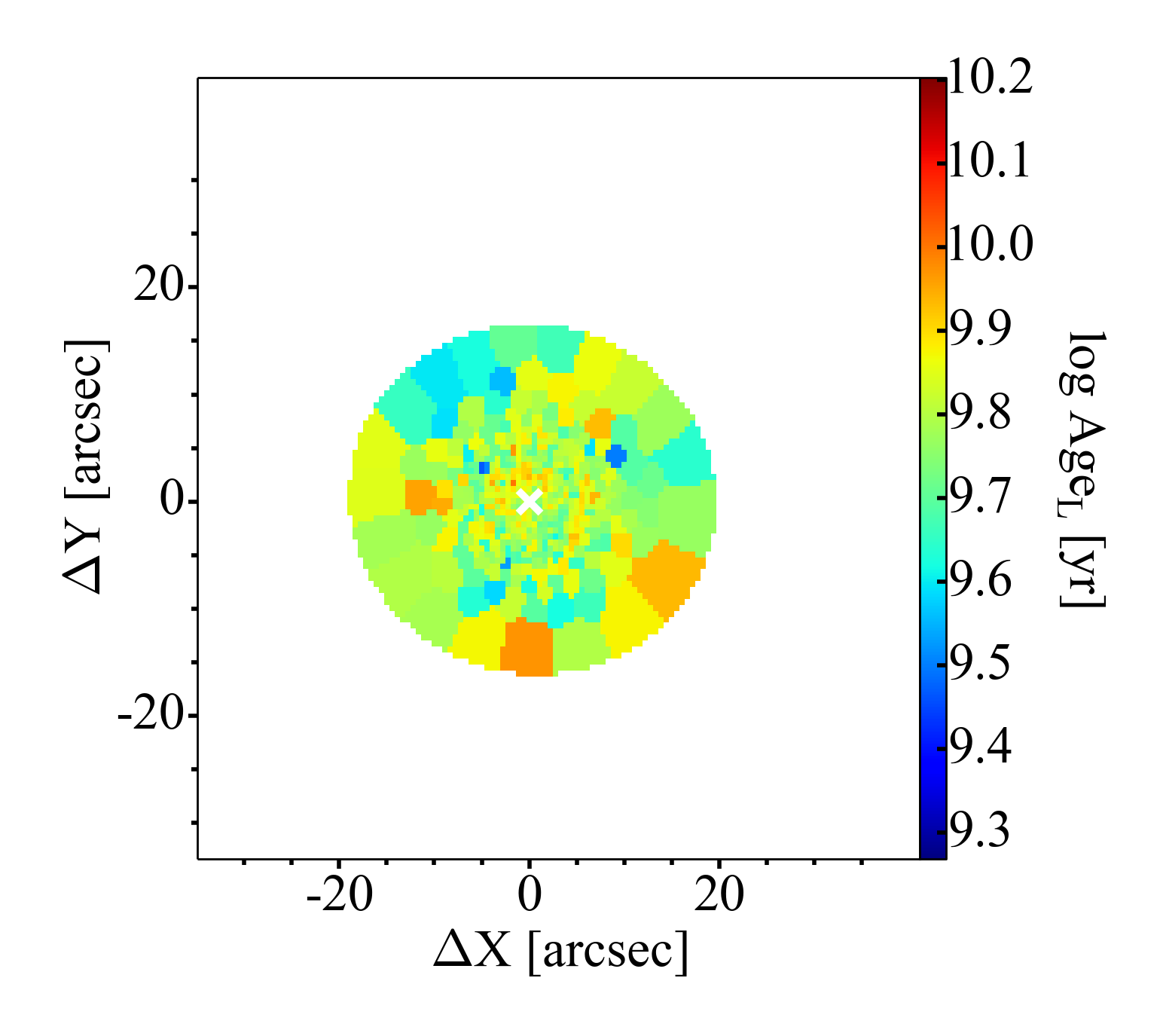}
\includegraphics[width=0.20\columnwidth]{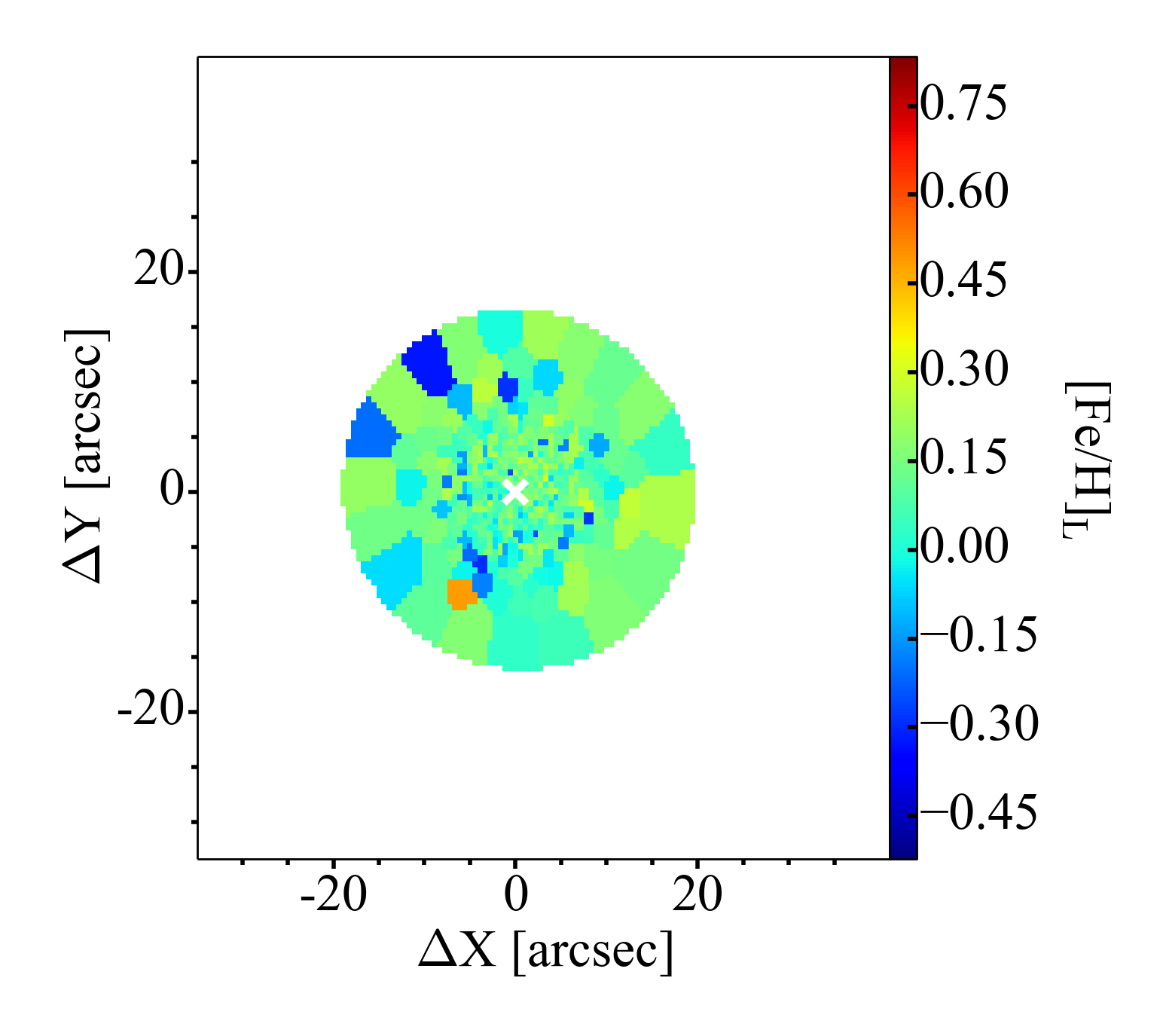}
\includegraphics[width=0.20\columnwidth]{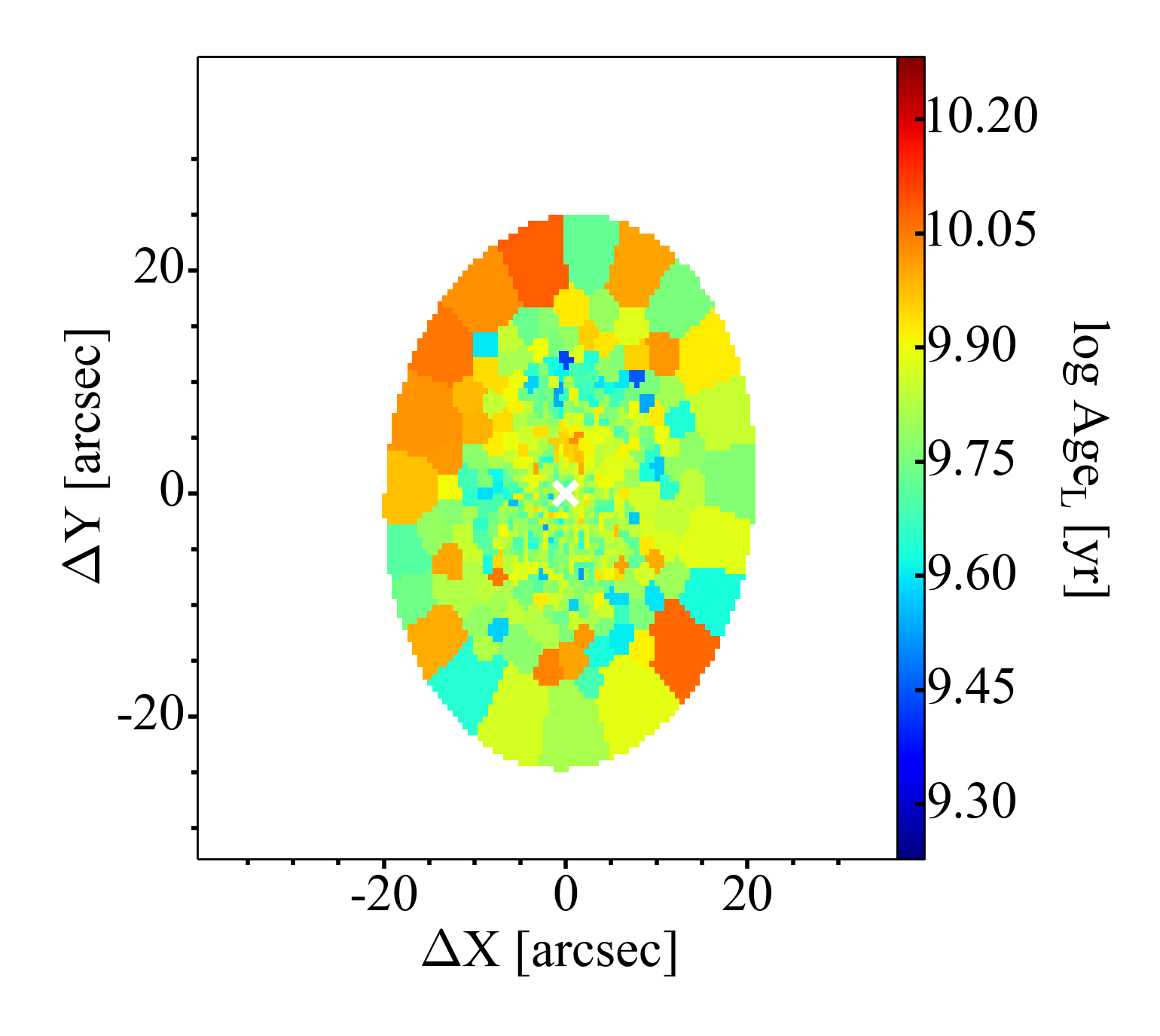}
\includegraphics[width=0.20\columnwidth]{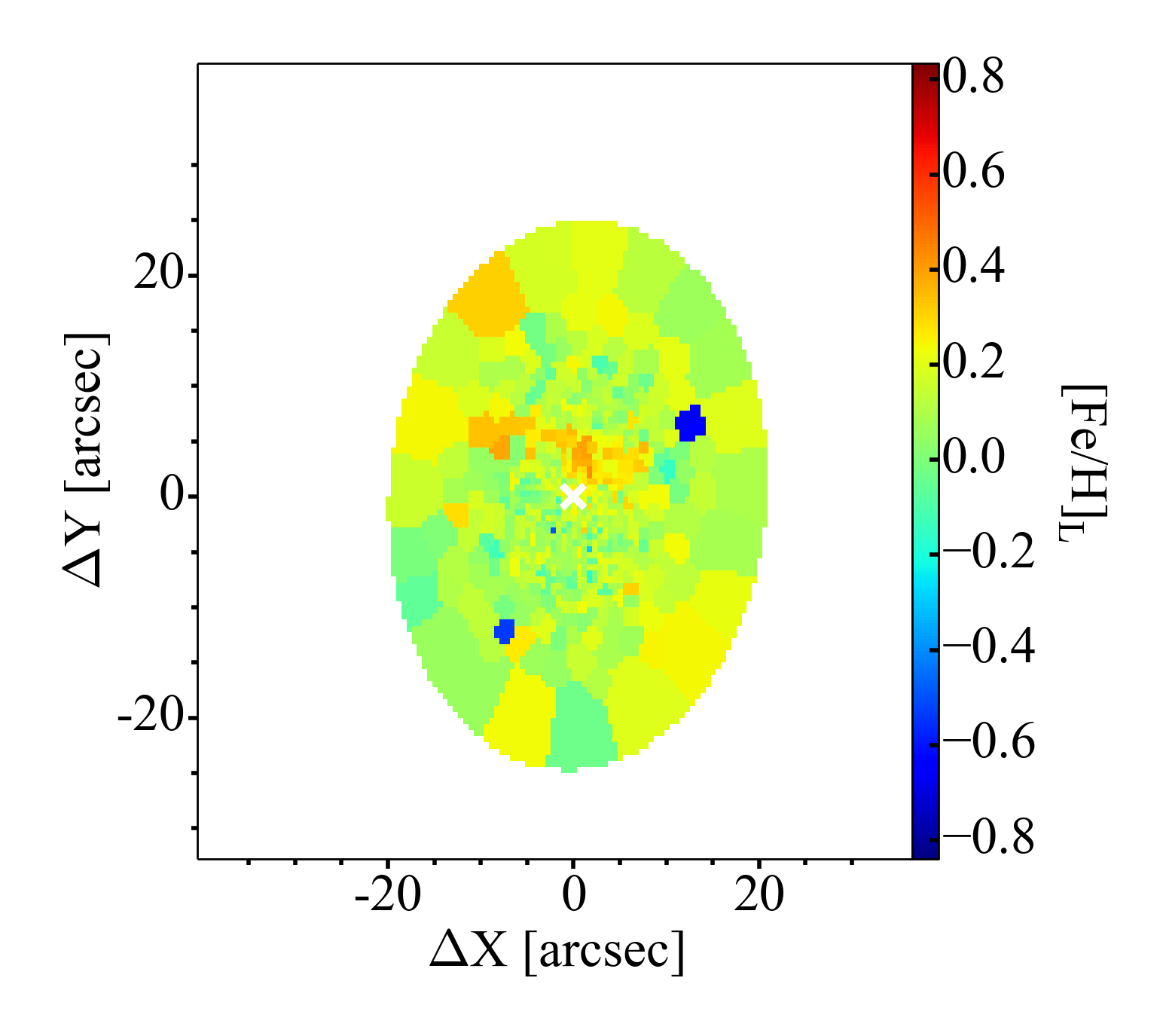}

\includegraphics[width=0.15\columnwidth, angle=90]{CALIFA_STARLIGHT.jpg}
\includegraphics[width=0.20\columnwidth]{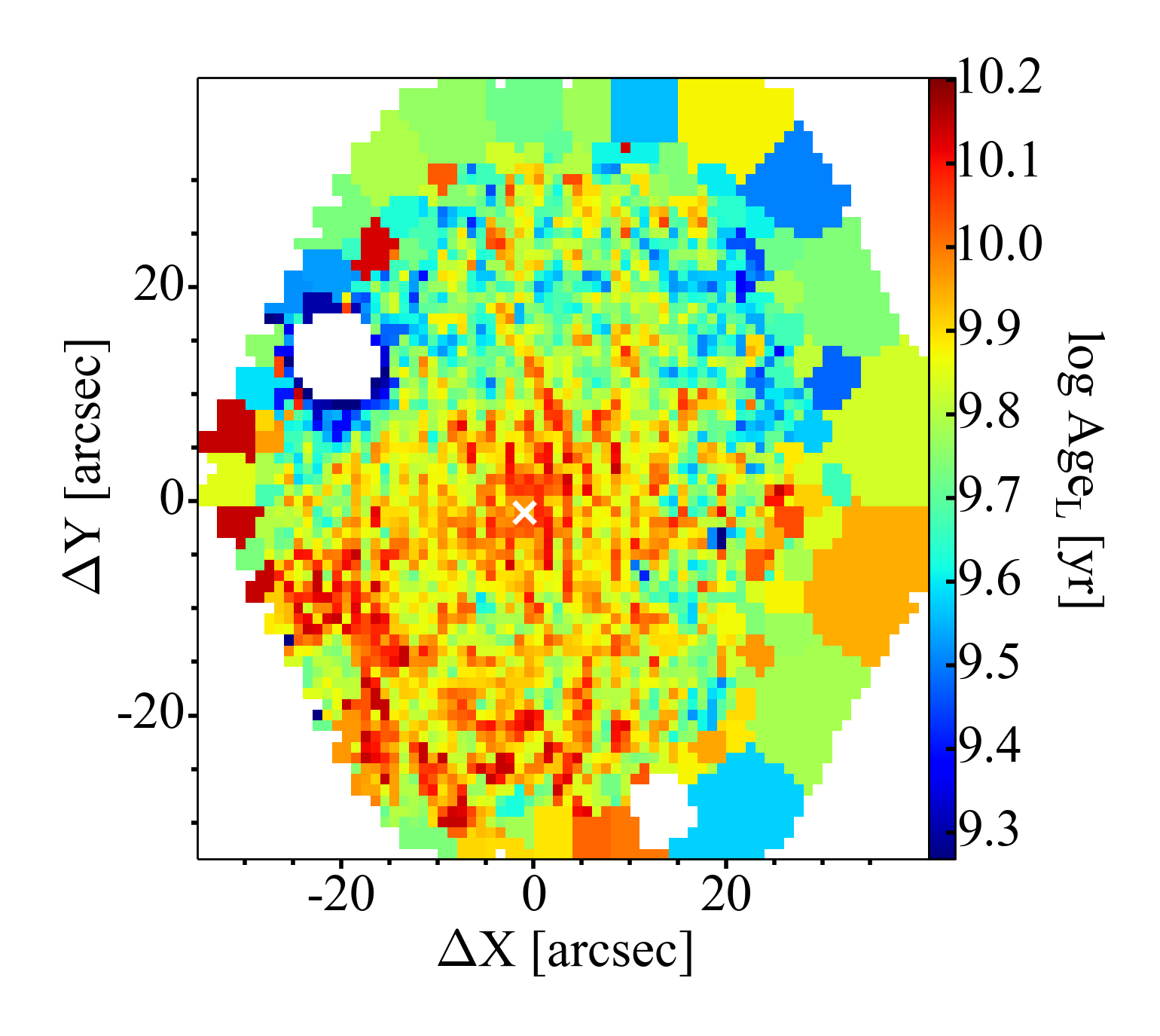}
\includegraphics[width=0.20\columnwidth]{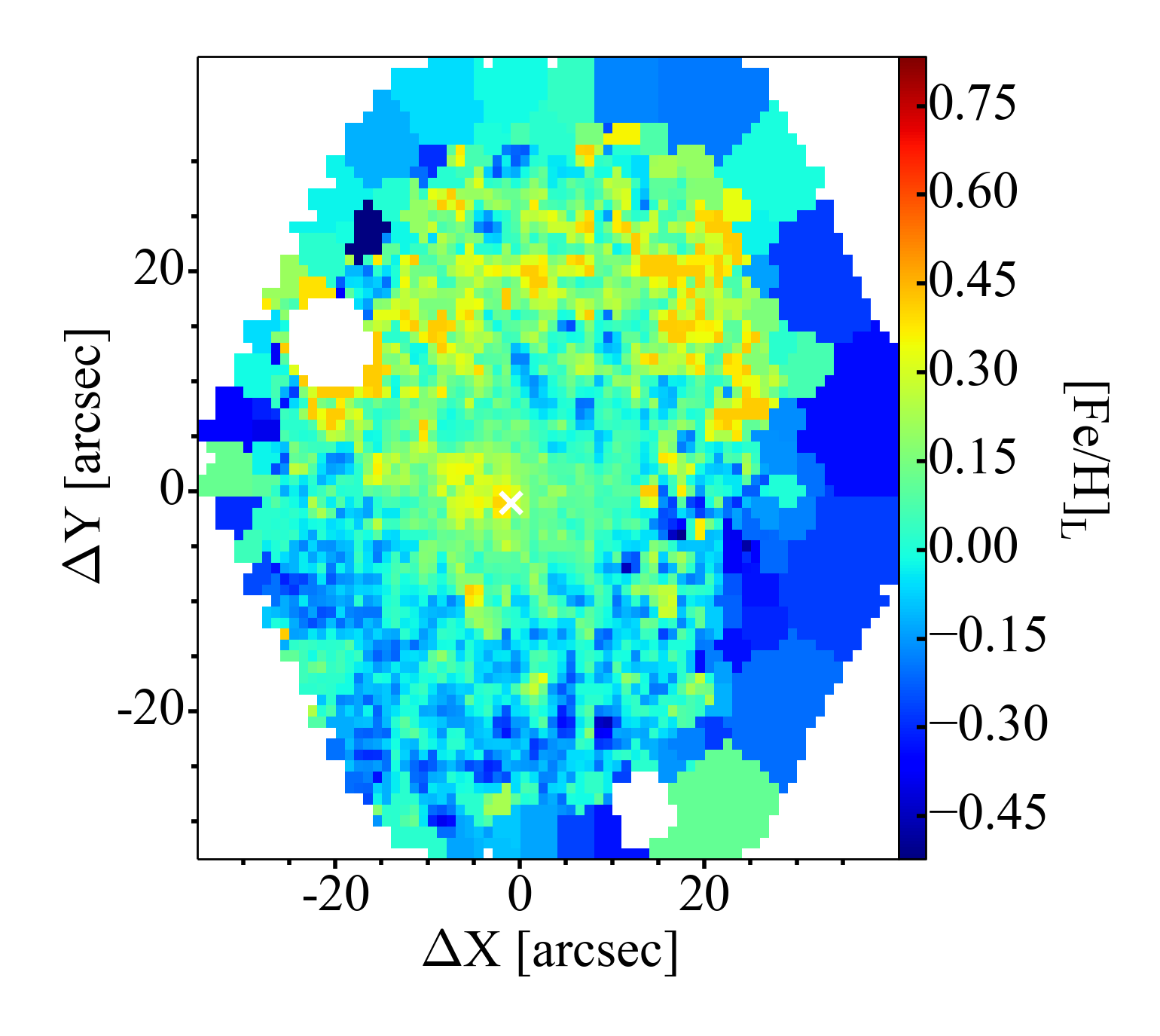}
\includegraphics[width=0.20\columnwidth]{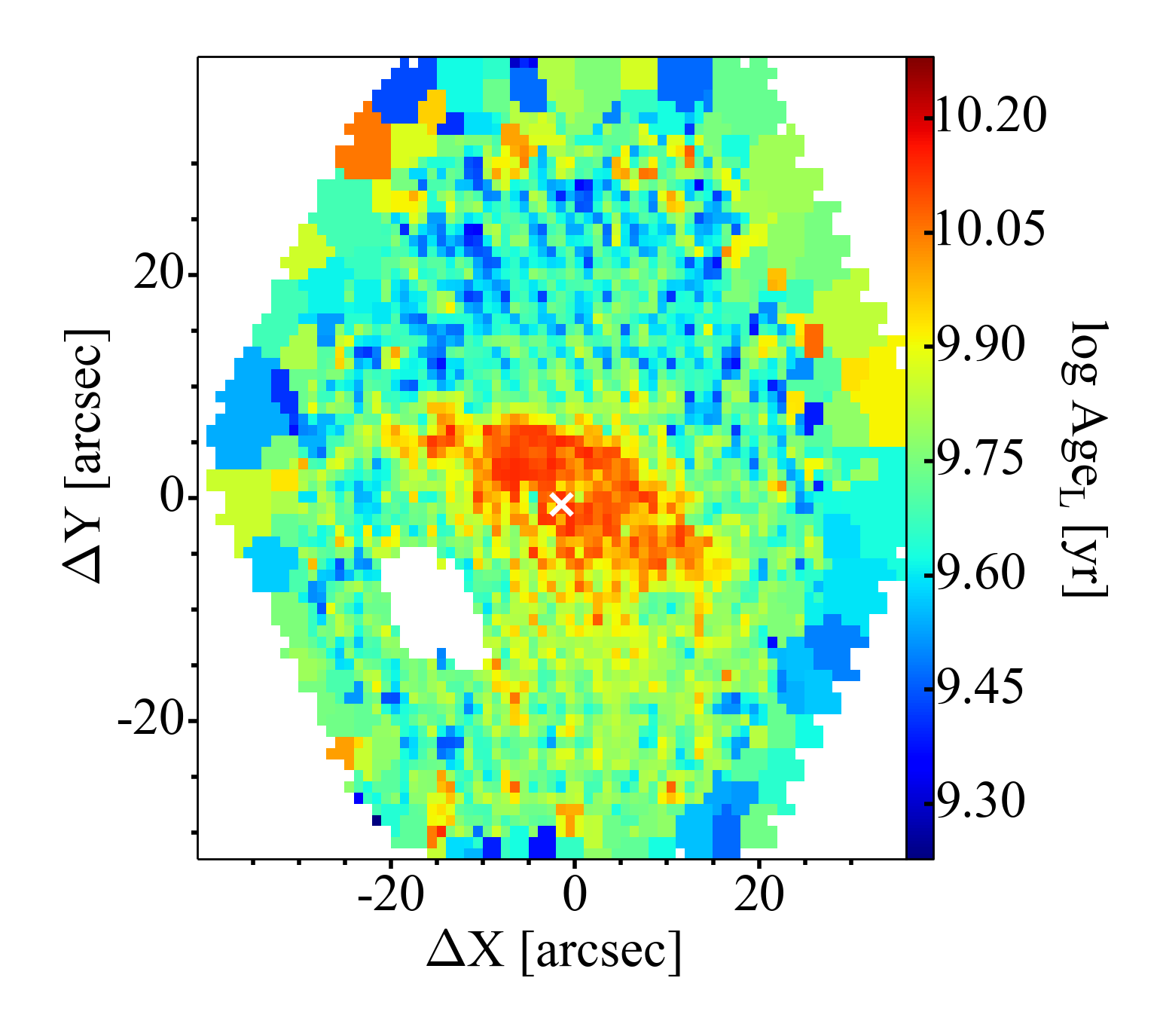}
\includegraphics[width=0.20\columnwidth]{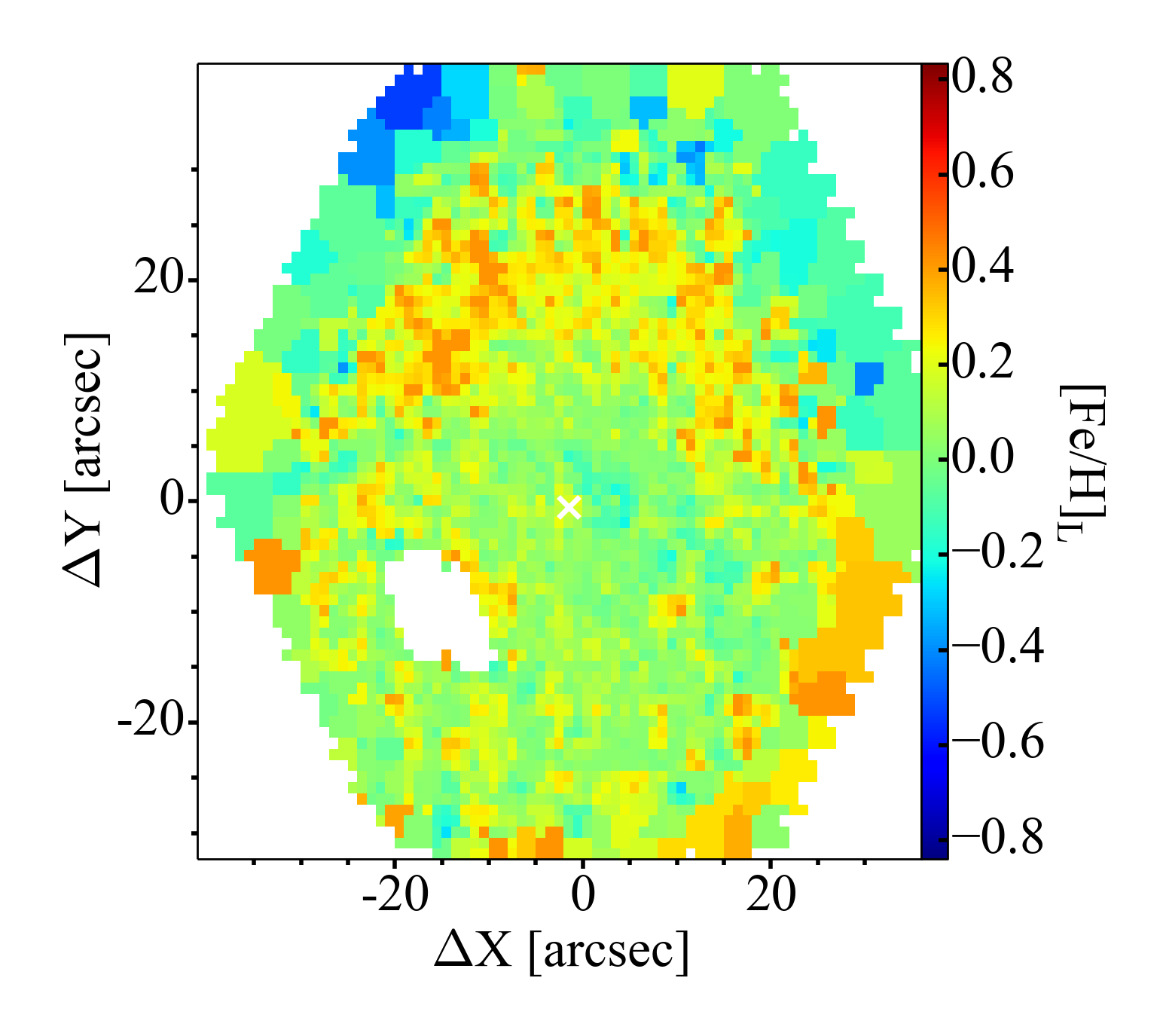}

\includegraphics[width=0.15\columnwidth, angle=90]{CALIFA_STECKMAP.jpg}
\includegraphics[width=0.20\columnwidth]{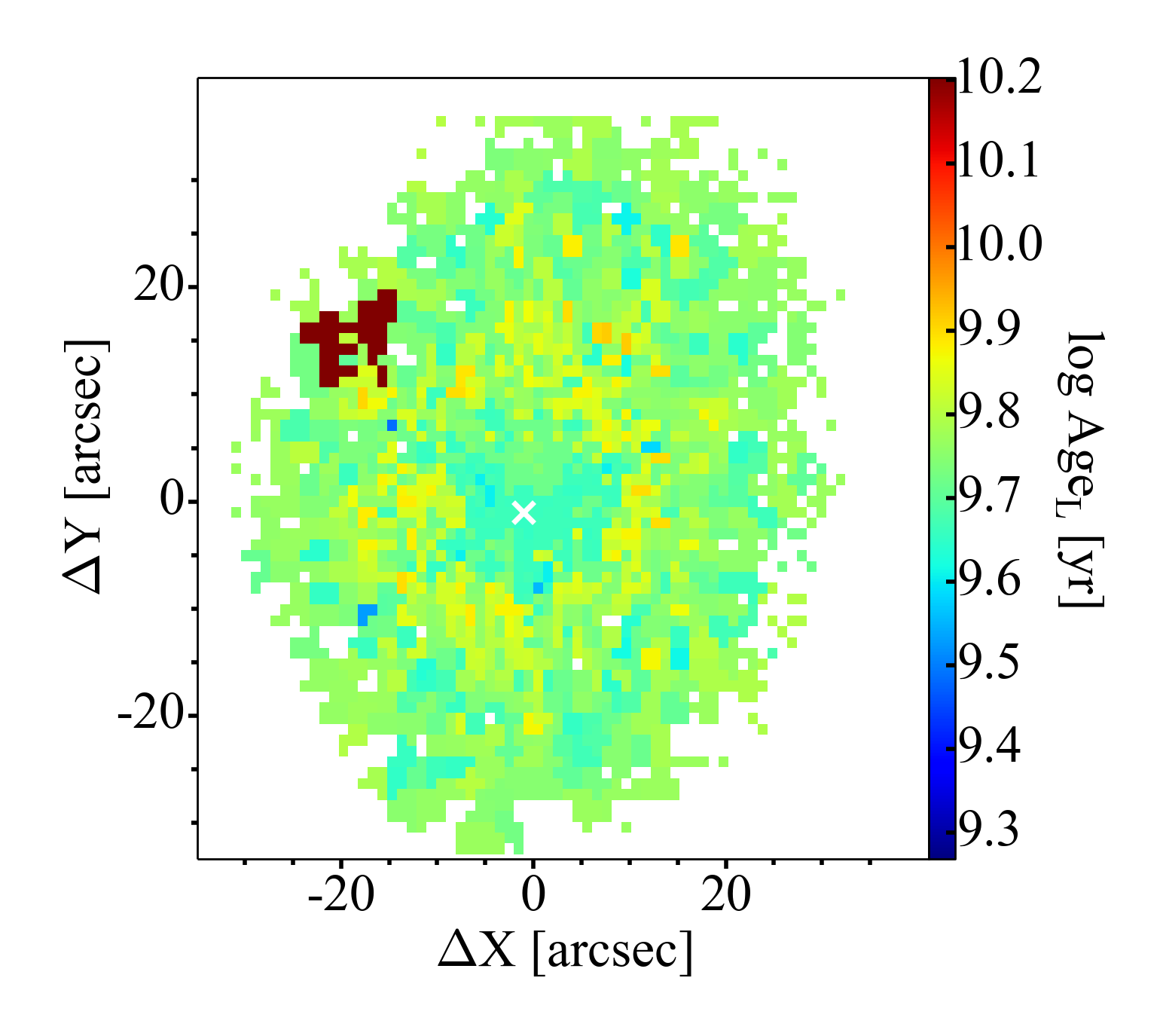}
\includegraphics[width=0.20\columnwidth]{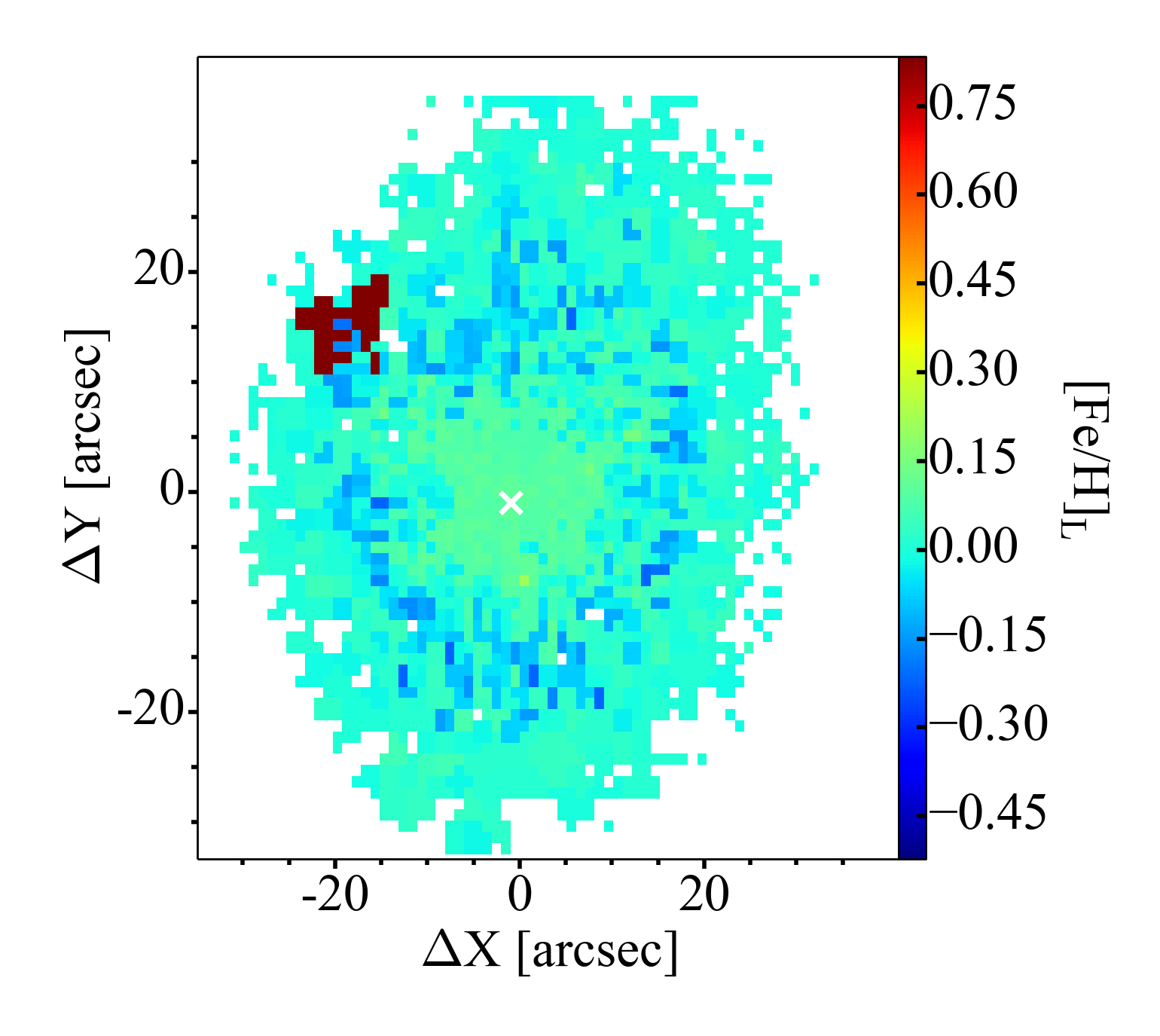}
\includegraphics[width=0.20\columnwidth]{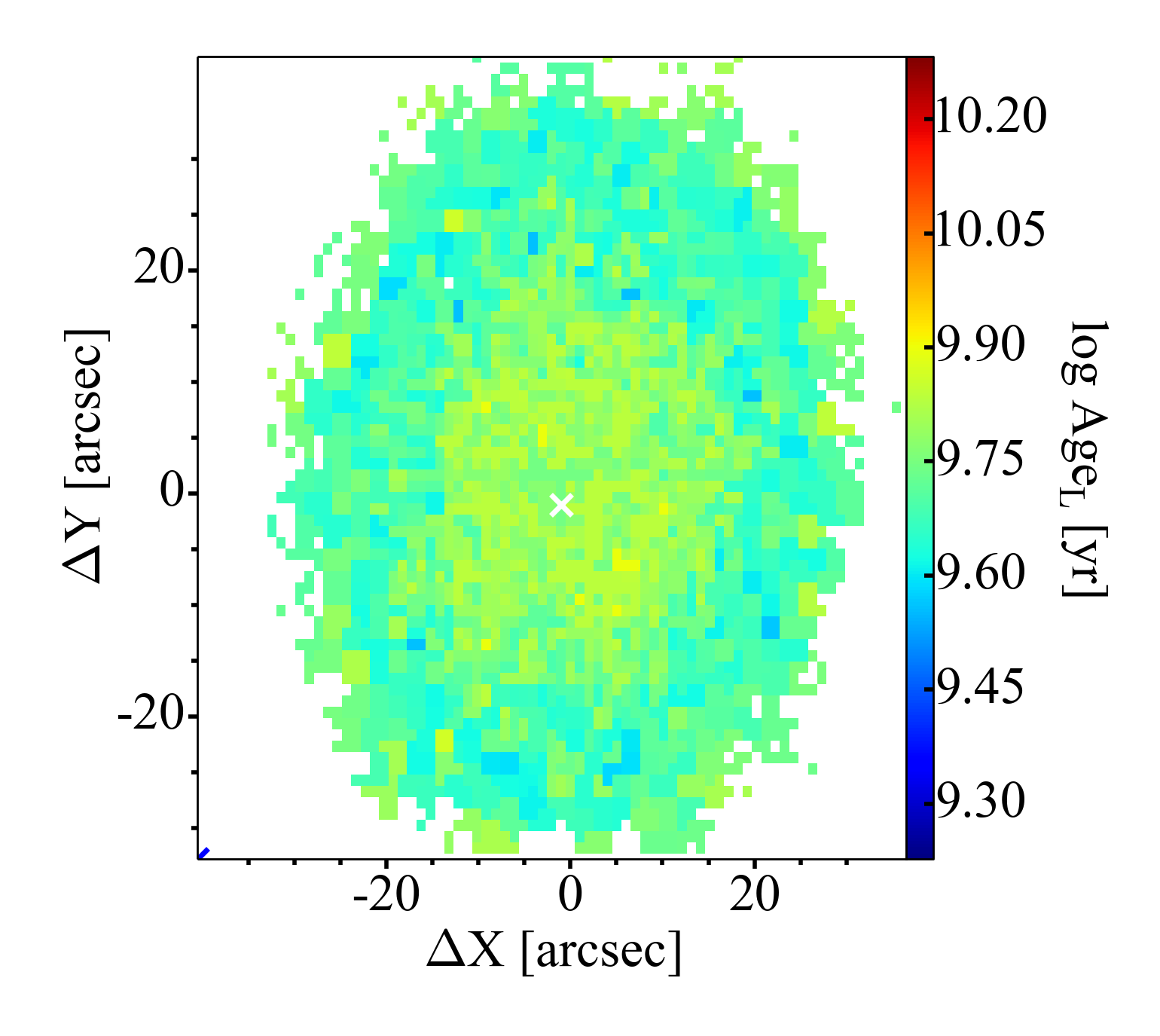}
\includegraphics[width=0.20\columnwidth]{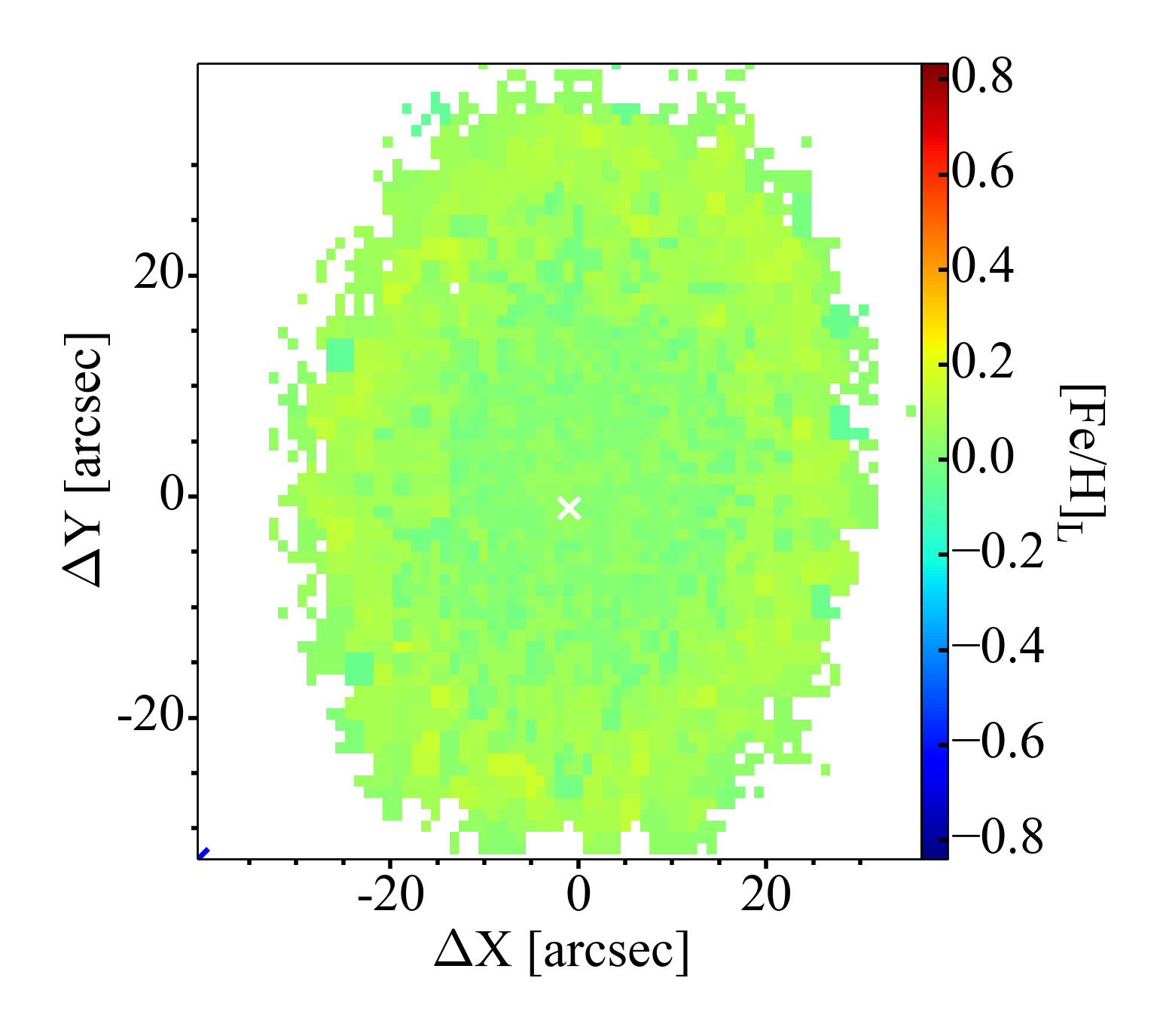}

\caption{Luminosity-weighted stellar population properties maps for NGC5473 (first two columns) and NGC5485 (last two columns) determined by J-PLUS/MUFFIT (first row), CALIFA/STARLIGHT (second row) and CALIFA/STECKMAP (third row). The color range is the same for the different methods. The center of the galaxy is marked with a white cross in each panel.} 
\label{fig:appex}
\end{center}
\end{sidewaysfigure*}


\end{document}